\pdfoutput=1
\documentclass[runningheads,openany]{svmult}

\usepackage{palatino}

\usepackage{euler}

\usepackage{helvet}

\usepackage{graphicx}  

\usepackage{physprbb}  

\usepackage{mathrsfs}

\usepackage{proc}  

\usepackage{epsfig}

\usepackage{amssymb}

\makeindex             

\usepackage{amsmath}   
\usepackage{multicol}
\usepackage{bm}

\usepackage[dvips,cam]{crop}

\let\tocauthor\authorrunning

\begin{document}

\frontmatter



\thispagestyle{empty}
\parindent=0pt

{\Large\sc Blejske delavnice iz fizike \hfill Letnik~9, \v{s}t. 2}

\smallskip

{\large\sc Bled Workshops in Physics \hfill Vol.~9, No.~2}

\smallskip

\hrule

\hrule

\hrule

\vspace{0.5mm}

\hrule

\medskip
{\sc ISSN 1580-4992}

\vfill

\bigskip\bigskip
\begin{center}

{\bfseries 
{\Large  Proceedings to the $11^\textrm{th}$ Workshop}\\
{\Huge What Comes Beyond the Standard Models\\}
\bigskip
{\Large Bled, July 15--25, 2008}\\
\bigskip
}

\vspace{5mm}

\vfill

{\bfseries\large
Edited by

\vspace{5mm}
Norma Manko\v c Bor\v stnik

\smallskip

Holger Bech Nielsen

\smallskip

Dragan Lukman

\bigskip

\vspace{12pt}

\vspace{3mm}

\vrule height 1pt depth 0pt width 54 mm}

\vspace*{3cm}

{\large {\sc  DMFA -- zalo\v{z}ni\v{s}tvo} \\[6pt]
{\sc Ljubljana, december 2008}}
\end{center}
\newpage

\thispagestyle{empty}
\parindent=0pt
\begin{flushright}
{\parskip 6pt
{\bfseries\large
                  The 11th Workshop \textit{What Comes Beyond  
                  the Standard Models}, 15.-- 25. July 2008, Bled}

\bigskip\bigskip

{\bfseries\large was organized by}

{\parindent8pt
\textit{Department of Physics, Faculty of Mathematics and Physics,
University of Ljubljana}

}

\bigskip

{\bfseries\large and sponsored by}

{\parindent8pt
\textit{Slovenian Research Agency}

\textit{Department of Physics, Faculty of Mathematics and Physics,
University of Ljubljana}


\textit{Society of Mathematicians, Physicists and Astronomers
of Slovenia}}}
\bigskip
\medskip

{\bfseries\large Organizing Committee}

\medskip

{\parindent9pt
\textit{Norma Manko\v c Bor\v stnik}

\textit{Holger Bech Nielsen}}

\end{flushright}

\setcounter{tocdepth}{0}

\tableofcontents

\cleardoublepage

\chapter*{Preface}
\addcontentsline{toc}{chapter}{Preface}

The series of workshops on "What Comes Beyond the Standard Model?" started
in 1998 with the idea of organizing a real workshop, in which participants
would spend most of the time in discussions, confronting different
approaches and ideas. The picturesque town of Bled by the lake of the
same name, surrounded by beautiful mountains and offering pleasant walks,
was chosen to stimulate the discussions.

The idea was successful and has developed into an annual workshop, which is
taking place every year since 1998.  Very open-minded and fruitful discussions
have become the trade-mark of our workshop, producing several published works.
It takes place in the house of Plemelj, which belongs to the Society of
Mathematicians, Physicists and Astronomers of Slovenia.

In this eleventh workshop, which took place from 15th to 25th of July
2008, we were discussing several topics, most of them presented in
this Proceedings mainly as talks. The main topic was this time the dark matter 
candidates. Is the approach unifying spins and charges, proposed by Norma, 
which is offering the mechanism for generating families and is accordingly 
predicting the fourth family to be possibly seen at LHC and the stable fifth family 
as the candidate to form the dark matter cluster, the right 
way beyond the standard model? Are the clusters of the fifth family members 
alone what constitute the dark matter? Can the fifth family baryons explain the 
observed properties of the dark matter with the direct measurements included? 
What if such a scenario is not confirmed by the direct measurements?

Talks and discussions in our workshop are not at
all talks in the usual way. Each talk or discussions lasted several
hours, divided in two hours blocks, with a lot of questions,
explanations, trials to agree or disagree from the audience or a
speaker side. 
Most of talks are "unusual" in the sense that they are trying to find out new 
ways of understanding and describing the observed  
phenomena.

New this year was the teleconference taking place during the Workshop on the
theme of dark matter. It was organized by the Virtual Institute for Astrophysics
(wwww.cosmovia.org) of Maxim Khlopov with able support by Didier Rouable.
We managed to have ample discussions and we thank in particular Jeffrey Fillipini 
of Berkeley University for discussions on the CDMS experiment.

What science has learned up to now are several effective theories
 which, after making several
starting assumptions, lead to theories (proven or not to be consistent
in a way that they do not run into obvious contradictions), and which
some of them are within the accuracy of calculations and experimental
data, in agreement with the observations, the others might agree with
the experimental data in future, and might answer at least some of the
open questions, left open by the scientific community accepted
effective theories.
We never can say that there is no other theory which generalizes the
accepted "effective theories", and that the assumptions made to come
to an effective theory in (1+3)-dimensions are meaningful also if we
allow larger number of dimensions. It is a hope that the law of Nature
is simple and "elegant", whatever the "elegance" might mean (besides
simplicity also as few assumptions as possible), while the observed 
states are usually not, suggesting that the "effective theories, laws, models" are
usually very complex.

We have tried accordingly also in this workshop to answer some of the
open questions which the two standard models (the electroweak and the
cosmological) leave unanswered, like:

\begin{itemize}
\item Why has Nature made a choice of four (noticeable) dimensions
    while all the others, if existing, are hidden? And what are the
    properties of space-time in the hidden dimensions?

\item How could "Nature make the decision" about breaking of
    symmetries down to the noticeable ones, coming from some higher
    dimension d?

\item Why is the metric of space-time Minkowskian and how is the choice
 of metric connected with the evolution of our universe(s)?

\item Why do massless fields exist at all? Where does the weak scale
    come from?

\item Why do only left-handed fermions carry the weak charge? Why does
    the weak charge break parity?
    
\item Where do families come from?

\item What is the origin of Higgs fields? Where does the Higgs mass
    come from?

\item Can all known elementary particles be understood as different
    states of only one particle, with a unique internal space of
    spins and charges?

\item How can all gauge fields (including gravity) be unified and
    quantized?

\item What is our universe made out of (besides the baryonic matter)?

\item What is the role of symmetries in Nature?
\end{itemize}

We have discussed these and other questions for ten days.  The reader
can see our progress in some of these questions in this proceedings.
Some of the ideas are treated in a very preliminary way.  Some ideas
still wait to be discussed (maybe in the next workshop) and understood
better before appearing in the next proceedings of the Bled workshops.

The organizers are grateful to all the participants for the lively discussions
and the good working atmosphere. This year Workshop mostly took place at 
the neighboring 
Bled School of Management (www.iedc.si). 
We thank it's director, Mr. Metod Dragonja, for kindly offering us 
the use of their facilities. We also thank their staff, particularly 
chief librarian Ms. Tanja Ovin, for providing us with excellent support.\\[1cm]

\parbox[b]{\textwidth}{%
   \textit{Norma Manko\v c Bor\v stnik, Holger Bech Nielsen,}\\
   \textit{Dragan Lukman} 
\hfill\textit{Ljubljana, December 2008}}


\newpage

\cleardoublepage


\mainmatter

\parindent=20pt

\setcounter{page}{1}


\title{Does the Dark Matter Consist of Baryons of New Heavy Stable 
Family Predicted by the Approach Unifying Spins and 
Charges?~\thanks{This talk was sent in a shorter version to Phys. Rev. Lett. 
at $17^{th}$ of Nov. 2008.}}
\author{G. Bregar and N.S. Manko\v c Bor\v stnik}
\institute{%
Department of Physics, FMF, University of Ljubljana\\
Jadranska 19, 1000 Ljubljana, Slovenia}

\authorrunning{G. Bregar and N.S. Manko\v c Bor\v stnik}
\titlerunning{Does the Dark Matter Consist of Baryons of New \ldots} 
\maketitle 
 
\begin{abstract} 
We investigate the possibility that clusters of the heavy family of quarks and leptons with  
zero Yukawa couplings to the 
lower families constitute the dark matter. Such a family 
is predicted by the approach unifying spins and charges.   
We make a rough estimation of properties of  clusters of this new  
family members and study limitations on the family properties due to the cosmological 
and the direct experimental evidences. 
\end{abstract}

\section{Introduction}
Although the origin of the dark matter is 
unknown, its gravitational interaction with the known matter and other cosmological 
observations require that 
a candidate for the dark matter constituent has the following properties:\\
\begin{itemize} 
\item 
i. The scattering amplitude of a cluster of constituents with the ordinary matter 
and among the dark matter clusters themselves must be small enough to be in agreement 
with the observations (so that no effect of such scattering has 
been observed, except possibly in the DAMA experiments~\cite{rita0708}).\\
\item 
ii. Its  density distribution (obviously different from the ordinary matter density 
distribution) within a galaxy is approximately spherically symmetric 
and decreases approximately 
with the second power of the radius of the galaxy. It is extended also far out of the galaxy,   
manifesting the gravitational lensing by galaxy clusters. \\
\item 
iii. The dark matter constituents must be stable in comparison with the age of our universe,  
having obviously for many orders of magnitude different time scale for forming (if at all) 
matter than the ordinary matter.\\
\item 
iv. The dark matter constituents and accordingly also the clusters had to have a chance 
to be formed during the evolution of our universe 
so that they contribute today the main part of the matter in the universe. 
The ratio of the dark matter density and the baryon matter density is evaluated to be 5-7.\\
%
\end{itemize}
 
Candidates  for the dark matter constituents may give the explanation for the non 
agreement between the two direct measurements of the dark matter constituents~\cite{rita0708,cdms}, 
provided that they measure a particular candidate. 

There are several candidates for the massive dark matter constituents in the literature, 
known as WIMPs 
(weakly interacting massive particles), the references can be found in%
~\cite{dodelson,rita0708}. 

In this talk the possibility that the dark matter constituents are 
clusters of a stable (from the point of view of the age of the universe)
family of quarks and leptons is discussed. Such a family is predicted 
by the approach unifying spin and  charges~\cite{pn06,n92,n93,n07bled,gmdn07}, proposed by  
N.S.M.B.. 

There are several attempts in the literature to explain the origin of families, 
all  in one or another way just postulating that there are at least three families, as 
does the standard model of the electroweak and colour interactions. 
Proposing the (right) mechanism for generating families 
is therefore one of the most promising guides to understanding   
physics beyond the standard model. 
The approach unifying spins and charges is 
offering a mechanism for the appearance of families. It introduces the second 
kind~\cite{pn06,n92,n93,n07bled,hn02hn03} 
of the Clifford algebra objects, which generates families by defining the 
equivalent representations 
with respect to the Dirac spinor representation~\footnote{If the families can not be 
explained by the second kind of the Clifford algebra objects as predicted by the author 
of the approach (S.N.M.B.), it should then be showed, why do the Dirac 
Clifford algebra objects play the very essential role in the description of fermions, while 
the second kind of the Clifford algebra objects does not at all.}.  
The approach predicts more than the observed three families. It predicts 
two times  four families with masses several orders of magnitude bellow the 
unification scale of the three observed charges.  
Since due to the approach (if a particular way of breaking the starting symmetry is assumed) 
the fifth family decouples in the Yukawa couplings from 
the lower four families~\cite{gmdn07}, 
the quarks and the leptons of the fifth family are stable as required by the condition iii.. 
Since the masses of 
the fifth family lie much above the known three  and the predicted fourth family masses 
(the fourth family might according to the first very rough estimates  be even seen at LHC), 
the baryons made out of the fifth family are heavy, forming small enough  clusters, so that 
 their scattering amplitude among themselves and with the ordinary matter  
is small enough and also the number of clusters forming 
the dark matter is low enough to fulfil the conditions i. and iii..

We make a rough estimation of properties of clusters of the members of the fifth family 
($u_5,d_5,\nu_5,e_5$), which in the approach unifying spin and charges have all the 
properties of the lower four families: the same family members with the same charges 
$U(1), SU(2)$ and $SU(3)$, and interact correspondingly 
with the same gauge fields. 

We use a simple (the Bohr-like) model~\cite{gnBled07} to estimate
the size and the binding energy of the fifth family baryons, assuming that the fifth family 
quarks are heavy enough to interact mainly exchanging one gluon. 
We estimate the behaviour of quarks and anti-quarks of the fifth family under the assumption 
that %
during the evolution of the universe quarks and anti-quarks  mostly 
succeeded to form neutral (with respect to the colour and electromagnetic charge) clusters, which 
now form the dark matter. 
We also estimate the behaviour of 
our fifth family clusters if hitting the DAMA/NaI,  DAMA-LIBRA~\cite{rita0708} and CDMS~\cite{cdms} 
experiments.  

All estimates are very approximate and need serious additional studies. Yet we believe 
that such  rough estimations give a guide to further studies. 

\section{The approach unifying spin and charges}

The approach unifying spin and charges~\cite{pn06,n92,n93,n07bled,gmdn07} motivates  
the assumption that clusters of the fifth heavy stable (with respect to the age of the 
universe) family members form the dark matter. 
The approach assumes that in $d\ge(1+13)$-dimensional space a 
Weyl spinor carries nothing but two kinds of spins (no charges): The Dirac spin described by 
$\gamma^a$'s defines the ordinary spinor representation,  the second kind of 
spin~\cite{hn02hn03}  described by $\tilde{\gamma}^a$'s, anticommuting  
with the Dirac one, defines the families 
of spinors~\footnote{There is no  third kind of the Clifford algebra objects: If the Dirac one 
corresponds to the multiplication of any object (any product of the Dirac $\gamma^a$'s included) 
from the left hand side, then the second kind of the Clifford objects correspond (up to a factor) 
to the multiplication of any object from the right hand side.}.  
Spinors interact with  the gravitational gauge fields: vielbeins and two kinds 
of spin connections. 
 A simple starting Lagrange density for a spinor and for  gauge fields in $d=1+13$
manifests, after the appropriate breaks of symmetries, in $d=1+3$ all the properties of 
the spinors (fermions) and the gauge fields assumed 
by the standard model of the electroweak and colour interaction, with the 
Yukawa couplings included.  The  approach offers accordingly 
the explanation for the appearance of families (see ref.~\cite{pn06,n92,n93,n07bled,gmdn07}
and the references cited in these references)
and predicts two times four families with zero  (that is negligible with respect 
to the age of the universe) Yukawa couplings among the two groups of families at 
low energy region. In the very rough estimations~\cite{gmdn07}
the fourth family masses are predicted to be at rather low energies (at around 250 GeV or higher). 
so that it might be seen at LHC. 
The lightest of the next four  families is the candidate to form the dark matter~\footnote{
If the approach unifying spin and charges is, by using the second kind of 
the Clifford algebra objects, offering the right explanation for 
the appearance of families~\cite{pn06,n07bled,gmdn07,hn02hn03},  
as does the first kind  describe the spin and all the charges, then 
more than three observed families must exist and the fifth family appears as a natural 
explanation for the dark matter.}. The energy range, in which the masses of  the fifth family 
quarks might appear, is far above $300 $ GeV (say higher than $10^{4} \; {\rm GeV}$ and much lower than the 
scale of the break of $SO(4)\times U(1)$ to $SU(2) \times U(1)$, which might 
occur at $10^{13}$ GeV~\cite{hnproc02}). \\

\section{ Properties of clusters of a heavy family}

Let us  assume that there is a heavy family of quarks and leptons as 
predicted by the approach unifying spins and charges: i. It has masses several orders of 
magnitude greater than the known three families. ii.   The matrix elements with 
the lower families in the mixing matrix (the Yukawa couplings) are equal to zero. 
iii. All the charges ($SU(3), SU(2), U(1)$ and correspondingly after the break of the 
electroweak symmetry $SU(3), U(1)$) are those of the known families and so are 
accordingly also the couplings to the gauge fields.  
Families distinguish among themselves  
in the family index (in the quantum number, which in the approach is determined  
by the operators 
$\tilde{S}^{ab}=\frac{i}{4}(\tilde{\gamma}^a \tilde{\gamma}^b - \tilde{\gamma}^b 
\tilde{\gamma}^a)$), and (due to the Yukawa couplings)  in their masses.

For a heavy enough family the properties of baryons (neutrons $n_5$ $(u_5 d_5 d_5)$, 
protons $(u_5 u_5 d_5)$, $\Delta_{5}^{-}$ $(u_5 u_5 d_5)$, $\Delta_{5}^{++}$ $(u_5 u_5 u_5)$,
e.t.c), made out of the quarks $u_5$ and $d_5$ can be estimated by using the non 
relativistic Bohr-like model 
with the $\frac{1}{r}$ (radial) dependence of the potential  
between a pair of quarks  $V= - \frac{3 \alpha_c}{r}$, where $\alpha_c$ is in this case the 
colour (3 for three possible colour charges) coupling constant. 
Equivalently goes for anti-quarks. 
This is a meaningful approximation as long as   
one gluon exchange contribution to the interaction among quarks 
is a dominant contribution (which means: as long as excitations  of 
a cluster are not influenced by  the linearly rising part of the potential).

Which one of $p_5$, $n_5$ or maybe $\Delta^-$  or  $\Delta^{++}$
is a stable fifth family baryon, depends on the ratio of the bare masses 
$m_{u_5}$ and  $m_{d_5}$, as well as on the  weak and 
electromagnetic interactions among quarks. If $m_{d_5}$ is appropriately 
lighter than $m_{u_5}$ so that the repulsive 
weak and electromagnetic interactions favors the neutron $n_5$, then $n_5$ is 
a colour singlet electromagnetic chargeless stable cluster of quarks with the 
lowest mass among the nucleons of the fifth family, with the weak charge $-1/2$. 

If $m_{d_5}$ is heavier (enough, due to stronger electromagnetic repulsion among 
the two $u_5$ than among the two $d_5$) than $m_{u_5}$, the proton $p_5$, which is 
a colour singlet stable nucleon, needs the electron $e_5$ or $e_1$ to form  
an electromagnetic chargeless cluster. (Such an electromagnetic and colour  chargeless 
cluster has also the expectation value of the weak charge equal to zero.)

An atom made out of only fifth family members might be lighter or not than $n_5$, 
depending on the masses of the fifth family members. We shall for simplicity assume 
in this first rough estimations 
that $n_5$ is a stable baryon and equivalently also $\bar{n}_5$, leaving all the other 
possibilities for further studies~\cite{gmn08}. 

In the Bohr-like model, when neglecting more than one gluon exchange contribution 
(the simple bag model evaluation does not contradict such a simple model~\footnote{
A simple bag model 
with the potential $V(r)=0 $ for $r<R$ 
and $V(r)= \infty $ otherwise, supports our rough estimation. It, namely, 
predicts for the lowest energy $E$ (the mass) of a cluster of three quarks:  
$E= 3 \,m_{q_5} c^2 (1+ (x \hbar c /m_{q_5}c^2 R)^2),$ with $\tan x = x/ [1-
(m_{q_5}c^2 R/\hbar c) - \sqrt{x^2 + (m_{q_5}c^2 R/\hbar c)^2} \,], $ where $2.04< x <\pi$ 
for $0 < (m_{q_5}c^2 R/ \hbar c) < \infty$. For $\,(m_{q_5}c^2 R/ \hbar c) \approx 8$, 
for example, is $x$ close to $3$ and rises very slowly to $\pi$. Accordingly 
the mass of the three quark cluster is close to three masses of the quark, provided 
that $R $ is assumed to be as calculated by the Bohr-like model. }, while the electromagnetic 
and weak interaction contribution is  more than $10^{-3}$ times smaller) the  
binding energy and the average radius are equal to 
\begin{eqnarray}
\label{bohr}
E_{c_{5}} = -\frac{3}{2}\, m_{q_5} c^2 (3\alpha_{c})^2,
\quad r_{c_{5}} =  \frac{\hbar c}{3 \alpha_c m_{q_5} c^2}
\end{eqnarray}
The mass of the cluster is approximately $m_{c_5}c^2 = 
3 m_{q_5} c^2(1-\frac{1}{2} (3 \alpha_c)^2)$ (if $n_5$ is 
the stable baryon, since we take that the space part of the wave function 
is symmetric and also the spin and the weak charge part, each of a mixed symmetry,  
couple to symmetric wave function, we neglect the weak and the electromagnetic interaction). 
Assuming that the coupling constant   
of the colour charge  $\alpha_c$   runs with the kinetic energy $E$ of quarks as in 
ref.~\cite{greiner}  with the number of flavours $N_F=8$
($\,\alpha_c(E^2)=\frac{\alpha_c(M^2)}{1+\frac{\alpha_c(M^2)}{4 \pi} (11-\frac{2 N_F}{3}) 
\textrm{ln}(\frac{E^2}{M^2}) }$, with $\alpha_{c}((91 \; \textrm{GeV})^2)=0.1176(20)$ ) 
we estimated  the properties of a baryon as presented on Table~\ref{snmb2dmTableI}. 

\begin{table}
\begin{center}  
\begin{tabular}{||l||l|l|l|l|}
\hline
$\frac{m_{q_5} c^2}{{\rm TeV}}$& $\alpha_c$ & $\frac{E_{c_5}}{{\rm TeV}}$& 
$\frac{r_{c5}}{10^{-7}{\rm fm}}$ 
& $\frac{\pi r_{c5}^2}{(10^{-7}{\rm fm})^{2} }$
\\
\hline
\hline
$10^2$ & 0.09   & 5.4                & 150      & $6.8 \cdot 10^4$           \\
\hline
$10^4$ & 0.07   & $3 \cdot 10^2 $    &  1.9     & 12                                 \\
\hline
$10^6$ & 0.05   & $2 \cdot 10^4 $    & 0.024    & $1.9 \cdot 10^{-3}$    \\
\hline
\end{tabular}
\end{center}
\caption{\label{snmb2dmTableI} Properties of a cluster of the fifth family quarks 
within the Bohr-like model. 
$m_{q_5}$ in TeV/c$^2$ is the assumed fifth family quark mass,
$\alpha_c$ is the coupling constant 
of the colour interaction at $E\approx (- E_{c_{5}}/3)$ (Eq.(\ref{bohr})),  
which is the kinetic energy 
of the quarks in the cluster, 
$r_{c5}$ is the corresponding Bohr-like radius,  
$\sigma_{c5}=\pi r_{c5}^2 $  
is the corresponding scattering cross section for a chosen quark mass.} 
\end{table}
The binding energy is approximately of two orders of magnitude smaller than the mass 
of the cluster. The $n_5$ ($u_{q_5} d_{q_5} d_{q_5}$) cluster is lighter than cluster $p_{5}$ 
($u_{q_5} d_{q_5} d_{q_5}$) if $(m_{u_5}-m_{d_5})$ is smaller then 
$(0.6,60,600)$ GeV for the three 
values of the $m_{q_5}$ on Table~\ref{snmb2dmTableI}, respectively. We clearly see that the 
''nucleon-nucleon force''  among the fifth family baryons leads to for many orders of 
magnitude smaller scattering than among the first family baryons. 

 The scattering cross section between two clusters of the fifth family quarks is 
 determined by the weak interaction as soon as the mass   exceeds  several GeV.

If a cluster of the heavy (fifth family) quarks and leptons and  of the 
ordinary (the lightest) family is made, 
 then, since ordinary family   dictates the radius and the excitation energies  
 of a cluster, its 
 properties are not far from the properties of the ordinary hadrons and atoms, except that such a  
  cluster has the mass dictated by the heavy family members. \\

\section{ Dynamics of a heavy family clusters in our galaxy} 

The density of the dark 
matter $\rho_{dm}$ in the Milky way can be evaluated from the measured rotation velocity  
of  stars and gas in our galaxy, which is approximately constant (independent of the distance from the 
center of our galaxy). For our Sun this velocity 
is $v_S \approx (170 - 270)$ km/s. Locally $\rho_{dm}$ is known within a factor of 
10  to be 
$\rho_0 \approx 0.3 \,{\rm GeV} /(c^2 \,{\rm cm}^3)$, 
we put $\rho_{dm}= \rho_0\, \varepsilon_{\rho},$ 
with $\frac{1}{3} < \varepsilon_{\rho} < 3$. 
The local velocity of the dark matter cluster $\vec{v}_{dm}$ is model dependant. 
In a simple model that all the clusters at any radius $r$ from the center 
of our galaxy rotate in circles way around the center, so that the paths are 
spherically symmetrically distributed, the velocity of a cluster at the position of 
the Earth is equal to $v_{S}$, the velocity of our Sun in the absolute value,
but has all possible orientations perpendicular to the radius $r$ with  equal probability.
In the model 
that all the clusters oscillate through the center of the galaxy, 
the velocities of the dark matter clusters at the Earth position have values from 
zero to the escape velocity, each one weighted so that all the contributions give  
$ \rho_{dm} $. Also the model  that clusters make all possible paths 
from the oscillatory one to the circle, weighted so that they reproduce the $\rho_{dm}$,
seems acceptable. Many other possibilities are presented in the references of~\cite{rita0708}. 

The velocity of the Earth around the center of the galaxy is equal to:  
$\vec{v}_{E}= \vec{v}_{S}+ \vec{v}_{ES} $, with $v_{ES}= 30$ km/s. 
Then the velocity with which the dark matter hits the Earth is equal to:  
$\vec{v}_{dmE\,i}= \vec{v}_{dm\,i} - \vec{v}_{E}$, where the index $i$ stays for  
  the $i$-th velocity class. 

Let us evaluate the cross section for a  heavy dark matter cluster to elastically scatter 
on an ordinary nucleus with $A$ nucleons  in the 
Born approximation: 
$\sigma_{c_5 A} = 
\frac{m_{A}^2}{\pi \hbar^2} <|M_{c_5 A}|>^2$. 
For our heavy dark matter cluster 
with a small cross section from Table~\ref{snmb2dmTableI} is  
$m_{A}  $  approximately the mass of the ordinary nucleus. 
If the mass of the cluster is around $1$ TeV or more and its velocity 
$\approx v_{S}$, is $\lambda= \frac{h}{p_A}$  for a nucleus large enough to make 
scattering totally coherent.  The cross section is  almost independent of the recoil 
velocity of the nucleus. (We are studying this problem intensively.) 
For masses of quarks $m_{q_5}< 10^4$ TeV
(when the ''nucleon-nucleon force'' dominates) is the cross section proportional to $(3A)^2$ 
(due to the square of the matrix element) times $(A)^2$ (due to the mass of the nuclei 
$m_A\approx 3 A \,m_{q_1}$, with $m_{q_1}$  which is  the first family dressed quark mass), 
so that $\sigma(\vec{v}_{dmE i}, A)= \sigma(A) \propto A^4$. Estimated with the Bohr-like model 
(Table~\ref{snmb2dmTableI}) $\sigma(\vec{v}_{dmE i}, A)= \sigma_{0}\, 
\varepsilon_{\sigma}\, A^4,$ 
with  $\frac{1}{30} < \varepsilon_{\sigma} < 30$ and $\sigma_{0}= 9\,\pi r_{c_5}^2 $. 
For masses of the fifth family quarks $m_{q_5}> 10^4$ TeV, 
the weak interaction starts to dominate. In this case the scattering cross section 
is $\sigma(\vec{v}_{dmE i}, A)=   (\frac{m_{n_1} \,A\,(A-Z) G_F}{\sqrt{2 \pi}} )^2
\varepsilon_{\sigma_{weak}}$
($\approx ( 10^{-6}\,  {\rm fm} \, \, A^2 \, \frac{A-Z}{ A})^2 \,
\varepsilon_{\sigma_{weak}}$) $=\sigma_{0} \, A^4 \, \varepsilon_{\sigma_{weak}}$,
with $\sigma_0 = (\frac{m_{n_1} \,(A-Z) G_F}{A\, \sqrt{2 \pi}} )^2$ and 
$ \varepsilon_{\sigma_{weak}} \approx 1$ (the weak force is pretty accurately 
evaluated, however,  taking into account the threshold of a measuring apparatus may change the 
results obtained with the averaging assumptions presented above).

We find accordingly for the flux per unit time and unit surface of our 
(any heavy with the small enough cross section) dark matter clusters 
hitting the Earth   
$\Phi_{dm} = \sum_i \,\frac{\rho_{dm i}}{m_{c_5}}  \,
|\vec{v}_{dm i} - \vec{v}_{E}|  $ 
to be approximately  (as long as $\frac{v_{ES}}{|\vec{v}_{dm i}- \vec{v}_S|}$ is small
) equal to: 
\begin{eqnarray}
\label{flux}
\Phi_{dm}\approx \sum_i \,\frac{\rho_{dm i}}{m_{c_5}}  \,
\{|\vec{v}_{dm i} - \vec{v}_{S}| - \vec{v}_{ES} \cdot \frac{\vec{v}_{dm i}- \vec{v}_S}{
|\vec{v}_{dm i}- \vec{v}_S|} \}.
\end{eqnarray}
We neglected further terms. The flux is very much model dependent. 
We shall approximately take that
$$\sum_i \, |\vec{v_{dm i}}- \vec{v_S}| \,\rho_{dm i} = \varepsilon_{v_{dmS}} 
\, \varepsilon_{\rho}\,  v_S\, \rho_0 ,$$ 
(with $\rho_0 = 0.3 \, {\rm GeV}/(c^2\, {\rm cm}^3) $) 
while we estimate 
 $$ \sum_i \, \vec{v}_{ES}  \cdot \frac{\vec{v}_{dm i}- \vec{v}_S}{
|\vec{v}_{dm i}- \vec{v}_S|} = v_{ES} \varepsilon_{v_{dmES}}
\cos \theta \, \sin \omega t ,$$ 
$\theta = 60^0$, 
$\frac{1}{3} < \frac{\varepsilon_{v_{dmES}}}{\varepsilon_{v_{dmS}}} < 3$ 
and $\omega $
determined by one year rotation of our Earth around our Sun.

\section{Direct measurements of the fifth family  baryons as dark matter constituents} 

Assuming that the  
 DAMA~\cite{rita0708} and CDMS~\cite{cdms} experiments are measuring  
 the fifth family neutrons, we are estimating properties 
 of $q_5$ ($u_5,d_5$). 
 We discussed our rough estimations with Rita 
 Bernabei~\cite{privatecommRBJF} and Jeffrey Filippini~\cite{privatecommRBJF} and both were very 
 clear that one can hardly compare both experiments (R.B. in particular), 
 and that the details about the way how do the dark matter constituents scatter on the nuclei and 
 with which velocity 
 do they scatter (in ref.~\cite{rita0708} such studies were done)  
 as well as how does a particular 
 experiment measure events are very important, 
 and that 
 the results depend very significantly on the details, which might change the 
 results for orders of magnitude. 
 We are completely aware of how rough our estimation is, %
 yet we see, since the number of measuring events is inversely proportional  
 to the third power of clusters' mass when the ''nuclear force'' dominates for 
 $m_{q_5}< 10^4$ TeV,  that even such rough  estimations   
 may in the case of our (any) heavy dark matter clusters say, whether both experiments
 do at all measure our (any) heavy family clusters, if one experiment 
 clearly sees  the dark matter signals and the 
 other does not (yet?).

 Let $N_A$ be the number of nuclei of type $A$ in the detectors  
 (of either DAMA~\cite{rita0708}, which has $4.0 \cdot 10^{24}$ nuclei of $I$, with $A_I=127$ nuclei 
 per kg and the same number 
 of $Na$, with $A_{Na}= 23$ or of CDMS~\cite{cdms}, which has $8.3 \cdot 10^{24}$ $Ge$ nuclei, 
 with $A_{Ge}=73$ per kg). 
 At velocities  of a dark matter cluster  $v_{dmE}$ $\approx$ $200$ km/s  
 are the $3A$ scatterers strongly bound in the nucleus,    
 so that if hitting one quark the whole nucleus with $A$ nucleons recoils  
 and accordingly  elastically scatters on a 
 heavy dark matter cluster.  
Then the number of events per second  ($R_A$) taking place 
in $N_A$ nuclei   is  due to Eq.~\ref{flux} and the recognition that the cross section 
is at these energies almost independent 
of the velocity (and depends accordingly only  on $A$ of the nucleus),  equal to
\begin{eqnarray}
\label{ra}
R_A = \, N_A \,  \frac{\rho_{0}}{m_{c_5}} \;
\sigma(A) \, v_S \, \varepsilon_{v_{dmS}}\, \varepsilon_{\rho} \, ( 1 + 
\frac{\varepsilon_{v_{dmES}}}{\varepsilon_{v_{dmS}}} \, \frac{v_{ES}}{v_S}\, \cos \theta
\, \sin \omega t).
\end{eqnarray}
Let $\Delta R_A$ mean the amplitude of the annual modulation of $R_A$ 
$\Delta R_A = R_A(\omega t = \frac{\pi}{2}) - R_A(\omega t = 0)$. Then 
$R_A (\sin \omega t =1)= N_A \, R_0 \, A^4\, 
\frac{\varepsilon_{v_{dmES}}}{\varepsilon_{v_{dmS}}}\, \frac{v_{ES}}{v_S}\, \cos \theta$, where 
$ R_0 = \sigma_{0} \, \frac{\rho_0}{3  \,
m_{q_5}} \,  v_S\, \varepsilon$, and $R_0$ is for the case that the ''nuclear force'' 
dominates $R_0 =  \pi\, (\frac{\hbar\, c}{\alpha_c \, m_{q_5}\, c^2})^2\, 
\frac{\rho_0}{m_{q_5}} 
\, v_S\, \varepsilon$, with $\varepsilon = 
\varepsilon_{\rho} \, \varepsilon_{v_{dmES}} \varepsilon_{\sigma} $. $R_0$ is therefore 
proportional to $m_{q_5}^{-3}$. We estimated  $\frac{1}{300} < \varepsilon < 300$,    
which demonstrates both, the uncertainties in the knowledge about the dark matter dynamics 
in our galaxy and our approximate treating of the dark matter properties.  
When for $m_{q_5} \, c^2 > 10^4$ TeV the weak interaction determines the cross section, 
$R_0 $ is in this case proportional to $m_{q_5}^{-1}$. 

We estimate that an experiment with $N_A$ scatterers  should  measure 
$R_A \varepsilon_{cut}$, with $\varepsilon_{cut}$ determining  the efficiency  of 
a particular experiment to detect a dark matter cluster collision. $\varepsilon_{cut}$ takes 
into account the threshold of a detector. %
Although the scattering cross section is independent of 
the energy, the number of detected events depends on the velocity of 
the dark matter clusters, due to the angular distribution of the scattered nuclei and due to 
the energy threshold of the detector, which is not included in $\varepsilon_{v_{dmS}}$. 
For small enough $\frac{\varepsilon_{v_{dmES}}}{\varepsilon_{v_{dmS}}}\, 
\frac{v_{ES}}{v_S}\, \cos \theta$ we have 
\begin{eqnarray}
R_A \, \varepsilon_{cut}  \approx  N_{A}\, R_0\, A^4\, 
 \varepsilon_{cut} = \Delta R_A \varepsilon_{cut} \,
 \frac{\varepsilon_{v_{dmS}}}{\varepsilon_{v_{dmES}}} \, \frac{v_{S}}{v_{ES}\, \cos \theta}. 
\label{measure}
\end{eqnarray}

If DAMA~\cite{rita0708}   is measuring 
our (any) heavy  dark matter clusters  
scattering mostly on $I$ (we shall neglect the same number of $Na$,  with $A =23$),  
then 
$$R_{I}\,\varepsilon_{cut\,dama} \approx  \Delta R_{ I} \; \varepsilon_{cut\,dama}\,
\frac{\varepsilon_{v_{dmS}}}{\varepsilon_{v_{dmES}}}\,
\frac{v_{S}  }{v_{SE}\, \cos \theta } .$$  
In this rough estimation 
most of unknowns, except the local velocity of our Sun,  
the cut off procedure ($\varepsilon_{cut\, dama}$) and 
$\frac{\varepsilon_{v_{dmS}}}{\varepsilon_{v_{dmES}}}$,
 are hidden in $\Delta R_{ 0}$. If we assume that the 
Sun's velocity is 
$v_{S}=100, 170, 220, 270$ km/s,  we find   
$\frac{v_S}{v_{SE} \cos \theta}= 7$, $10$, $14$, $18$, 
respectively. The recoil energy of the nucleus $A=I$ changes correspondingly 
with the square of   $v_S $.
DAMA~\cite{rita0708} publishes 
$\varepsilon_{cut \,dama} \, \Delta R_I= 0,052  $ counts per day and per kg of NaI. 
Correspondingly  is $R_I \, \varepsilon_{cut\, dama}  = 
 0,052 \, \frac{\varepsilon_{v_{dmS}}}{\varepsilon_{v_{dmES}}}\, \frac{v_S}{v_{SE} \cos \theta} $ 
counts per day and per kg. 

CDMS should then in $121$ days with 1 kg of Ge ($A=73$) detect   
$R_{Ge}\, \varepsilon_{cut\, cdms}$
$\approx \frac{8.3}{4.0} \, 
 (\frac{73}{127})^4 \; \frac{\varepsilon_{cut\,cdms}}{\varepsilon_{cut \,dama}}\, 
 \frac{\varepsilon_{v_{dmS}}}{\varepsilon_{v_{dmES}}}\;
 \frac{v_S}{v_{SE} \cos \theta} \;  0.052 \cdot 
 121 \;$ events, 
which is for the above measured velocities equal to $(10,16,21,25)
\, \frac{\varepsilon_{cut\,cdms}}{\varepsilon_{cut\,dama}}\;
\frac{\varepsilon_{v_{dmS}}}{\varepsilon_{v_{dmES}}}$. CDMS~\cite{cdms} 
has found no event.

The approximations we made might cause that the expected  numbers 
$(10$, $16$, $21$, $25)$ multiplied by $\frac{\varepsilon_{cut\,cdms}}{\varepsilon_{cut\,dama}}\;
\frac{\varepsilon_{v_{dmS}}}{\varepsilon_{v_{dmES}}}$  
are too high for a factor let us say $4$ or $10$. (But they also might be too low for the same 
factor!)  
If in the near future  
CDMS (or some other equivalent experiment) 
will measure the above predicted events, then there might be  heavy 
family clusters which form the dark matter. In this case the DAMA experiment 
put the limit on our heavy family masses (Eq.(\ref{measure})).
In this case the DAMA experiments  
puts the limit on our heavy family masses (Eq.(\ref{measure})). 
Taking into account the uncertainties in the ''nuclear force'' cross section, we evaluate 
the lower limit for the mass $m_{q_5}\, c^2> 200$ TeV. Observing that 
for $m_{q_5} \, c^2> 10^4$ TeV 
the weak force starts to dominate, we estimate the upper limit $m_{q_5}\, c^2< 10^5$ TeV. 
In the case that the weak 
interaction determines the $n_5$ cross section we find for the mass range    
$10 \,{\rm TeV} < m_{q_5} \, c^2< 10^5$ TeV.

\section{ Evolution of the abundance of the fifth family members in the universe}

There are several questions to which we would need the answers before estimating the 
behaviour of our heavy family in the expanded universe, like:  What is the 
particle---anti-particle asymmetry for the fifth family? What are the fifth family masses? 
How are gluons and quark---anti-quark pairs of the fifth family members (nonperturbatively) 
''dressing'' the quarks of the fifth family, after the quarks decouple from the rest 
of the cosmic plasma in the expanding 
universe and how do quarks form baryons? And others.
We are not yet able to answer these questions. 
(These difficult studies are under considerations.)

We shall  simply assume 
that there are clusters of baryons and anti-baryons of the fifth family quarks constituting 
the dark matter. 
We estimate a possible  
evolution of the fifth family members' abundance when $q_5$ and $\bar{q}_5$ 
are in the equilibrium with the cosmic 
plasma (to which all the families with lower masses and all the gauge fields contribute) 
decoupling from the plasma at the temperature $T_1\approx m_{q_5} c^2/k_b$, with $k_b$ the 
Boltzmann constant, of the fifth family quarks and anti-quarks, assuming that the 
quarks and anti-quarks 
form (recombine into) the baryons $n_5$ and $\bar{n}_5$. 
(Namely, when at $T_1$ the fifth family
quarks' (as well as the anti-quarks') scattering amplitude is too low to keep the quarks 
at equilibrium with the plasma, quarks loose the contact with the plasma. 
The gluon interaction, 
however, sooner or later either causes the annihilation of quarks and anti-quarks, or 
forces the quarks 
and anti-quarks to form baryons and anti-baryons, which are colour neutral (reheating the plasma). 
We are studying these  possibilities in more details in~\cite{gmn08}. Here we present the 
results of the very rough estimations, which need to be studied in more details.)

To estimate the number of the fifth family quarks and anti-quarks clustered into 
$n_5$ and $\bar{n}_5$ we follow the ref.~\cite{dodelson}, chapter 3. 
Let $\Omega_5= \frac{\rho_{c_{5}}}{\rho_{cr}}$ ($\rho_{cr}= \frac{3 H_{0}^2}{8 \pi G}$, 
$H_0$ is the present Hubble constant and $G$ is the gravitational constant) 
denote the ratio between the abundance of the fifth family clusters  and 
the ordinary baryons (made out of the first family quarks), 
which is estimated to be 
$ \approx 0.1$. It follows
\begin{eqnarray}
\label{omega}
\Omega_5 &=&  \frac{1}{\beta}\,\frac{T_1 k_B}{m_{c_5} c^2} \, \sqrt{g^*}\,
(\frac{a(T_1) T_1}{a(T_0) T_0})^3 \, \sqrt{\frac{4 \pi^3 G}{45 (\hbar \, c)^3}}\, 
\frac{(T_{0}\, k_b)^3}{\rho_{cr} c^4}\frac{1}{<\sigma_5 \,v/c>}
\end{eqnarray}
where we evaluated $ \frac{T_1 k_B}{m_{c_5} c^2}\,
(\frac{a(T_1) T_1}{a(T_0) T_0})^3 \approx 10^{-3} $.  
$T_0$ is the today's 
black body radiation temperature, $a(T_0) =1$ and $a(T_{1})$ is the metric tensor component in 
the expanding flat universe, the Friedmann-Robertson-Walker metric:  
$$g_{\mu \nu} = {\rm diag} 
(1, - a(t)^2, - a(t)^2, - a(t)^2), \, T=T(t).$$
We evaluate $\frac{m_{c_5} c^2}{T_1 k_B} \approx 10$
$\sqrt{g^*} \approx \sqrt{200}$ ($g^*$ measures the number of degrees of freedom of 
our families and all gauge fields), 
$0.1 < \beta <10 $ stays for uncertainty in the evaluation of $\frac{m_{c_5} c^2}{T_1 k_B}$ and
$\frac{a(T_1) T_1}{a(T_0) T_0}$ 
($\beta $ determines the contribution of all the 
degrees of freedom to the temperature of the cosmic plasma after the fifth family members 
''freezed out'' at $T_1$ forming clusters: $\frac{1}{\beta}= \frac{a(T_1) T_1}{a(T_0) T_0}$).  
The dependence of $\Omega_{5}$ on the mass of the fifth family 
quarks is accordingly mainly in $\beta \, \sigma_5$.

Evaluating 
$\sqrt{\frac{4 \pi^3 \,G}{45 (\hbar c)^3}} \, \frac{(T_{0}\, k_b)^3}{\rho_{cr} c^4} = 200 \, 
(10^{-7} {\rm fm})^2$ we estimate  
(for $\frac{v}{c}\approx 1$) 
that the scattering cross section at the relativistic energies (when $k_b T_1 
\approx m_{c_{5}} c^2$) is
 $( 10^{-7} {\rm fm} )^2 < \sigma_5 <  (10^{-6} {\rm fm})^2  $.
 Taking into account 
  the relation for the relativistic scattering of quarks, with the 
  one gluon exchange contribution dominating 
  $\sigma= 8 \, \pi (\frac{3 \alpha_c(E)}{E})^2$, we obtain the mass limit 
  $10^2$ TeV $< m_{q_5}\, c^2< 10^3 $ TeV.  
\noindent
\section{ Concluding remarks}

We estimated in this talk a possibility that a new  stable family,   
distinguishable in masses from the families of lower masses 
and having the matrix elements of the Yukawa couplings to the lower mass 
families equal to zero, 
forms clusters, which are the dark matter constituents. Such a family is 
predicted by the approach unifying spins and 
charges~\cite{pn06,n92,n93,n07bled,gmdn07} as the fifth family,  together with the lower fourth 
family with the quark mass around 250 GeV or higher, which might be measured at LHC. 
The approach, which is offering a mechanism for generating families (and is accordingly 
offering the way beyond the standard model of the electroweak and colour interactions), 
predicts within the rough 
estimations that the fifth family lies several orders of magnitude above the fourth 
family and also several orders of magnitude bellow the unification scale of the standard model 
(that is the observed) charges. 

In this talk it is  assumed that heavy enough quarks form clusters,  interacting
with the one gluon exchange 
potential predominantly. We use the simple Bohr-like model to evaluate the 
properties of these heavy baryons. 
We assume further (with no justification yet) that in the evolution of our 
universe the asymmetry of $q_5$ and $\bar{q}_5$, if any, resembles in neutral (with respect to the 
colour and electromagnetic charge) clusters of 
baryons and anti-baryons, say neutrons and anti-neutrons made out of the fifth family 
($n_5,\bar{n}_5$), provided that the fifth family $u_5$ quark mass is 
(appropriately with respect to 
the weak and electromagnetic charge interaction) heavier than the $d_5$ quark, so that $n_5$ 
($u_5 d_5 d_5$) and not $p_5$ or $d_5 d_5 d_5$ is the lightest colour chargeless cluster made 
out of the fifth family quarks. Other possibilities (which might dominate and even 
offer the chargeless, with respect to the electromagnetic, weak and colour charge, 
and spinless clusters) will be studied in the future.

While the measured density of  the  dark matter 
does not put much limitation on the properties of heavy enough clusters,  
the DAMA/NaI 
experiments~\cite{rita0708} does if they measure 
our heavy fifth family clusters and also does the cosmological evolution.   
DAMA limits our fifth family quark mass 
to  $ 200 \,\rm{TeV} < m_{5}c^2 < 10^5\, \rm{TeV}$
(in the case that the weak 
interaction determines the $n_5$ cross section we find     
$10 \,{\rm TeV} < m_{q_5} \, c^2< 10^5$ TeV). 
The cosmological evolution 
suggests for the relativistic scattering cross sections 
$( 10^{-7} {\rm fm} )^2 < \sigma_5 <   (10^{-6} {\rm fm})^2  $ and the mass limit
$200$ TeV $< m_{q_5} \, c^2 < 2 \cdot 10^3$ TeV.

Let us add that in this case our Earth would contain for $m_{q_5}< 100$ TeV 
a mass part  $\approx 10^{-9}$ 
of dark matter clusters, 
the mean time between two collisions among the dark matter clusters in our galaxy would be 
from $10^{22} $ years on. 

Our rough estimations predict that if the DAMA/NaI experiment~\cite{rita0708} 
is measuring our heavy family clusters (or any heavy enough family cluster with 
small enough cross section),  the CDMS experiments~\cite{cdms} would observe  
a few events as well in the near future. 
 CDMS itself allows the heavy family mass
to be $8\cdot 10^3$ TeV or higher. 
 
The ref.~\cite{maxim} studies  
the possibility that a heavy family cluster absorbs a light (first) family member, 
claiming that such clusters would survive in the evolution of the universe. They found out 
that such families would be seen by the DAMA experiment but not by the CDMS experiment. 

If   future results from CDMS and 
DAMA will confirm our heavy family clusters with no light family quarks contributing,   
then we shall soon know, what is the origin of the dark matter.  

 Let us conclude this talk  with the recognition:   
 If the approach unifying spins and charges is the right way beyond the 
 standard model of the electroweak and colour interactions, 
 then more than three 
 families of quarks and leptons do exist, and the stable fifth family of quarks and leptons 
 is the candidate to form the dark matter, in agreement with the observed DAMA data and our 
 very rough cosmological estimations, but not yet in agreement with the CDMS experiments. The 
 contradiction, if it is at all a contradiction, between the DAMA and CDMS experiments 
 will be resolved with future statistics.

 In the case that the  fifth family alone forms the dark matter, much more accurate 
 calculations are needed 
 to say more about the family and its evolution during the history of our universe. 
 
 If our fifth family members do form the dark matter clusters, further studies of all the other 
 possibilities, like what could $\textrm{neutrino}_5$, $H_5$ ($u_5 u_5 d_5 e_5$, which has all the charges 
 of the standard model, with the weak charge included, equal to zero, so that only the nuclear 
 force and the corresponding forces among neutral clusters with respect to the $U(1)$ and the 
 weak charge play the role) and other clusters of the fifth family contribute to the dark matter. 
 
 Much more demanding studies are needed to understand  possible behaviour of the members of 
 the fifth family and the corresponding clusters in the early expanding universe.
 
 It would also be very interesting to estimate the 
 properties of the matter the dark matter would form far in the future, if the dark matter 
 baryons and anti-baryons of the fifth family members have the asymmetry. 

\section*{Acknowledgments } 

The authors  would like to thank  all the participants 
of the   workshops entitled 
"What comes beyond the Standard model", 
taking place  at Bled annually (usually) in  July, starting at 1998, and in particular to 
M. Khlopov an H.B. Nielsen. Also discussions with M. Rosina are appreciated.

%
%
\author{V.V. Dvoeglazov}
\title{Lorentz Transformations for a Photon}
\institute{
Universidad de Zacatecas\\
Ap.P. 636, Suc. 3 Cruzes, C. P. 98064, Zacatecas, Zac., M\'exico\\
E-mail: valeri@planck.reduaz.mx\\
URL: http://planck.reduaz.mx/\~~valeri/}

\titlerunning{Lorentz Transformations for a Photon}
\authorrunning{V.V. Dvoeglazov}
\maketitle

\begin{abstract} 
We discuss transormation laws of electric and magnetic fields
under Lorentz transformations, deduced from the Classical Field Theory. It
is found that we can connect the resulting expression for a bivector formed
with those fields, with the expression deduced from the Wigner
transformation rules for spin-1 functions of {\it massive} particles. This
mass parameter should be interpreted because the constancy of speed of light
forbids the existence of the photon mass.
\end{abstract}

\section{Introduction}

Within the Classical Electrodynamics (CED) we can obtain transformation
rules for electric and magnetic fields when we pass from one frame to
another which is moving with respect to the former with constant velocity;
in other words, we can obtain the relationships between the fields under
Lorentz transformations or {\it boosts}. On the other hand we have that
electromagnetic waves are constituted of ``quanta" of the fields, which are
called {\it photons}. It is usually accepted that photons do not have mass.
Furthermore, the photons are the particles which can be in the eigenstates
of helicities $\pm 1$. The dynamics of such fields is described by the
Maxwell equations on the classical level. On the other hand, we know the 
{\it Weinberg-Tucker-Hammer formalism}~\cite{vd1wein,vd1dvoe} which describe
spin-1 massive particles. The massless limit of the Weinberg-Tucker-Hammer
formalism can be well-defined in the light-cone basis~\cite{vd1saw}.

In this work we show how the classical-electrodynamics  reasons can be related
with the Lorentz-group (and quantum-electrodynamics) reasons.

\section{The Lorentz Transformations for Electromagnetic Field Presented by
Bivector}

In the Classical Electrodynamics we know the following equations to
transform electric and magnetic fields under Lorentz transformations~\cite%
{vd1landau} 
\begin{eqnarray}
{\bf E^{\prime}}&=&\gamma({\bf E}+{{\boldsymbol \beta} \times {\bf B}})-\frac{\gamma^2}{%
\gamma+1}{\boldsymbol \beta}({{\boldsymbol \beta} \cdot {\bf E}})  \label{vd1campoE} \\
{\bf B^{\prime}}&=&\gamma({\bf B}-{{\boldsymbol \beta} \times {\bf E}})-\frac{\gamma^2}{%
\gamma+1}{\boldsymbol \beta}({{\boldsymbol \beta} \cdot {\bf B}})  \label{vd1campoB}
\end{eqnarray}
where $\gamma=1/\sqrt{1-\frac{v^2}{c^2}}$, ${\boldsymbol\beta}= {\bf v}/c$; ${\bf E%
}, {\bf B}$ are the field in the original frame of reference, ${\bf E}%
^\prime$, ${\bf B}^\prime$ are the fields in the transformed frame of
reference. In the Cartesian component form we have 
\begin{eqnarray}
E^{i^\prime}&=&\gamma(E^i+\epsilon^{ijk}\beta^j B^k)-\frac{\gamma^2}{\gamma+1%
} \beta^i \beta^j E^j  \label{vd1campoE2} \\
B^{i^\prime}&=&\gamma(B^i-\epsilon^{ijk}\beta^j E^k)-\frac{\gamma^2}{\gamma+1%
} \beta^i \beta^j B^j  \label{vd1campoB2}
\end{eqnarray}
Now we introduce a particular representation of ${\bf S}$ matrices
(generators of rotations for spin 1): $({\bf S}^i)^{jk}=-i\epsilon^{ijk}$,
i.e. 
\begin{equation}
S_x=\left(%
\begin{array}{ccc}
0 & 0 & 0 \\ 
0 & 0 & -i \\ 
0 & i & 0%
\end{array}%
\right),\hspace{2mm} S_y=\left(%
\begin{array}{ccc}
0 & 0 & i \\ 
0 & 0 & 0 \\ 
-i & 0 & 0%
\end{array}%
\right),\hspace{2mm} S_z=\left(%
\begin{array}{ccc}
0 & -i & 0 \\ 
i & 0 & 0 \\ 
0 & 0 & 0%
\end{array}%
\right).
\end{equation}
Using the relation $\epsilon^{ijk}\epsilon^{lmk}=\delta^{il}\delta^{jm}-%
\delta^{im}\delta^{jl}$ (the Einstein sum rule on the repeated indices is
assumed), we have for an arbitrary vector ${\bf a}$: 
\begin{equation}
({\bf S\cdot a})^2_{ij}={\bf a}^2\delta^{ij}-a^i a^j\,.  \label{vd1Spuntoa}
\end{equation}
So with the help of the ${\bf S}$ matrices we can write (\ref{vd1campoE2},\ref%
{vd1campoB2}) like 
\begin{eqnarray}
E^{i^\prime}=\gamma(E^i-i(S^j)^{ik}\beta^j B^k)-\frac{\gamma^2}{\gamma+1}[{%
\boldsymbol\beta}^2 \delta^{ij}-({\bf S\cdot {\boldsymbol\beta})}^2_{ij}] E^j \,,
\label{vd1EconS} \\
B^{i^\prime}=\gamma(B^i+i(S^j)^{ik}\beta^j E^k)-\frac{\gamma^2}{\gamma+1}[{%
\boldsymbol\beta}^2 \delta^{ij}-({\bf S\cdot {\boldsymbol\beta})}^2_{ij}] B^j\,,
\label{vd1BconS}
\end{eqnarray}
or 
\begin{eqnarray}
{\bf E^{\prime}}=\{\gamma-\frac{\gamma^2}{\gamma+1}[{\boldsymbol\beta}^2-({\bf %
S\cdot {\boldsymbol\beta})}^2]\}{\bf E}-i\gamma({\bf S\cdot {\boldsymbol\beta}}){\bf B}
\label{vd1EconS2} \\
{\bf B^{\prime}}=\{\gamma-\frac{\gamma^2}{\gamma+1}[{\boldsymbol\beta}^2-({\bf %
S\cdot {\boldsymbol\beta})}^2]\}{\bf B}+ i\gamma({\bf S\cdot {\boldsymbol\beta}}){\bf E}%
\,.  \label{vd1BconS2}
\end{eqnarray}
In the matrix form we have: 
\begin{equation}
\left(%
\begin{array}{c}
{\bf E^{\prime}} \\ 
{\bf B^{\prime}}%
\end{array}%
\right)= \left(%
\begin{array}{cc}
\gamma-\frac{\gamma^2}{\gamma+1}[{\boldsymbol\beta}^2-({\bf S\cdot {\boldsymbol\beta})}%
^2] & -i\gamma({\bf S\cdot {\boldsymbol\beta}}) \\ 
i\gamma({\bf S\cdot {\boldsymbol\beta}}) & \gamma-\frac{\gamma^2}{\gamma+1}[{\boldsymbol%
\beta}^2-({\bf S\cdot {\boldsymbol\beta})}^2]%
\end{array}%
\right) \left(%
\begin{array}{c}
{\bf E} \\ 
{\bf B}%
\end{array}%
\right).  \label{vd1matrizEB}
\end{equation}
Now we introduce the unitary matrix $U=\frac{1}{\sqrt{2}}\left(%
\begin{array}{rr}
1 & i \\ 
1 & -i%
\end{array}%
\right)$ which satisfies $U^{\dagger}U=1$. Multiplying the equation (\ref%
{vd1matrizEB}) by this matrix we have 
\begin{equation}
U\left(%
\begin{array}{c}
{\bf E^{\prime}} \\ 
{\bf B^{\prime}}%
\end{array}%
\right)= U\left(%
\begin{array}{cc}
\gamma-\frac{\gamma^2}{\gamma+1}[{\boldsymbol\beta}^2-({\bf S\cdot {\boldsymbol\beta})}%
^2] & -i\gamma({\bf S\cdot {\boldsymbol\beta}}) \\ 
i\gamma({\bf S\cdot {\boldsymbol\beta}}) & \gamma-\frac{\gamma^2}{\gamma+1}[{\boldsymbol%
\beta}^2-({\bf S\cdot {\boldsymbol\beta})}^2]%
\end{array}%
\right)U^{\dagger}U \left(%
\begin{array}{c}
{\bf E} \\ 
{\bf B}%
\end{array}%
\right),  \label{vd1matrizEB2}
\end{equation}
which can be reduced to 
\begin{eqnarray}
\left(%
\begin{array}{c}
{\bf E^{\prime}}+i{\bf B^{\prime}} \\ 
{\bf E^{\prime}}-i{\bf B^{\prime}}%
\end{array}%
\right)=\label{vd1matrizEB3}\\ 
&&\hspace*{-30mm}\left(%
\begin{array}{cc}
1-\gamma({\bf S\cdot {\boldsymbol\beta}})+\frac{\gamma^2}{\gamma+1}({\bf S\cdot {%
\boldsymbol\beta})}^2 & 0 \\ 
0 & 1+\gamma({\bf S\cdot {\boldsymbol\beta}})+\frac{\gamma^2}{\gamma+1}({\bf %
S\cdot {\boldsymbol\beta})}^2%
\end{array}%
\right) \left(%
\begin{array}{c}
{\bf E}+i{\bf B} \\ 
{\bf E}-i{\bf B}%
\end{array}%
\right).  \nonumber
\end{eqnarray}
Now, let us take into account that ${\boldsymbol\beta}$-parameter is related to
the momentum and the energy in the following way: when we differentiate $E^2-%
{\bf p}^2 c^2=m^2 c^4$ we obtain $2E\mbox{d}E-2c^2{\bf p}\cdot\mbox{d}{\bf p}%
=0$, hence $\frac{\partial E}{\partial {\bf p}}=c^2 \frac{{\bf p}}{E}={\bf v}=
c{\boldsymbol \beta}$. Then, we set $\gamma =\frac{E}{mc^2}$, where we must
interpret $m$ as some mass parameter (as in~\cite[p.43]{vd1ryder}). It is
rather related not to the photon mass but to the particle mass, with which
we associate the second frame (the energy and the momentum as well). So, we
have 
\begin{equation}
\left(%
\begin{array}{c}
{\bf E^{\prime}}+i{\bf B^{\prime}} \\ 
{\bf E^{\prime}}-i{\bf B^{\prime}}%
\end{array}%
\right)= \left(%
\begin{array}{cc}
1-\frac{({\bf S\cdot p})}{mc}+\frac{({\bf S\cdot p)}^2}{m(E+mc^2)} & 0 \\ 
0 & 1+\frac{({\bf S\cdot p})}{mc}+\frac{({\bf S\cdot p)}^2}{m(E+mc^2)}%
\end{array}%
\right) \left(%
\begin{array}{c}
{\bf E}+i{\bf B} \\ 
{\bf E}-i{\bf B}%
\end{array}%
\right).  \label{vd1ecbivector}
\end{equation}
Note that we have started from the transformation equations for the fields,
which do not involve any mass and, according to the general wisdom, they
should describe massless particles. So, here the mass parameter is an auxiliary 
concept, which is possible to be used.

\section{The Lorentz Transformations for Massive Spin-1 Particles in the
Weinberg-Tucker-Hammer Formalism}

When we want to consider Lorentz transformations and derive relativistic
quantum equations for quantum-mechanical state functions, we first have to
work with the representations of the quantum-mechanical Lorenz group. These
representations have been studied by E. Wigner~\cite{vd1wigner}. In order to
consider the theories with definite-parity solutions of the corresponding
dynamical equations (the 'definite-parity' means that the solutions are the
eigenstates of the space-inversion operator), we have to look for a
function formed by two components (called the ``right" and ``left"
components), ref.~\cite{vd1ryder}. According to the Wigner rules, we have the
following expressions 
\begin{eqnarray}
\phi_R(p^{\mu})=\Lambda_R({p}^{\mu}\leftarrow\stackrel{0}{p^{\mu}})\phi_R(%
\stackrel{0}{p^{\mu}}) ,  \label{vd1phiR} \\
\phi_L(p^{\mu})=\Lambda_L({p}^{\mu}\leftarrow\stackrel{0}{p^{\mu}})\phi_L(%
\stackrel{0}{p^{\mu}})\,  \label{vd1phiL}
\end{eqnarray}
where $\stackrel{0}{p^{\mu}} =(E, {\bf 0})$ is the 4-momentum at rest, $p^\mu
$ is the 4-momentum in the second frame (where a particle has 3-momentum ${\bf p%
}$, $c=\hbar =1$). In the case of spin $S$, $\psi=\begin{pmatrix}\phi_R(p^{\mu})%
\cr \phi_L(p^{\mu})\end{pmatrix}$ is called the Weinberg $2(2S+1)$ function~\cite{vd1wein}.
Let us consider the case of $S=1$. The matrices $\Lambda_{R,L}$ are then the
matrices of the $(1,0)\oplus (0,1)$ representations of the Lorentz group. 
Their explicit forms are (${\boldsymbol \phi} = {\bf n} \phi$) 
\begin{eqnarray}
\Lambda_{R,L}=\exp(\pm {\bf S\cdot {\boldsymbol\phi}) =1+(S\cdot\hat{n})^2\left[%
\frac{\phi^2}{2!}+\frac{\phi^4}{4!}+\frac{\phi^6}{6!}+...\right]} \label{vd1lambdas}\\
&&\hspace*{-40mm}\pm{\bf S\cdot\hat{n}}\left[\frac{\phi}{1!} +\frac{\phi^3}{3!}+\frac{\phi^5}{%
5!}+...\right]\,,  \nonumber
\end{eqnarray}
or 
\begin{equation}
\exp(\pm{\bf S\cdot \phi})=1+({\bf S\cdot\hat{n}})^2(\cosh\phi -1) \pm({\bf %
S\cdot\hat{n}})\sinh\phi\,.
\end{equation}

If we introduce the parametrizations $\cosh\phi =\frac{E}{m},\hspace{2mm}%
\sinh\phi =\frac{\mid{\bf p}\mid}{m},\hspace{2mm}{\bf \hat{n}}=\frac{{\bf p}%
}{\mid{\bf p}\mid}$, see~\cite[p.39-43]{vd1ryder}, $c=\hbar=1$, we obtain 
\begin{eqnarray}
\Lambda_R ({p}^{\mu}\leftarrow\stackrel{0}{p^{\mu}}) &=& 1+\frac{{\bf S\cdot
p}}{m}+\frac{({\bf S\cdot p})^2}{m(E+m)}, \\
\Lambda_L ({p}^{\mu}\leftarrow\stackrel{0}{p^{\mu}}) &=& 1-\frac{{\bf S\cdot
p}}{m}+\frac{({\bf S\cdot p})^2}{m(E+m)}\,.  \label{vd1lambdas2}
\end{eqnarray}
Thus, the equations (\ref{vd1phiR}, \ref{vd1phiL}) are written as 
\begin{eqnarray}
\phi_R(p^{\mu})=\left\{1+\frac{{\bf S\cdot p}}{m}+\frac{({\bf S\cdot p})^2}{%
m(E+m)}\right\}\phi_R(\stackrel{0}{p^{\mu}}) ,  \label{vd1phiR1} \\
\phi_L(p^{\mu})=\left\{1-\frac{{\bf S\cdot p}}{m}+\frac{({\bf S\cdot p})^2}{%
m(E+m)}\right\}\phi_L(\stackrel{0}{p^{\mu}}).  \label{vd1phiL1}
\end{eqnarray}
If we compare the equations (\ref{vd1phiR1},\ref{vd1phiL1}) with the equation (\ref%
{vd1ecbivector}) we see that ${\bf E}-i{\bf B}$ can be considered as $\phi_R$, $%
{\bf E}+i{\bf B}$ can be considered as $\phi_L$.

\section{Conclusions}

We have found that when we introduce a mass parameter in the equation (\ref%
{vd1matrizEB3}) we can make the equation (\ref{vd1ecbivector}) and the equations (%
\ref{vd1phiR1},\ref{vd1phiL1}) to coincide. This result suggests we have to
attribute the mass parameter to the frame and not to the
electromagnetic-like fields.\footnote{Several authors (including de Broglie and Vigier)
argued in favor of the photon mass and modifications of the Maxwell equations.
However, at the present time, the constraints on the possible photon mass are 
very tight, $m<6\times 10^{-17}\mbox{eV}$, ref.~\cite{vd1pdg}.} This should be done in order to preserve the
postulate which states that all inertial observers must measure the same
speed of light. Moreover, our consideration illustrates a sutuation in which
we have to distinguish between passive and active transformations. The answer on the question, 
whether the similarity between (\ref{vd1ecbivector}) and (\ref{vd1phiR1},\ref{vd1phiL1}) is just 
a mere coincidence or not, should be answered after full understanding of the nature of the mass.

\section*{Acknowledgements} We are grateful to Sr. Alfredo Casta\~neda for discussions in the classes of
quantum mechanics at the University of Zacatecas.

%
\author{V.V. Dvoeglazov}
\title{Is the Space-Time Non-commutativity Simply Non-commutativity of Derivatives?}
\institute{
Universidad de Zacatecas\\
Ap.P. 636, Suc. 3 Cruzes, C. P. 98064, Zacatecas, Zac., M\'exico\\
E-mail: valeri@planck.reduaz.mx\\
URL: http://planck.reduaz.mx/\~~valeri/}

\titlerunning{Is the Space-Time Non-commutativity \ldots}
\authorrunning{V.V. Dvoeglazov}
\maketitle

\begin{abstract}
Recently, some problems have been found in the definition of the partial derivative in the case of  the presence of both explicit and implicit
functional dependencies in the classical analysis. In this talk we investigate the influence of this observation on the quantum mechanics and classical/quantum  field theory. Surprisingly, some commutators of the coordinate-dependent operators  are not equal to zero. Therefore, we try to provide mathematical  foundations to the modern non-commutative theories. We also indicate
possible applications in the Dirac-like theories.
\end{abstract}

The assumption that operators of
coordinates do {\it not} commute $[\hat{x}_{\mu },\hat{x}_{\nu }]_{-} = i\theta_{\mu\nu}$ (or, alternatively, $[\hat{x}_{\mu },\hat{x}_{\nu }]_{-}= iC_{\mu\nu}^\beta x_\beta$)
has been first made by H. Snyder~\cite{vd2snyder}. Later it was shown that such an anzatz may lead to non-locality. Thus, the Lorentz symmetry may be
broken. Recently, some attention has again been paid to this 
idea~\cite{vd2noncom} in the context of ``brane theories''.

On the other hand, the famous Feynman-Dyson proof of Maxwell equations~\cite{vd2FD}
contains intrinsically the non-commutativity of velocities. While therein 
$[ x^i, x^j ]_-=0$, but  $[ \dot x^i (t), 
\dot x^j (t) ]_- = {i\hbar\over m^2} \epsilon^{ijk} B_k \neq 0$ (at the same time with $[x^i, \dot x^j]_- = {i\hbar \over m} \delta^{ij}$) that also may be considered as a contradiction with
the well-accepted theories. Dyson wrote in a very clever way: ``Feynman in 1948 was not alone in trying to build theories outside the framework of conventional physics... All these radical programms, including Feynman's, failed... I venture to disagree with Feynman now, as I often did while he was alive..."

Furthermore, it was recently shown that notation and terminology, 
which physicists used when speaking about partial
derivative of many-variables functions, are sometimes 
confusing~\cite{vd2chja} (see also the discussion in~\cite{vd2eld}).
They referred to books~\cite{vd2Arnold}: ``...one identifies sometime $f_1$ and $f$, saying, that is the same function represented with the help of variables $x_1$ instead of $x$. Such a simplification is very dangerous and may result in very serious contradictions" (see the text after Eq. (1.2.5) in~[6b]; $f=f(x)$, $f_1 = f (u(x_1))$). In~\cite{vd2chja} the basic question was: how should one define correctly the time derivatives of the  functions $E [ x_1(t), \ldots x_{n-1} (t), t ]$ and $E ( x_1, \ldots x_{n-1}, t )$? Is there any sense in 
${\partial \over \partial t} E [{\bf r} (t), t]$ and ${d\over dt} 
E ({\bf r}, t)$?\footnote{The quotation from [4c, p. 384]: ``the [above] symbols
are meaningless, because the process denoted by the operator of {\it partial} differentiation can be applied only to functions of several {\it independent} variables and ${\partial \over \partial t} E [{\bf r} (t), t]$
is not {\it such} a function."} Those authors claimed that even well-known formulas 
\begin{equation}
{df \over dt} = \{ {\cal H}, f \} +{\partial f \over \partial t}\,,\quad
\mbox{and}\,\quad
{dE \over dt} = ({\bf v} \cdot {\bf \nabla} ) E + {\partial E\over \partial t}\,
\label{vd2oldeq}
\end{equation}
can be confusing unless additional definitions present.\footnote{As for these formulas the authors of~\cite{vd2chja} write:``this equation [cannot be correct] because the partial differentiation would involve increments of the functions ${\bf r} (t)$ in the form ${\bf r} (t) +\Delta {\bf r} (t)$ and we do not know how we must interpret this increment because we have two options: {\it either} $\Delta {\bf r} (t) = {\bf r} (t) - {\bf r}^\ast (t)$, {\it or} $\Delta {\bf r} (t) = {\bf r} (t) - {\bf r} (t^\ast)$. Both are different processes because the first one involves changes in the functional form of the functions ${\bf r} (t)$, while the second involves changes in the position along the path defined by ${\bf r} = {\bf r} (t)$ but preserving the same functional form." Finally, they gave the correct form, in their opinion, of (\ref{vd2oldeq}). See in~[4d].}

Another well-known physical example of the situation, when we have both explicite and implicite dependences of the function which derivatives act upon, is the field of an accelerated charge~ \cite{vd2landau}.
First, Landau and Lifshitz wrote that the functions depended on the retarded time $t^{\prime }$
and only through $t^{\prime }+R(t^{\prime })/c=t$ they depended implicitly
on $x,y,z,t$. However, later they used
the explicit dependence of $R$ and fields on the space coordinates of the
observation point too. Of course! Otherwise, the ``simply" retarded fields do not satisfy the Maxwell equations~[4b]. In the same work Chubykalo and Vlayev claimed that the time derivative
and curl did {\it not} commute in their case. Jackson, in fact, disagreed
with their claim on the basis of the definitions (``the equations
representing Faraday's law and the absence of magnetic charges ... are satisfied automatically''; see his Introduction in~[5b]). But,
he agrees with~\cite{vd2landau} that one should find ``a contribution to the
spatial partial derivative for fixed time $t$ from explicit spatial
coordinate dependence (of the observation point)''. So, actually
the fields and the potentials are the functions of the following forms:
$$A^\mu (x, y, z, t' (x,y,z,t)), {\bf E} (x, y, z, t' (x,y,z,t)), {\bf B} (x, y, z, t' (x,y,z,t)).$$ 
 \v{S}kovrlj and Ivezi\'{c}~[5c] call this partial derivative as `{\it complete} partial
derivative'; Chubykalo and Vlayev~[4b], as `{\it total} derivative with respect to a given variable'; the terminology suggested by Brownstein~[5a] is
`the {\it whole}-partial derivative'. We shall denote below this whole-partial derivative operator as $\hat\partial \over \hat \partial 
x^i$, while still keeping the definitions of~[4c,d].

In~[5d] I studied the case when we deal with explicite and implicite dependencies  $f ({\bf p}, E ({\bf p}))$. It is well known that the energy in the
relativism is connected with the 3-momentum as $E=\pm \sqrt{{\bf p}^2 +m^2}$
; the unit system $c=\hbar=1$ is used. In other words, we must choose the
3-dimensional hyperboloid from the entire Minkowski space and the energy is 
{\it not} an independent quantity anymore. Let us calculate the commutator
of the whole derivative $\hat\partial /\hat\partial E$ and $\hat\partial / 
\hat\partial p_i$.\footnote{
In order to make distinction between differentiating the explicit function
and that which contains both explicit and implicit dependencies, the `whole
partial derivative' may be denoted as $\hat\partial$.} In the general case
one has 
\begin{equation}
{\frac{\hat\partial f ({\bf p}, E({\bf p})) }{\hat\partial p_i}} \equiv {
\frac{\partial f ({\bf p}, E({\bf p})) }{\partial p_i}} + {\frac{\partial f (
{\bf p}, E({\bf p})) }{\partial E}} {\frac{\partial E}{\partial p_i}}\, .
\end{equation}
Applying this rule, we surprisingly find 
\begin{eqnarray}
&&[{\frac{\hat\partial }{\hat\partial p_i}},{\frac{\hat\partial }{\hat
\partial E}}]_- f ({\bf p},E ({\bf p})) = {\frac{\hat\partial }{\hat\partial
p_i}} {\frac{\partial f }{\partial E}} -{\frac{\partial }{\partial E}} ({
\frac{\partial f}{\partial p_i}} +{\frac{\partial f}{\partial E}}{\frac{
\partial E}{\partial p_i}}) =  \nonumber \\
&=& {\frac{\partial^2 f }{\partial E\partial p_i}} + {\frac{\partial^2 f}{
\partial E^2}}{\frac{\partial E}{\partial p_i}} - {\frac{\partial^2 f }{
\partial p_i \partial E}} - {\frac{\partial^2 f}{\partial E^2}}{\frac{
\partial E}{\partial p_i}}- {\frac{\partial f }{\partial E}} {\frac{\partial
}{\partial E}}({\frac{\partial E}{\partial p_i}})\,.  \label{vd2com}
\end{eqnarray}
So, if $E=\pm \sqrt{m^2+{\bf p}^2}$ 
and one uses the generally-accepted 
representation form of $\partial E/\partial p_i
=  p^i/E$,
one has that the expression (\ref{vd2com})
appears to be equal to $$(p_i/E^2) {\frac{\partial f({\bf p}, E ({\bf p}))}{
\partial E}}.$$ Within the choice of the normalization the coefficient is the
longitudinal electric field in the helicity basis (the electric/magnetic
fields can be derived from the 4-potentials which have been presented in~ 
\cite{vd2hb}).\footnote{They are written in the following way:
\begin{eqnarray} &&\epsilon
_{\mu }({\bf p},\lambda =+1)={\frac{1}{\sqrt{2}}}{\frac{e^{i\phi }}{
p}}\begin{pmatrix} 0, {p_x p_z -ip_y p\over \sqrt{p_x^2 +p_y^2}}, {p_y p_z +ip_x
p\over \sqrt{p_x^2 +p_y^2}}, -\sqrt{p_x^2 +p_y^2}\end{pmatrix}\,, \\
&&\epsilon _{\mu }({\bf p},\lambda =-1)={\frac{1}{\sqrt{2}}}{\frac{e^{-i\phi }
}{p}}\begin{pmatrix} 0, {-p_x p_z -ip_y p\over \sqrt{p_x^2 +p_y^2}}, {-p_y p_z
+ip_x p\over \sqrt{p_x^2 +p_y^2}}, +\sqrt{p_x^2 +p_y^2}\end{pmatrix}\,, \\
&&\epsilon _{\mu }({\bf p},\lambda =0)={\frac{1}{m}}\begin{pmatrix} p, -{E \over p}
p_x, -{E \over p} p_y, -{E \over p} p_z \end{pmatrix}\,, \\
&&\epsilon _{\mu }({\bf p},\lambda =0_{t})={\frac{1}{m}}\begin{pmatrix}E , -p_x,
-p_y, -p_z \end{pmatrix}\,.
\end{eqnarray}
And,
\begin{eqnarray}
&&{\bf E}({\bf p},\lambda =+1)=-{\frac{iEp_{z}}{\sqrt{2}pp_{l}}}{\bf p}-{\frac{
E}{\sqrt{2}p_{l}}}\tilde{{\bf p}},\quad {\bf B}({\bf p},\lambda =+1)=-{\frac{
p_{z}}{\sqrt{2}p_{l}}}{\bf p}+{\frac{ip}{\sqrt{2}p_{l}}}\tilde{{\bf p}}, \\
&&{\bf E}({\bf p},\lambda =-1)=+{\frac{iEp_{z}}{\sqrt{2}pp_{r}}}{\bf p}-{\frac{
E}{\sqrt{2}p_{r}}}\tilde{{\bf p}}^{\ast },\quad {\bf B}({\bf p},\lambda
=-1)=-{\frac{p_{z}}{\sqrt{2}p_{r}}}{\bf p}-{\frac{ip}{\sqrt{2}p_{r}}}\tilde{
{\bf p}}^{\ast }, \\
&&{\bf E}({\bf p},\lambda =0)={\frac{im}{p}}{\bf p},\quad {\bf B}({\bf p}
,\lambda =0)=0,
\end{eqnarray}
with $\tilde{{\bf p}}=\begin{pmatrix}p_y\\ -p_x\\ -ip\end{pmatrix}$. It is easy seen that
the parity properties of these vectors are different comparing with the standard basis. The parity operator for polarization vectors coincides with the metric tensor of
the Minkowski 4-space.} 
On the other hand, the commutator 
\begin{equation}
[{\frac{\hat\partial}{\hat\partial p_i}}, {\frac{\hat\partial}{\hat\partial
p_j}}]_- f ({\bf p},E ({\bf p})) = {\frac{1}{E^3}} {\frac{
\partial f({\bf p}, E ({\bf p}))}{\partial E}} [p_i, p_j]_-\,.
\end{equation}
This may be considered to be zero unless we would trust to the genius
Feynman. He postulated that the velocity (or, of course, the 3-momentum)
commutator is equal to $[p_i,p_j]\sim i\hbar\epsilon_{ijk} B^k$, i.e., to
the magnetic field.

Furthermore, since the energy derivative corresponds to the operator of time
and the $i$-component momentum derivative, to $\hat x_i$, we put forward the
following anzatz in the momentum representation: 
\begin{equation}
[\hat x^\mu, \hat x^\nu]_- = \omega ({\bf p}, E({\bf p})) \,
F^{\mu\nu}_{\vert\vert}{\frac{\partial }{\partial E}}\,,
\end{equation}
with some weight function $\omega$ being different for different choices of
the antisymmetric tensor spin basis. In the modern literature, the idea of the broken Lorentz invariance by this method is widely discussed, see e.g.~\cite{vd2amelino}. 

Let us turn now to the application of the presented ideas to the Dirac case.
Recently, we analized Sakurai-van der Waerden method of derivations of the Dirac
(and higher-spins too) equation~\cite{vd2Dvoh}. We can start from
\begin{equation}
(E I^{(2)}-{\bf \sigma}\cdot {\bf p}) (E I^{(2)}+ {\bf\sigma}\cdot
{\bf p} ) \Psi_{(2)} = m^2 \Psi_{(2)} \,,
\end{equation}
or
\begin{equation}
(E I^{(4)}+{\bf \alpha}\cdot {\bf p} +m\beta) (E I^{(4)}-{\bf\alpha}\cdot
{\bf p} -m\beta ) \Psi_{(4)} =0.\label{vd2f4}
\end{equation}
Of course, as in the original Dirac work, we have
\begin{equation}
\beta^2 = 1\,,\quad
\alpha^i \beta +\beta \alpha^i =0\,,\quad
\alpha^i \alpha^j +\alpha^j \alpha^i =2\delta^{ij} \,.
\end{equation}
For instance, their explicite forms can be chosen 
\begin{eqnarray}
\alpha^i =\begin{pmatrix}\sigma^i& 0\\
0&-\sigma^i\end{pmatrix}\,,\quad
\beta = \begin{pmatrix}0&1_{2\times 2}\\
1_{2\times 2} &0\end{pmatrix}\,,
\end{eqnarray}
where $\sigma^i$ are the ordinary Pauli $2\times 2$ matrices.

We also postulate the non-commutativity
\begin{equation}
[E, {\bf p}^i]_- = \Theta^{0i} = \theta^i,,
\end{equation}
as usual. Therefore the equation (\ref{vd2f4}) will {\it not} lead
to the well-known equation $E^2 -{\bf p}^2 = m^2$. Instead, we have
\begin{equation}
\left \{ E^2 - E ({\bf \alpha} \cdot {\bf p})
+({\bf \alpha} \cdot {\bf p}) E - {\bf p}^2 - m^2 - i {\bf\sigma}\times I_{(2)}
[{\bf p}\times {\bf p}] \right \} \Psi_{(4)} = 0
\end{equation}
For the sake of simplicity, we may assume the last term to be zero. Thus we come to
\begin{equation}
\left \{ E^2 - {\bf p}^2 - m^2 -  ({\bf \alpha}\cdot {\bf \theta})
\right \} \Psi_{(4)} = 0\,.
\end{equation} 
However, let us make the unitary transformation. It is known~\cite{vd2Berg}
that one can\footnote{Of course, the certain relations for the components ${\bf a}$ should be assumed. Moreover, in our case ${\bf \theta}$ should not depend on $E$ and ${\bf p}$. Otherwise, we must take the noncommutativity $[E, {\bf p}^i]_-$ again.}
\begin{equation}
U_1 ({\bf \sigma}\cdot {\bf a}) U_1^{-1} = \sigma_3 \vert {\bf a} \vert\,.\label{vd2s3}
\end{equation}
For ${\bf \alpha}$ matrices we re-write (\ref{vd2s3}) to
\begin{eqnarray}
U_1 ({\bf \alpha}\cdot {\bf \theta}) U_1^{-1} = \vert {\bf \theta} \vert
\begin{pmatrix}1&0&0&0\\
0&-1&0&0\\
0&0&-1&0\\
0&0&0&1\end{pmatrix} = \alpha_3 \vert {\bf\theta}\vert\,.
\end{eqnarray}
applying the second unitary transformation:
\begin{eqnarray}
U_2 \alpha_3 U_2^\dagger =
\begin{pmatrix}1&0&0&0\\
0&0&0&1\\
0&0&1&0\\
0&1&0&0\end{pmatrix} \alpha_3 \begin{pmatrix}1&0&0&0\\
0&0&0&1\\
0&0&1&0\\
0&1&0&0\end{pmatrix} = \begin{pmatrix}1&0&0&0\\
0&1&0&0\\
0&0&-1&0\\
0&0&0&-1\end{pmatrix}\,.
\end{eqnarray}
The final equation is
\begin{equation}
[E^2 -{\bf p}^2 -m^2 -\gamma^5_{chiral} \vert {\bf \theta}\vert ] \Psi^\prime_{(4)} = 0\,.
\end{equation}
In the physical sense this implies the mass splitting for a Dirac particle over the non-commutative space. This procedure may be attractive for explanation of the mass creation and the mass splitting for fermions.

The presented ideas permit us to provide some
foundations for non-commu\-tative field theories and induce us to look for further
applications of the functions with explicit and implicit dependencies in physics and mathematics.
Perhaps, all this staff is related to the fundamental length concept~\cite{vd2Kad,vd2amelino}. Let see.

\section*{Acknowledgments}
I am grateful to the participants of recent conferences on mathematical physics. Particularly, I am obliged to Profs. R. Flores Alvarado, L. M. Gaggero Sager, M. Khlopov,  N. Manko\v c-Bor\v stnik, M. Plyushchay and S. Vlayev  for discussions.

\def\Journal#1#2#3#4{{#1} {\bf #2}, #3 (#4)}

\def\NCA{\em Nuovo Cimento}
\def\RNC{\em Rivista Nuovo Cimento}
\def\NIM{\em Nucl. Instrum. Methods}
\def\NIMA{{\em Nucl. Instrum. Methods} A}
\def\NPB{{\em Nucl. Phys.} B}
\def\PLB{{\em Phys. Lett.}  B}
\def\PRL{\em Phys. Rev. Lett.}
\def\PRD{{\em Phys. Rev.} D}
\def\ZPC{{\em Z. Phys.} C}
\def\GaC{\em Gravitation and Cosmology}
\def\GaCS{{\em Gravitation and Cosmology} Supplement}
\def\JETP{\em JETP}
\def\JETPL{\em JETP Lett.}
\def\PAN{\em Phys.Atom.Nucl.}
\def\CQG{\em Class. Quantum Grav.}
\def\APJ{\em Astrophys. J.}
\def\SCI{\em Science}
\def\MPLA{{\em Mod. Phys. Lett.}  A}
\def\IJTP{\em Int. J. Theor. Phys.}
\def\NJP{\em New J. of Phys.}
\def\JHEP{\em JHEP}
\def\IJMP{\em Int. J. Mod. Phys.}
\def\EPJC{{\em Eur. Phys. J.} C}

\def\st{\scriptstyle}
\def\sst{\scriptscriptstyle}
\def\mco{\multicolumn}
\def\epp{\epsilon^{\prime}}
\def\vep{\varepsilon}
\def\ra{\rightarrow}
\def\ppg{\pi^+\pi^-\gamma}
\def\vp{{\bf p}}
\def\ko{K^0}
\def\kb{\bar{K^0}}
\def\al{\alpha}
\def\ab{\bar{\alpha}}
\def\s{{\,\rm s}}
\def\g{{\,\rm g}}
\def\eV{\,{\rm eV}}
\def\keV{\,{\rm keV}}
\def\MeV{\,{\rm MeV}}
\def\GeV{\,{\rm GeV}}
\def\TeV{\,{\rm TeV}}
\def\sv{\left<\sigma v\right>}
\def\({\left(}
\def\){\right)}
\def\cm{{\,\rm cm}}
\def\K{{\,\rm K}}
\def\kpc{{\,\rm kpc}}
\def\beq{\begin{equation}}
\def\eeq{\end{equation}}
\def\bea{\begin{eqnarray}}
\def\eea{\end{eqnarray}}
\def\CPbar{\hbox{{\rm CP}\hskip-1.80em{/}}}
\author{M.Yu. Khlopov$^{1,2,3}$, A.G. Mayorov $^{1}$ and E.Yu. Soldatov $^{1}$}
\title{Composite Dark Matter: Solving the Puzzles of Underground Experiments?}
\institute{%
$^{1}$Moscow Engineering Physics Institute (National Nuclear Research University), 115409 Moscow, Russia \\
$^{2}$ Centre for Cosmoparticle Physics "Cosmion" 125047 Moscow, Russia \\
$^{3}$ APC laboratory 10, rue Alice Domon et L\'eonie Duquet \\75205
Paris Cedex 13, France}

\titlerunning{Composite Dark Matter}
\authorrunning{M.Yu. Khlopov, A.G. Mayorov and E.Yu. Soldatov}
\maketitle

\begin{abstract}

Particle physics candidates for cosmological dark matter are usually
considered as neutral and weakly interacting. However stable charged
leptons and quarks can also exist and, hidden in elusive atoms, play
the role of dark matter. Stable particles with charge -2 bind with
primordial helium in O-helium "atoms" (OHe), representing a specific
Warmer than Cold nuclear-interacting form of dark matter. O-helium
can influence primordial nucleosynthesis, giving rise, in
particular, to primordial heavy elements. Its excitation in
collisions in galactic bulge can lead to enhancement of positron
annihilation line, observed by Integral. Slowed down in the
terrestrial matter by elastic collisions, OHe is elusive for direct
methods of underground Dark matter detection like those used in CDMS
experiment, but its rare inelastic reactions with nuclei can lead to
annual variations of energy release in the interval of energy 2-6
keV in DAMA/NaI and DAMA/Libra experiments, being consistent with
the number of events registered in these experiments.

\end{abstract}
\section{Introduction}
The widely shared belief is that the dark matter, corresponding to
$25\%$ of the total cosmological density, is nonbaryonic and
consists of new stable particles. One can formulate the set of
conditions under which new particles can be considered as candidates
to dark matter (see e.g. \cite{book,Cosmoarcheology,Bled07} for
review and reference): they should be stable, saturate the measured
dark matter density and decouple from plasma and radiation at least
before the beginning of matter dominated stage. The easiest way to
satisfy these conditions is to involve neutral weakly interacting
particles. However it is not the only particle physics solution for
the dark matter problem. In the composite dark matter scenarios new
stable particles can have electric charge, but escape experimental
discovery, because they are hidden in atom-like states maintaining
dark matter of the modern Universe.

Elementary particle frames for heavy stable charged particles
include: (a) A heavy quark of fourth generation
\cite{I,lom,Khlopov:2006dk} accompanied by heavy neutrino \cite{N};
which can avoid experimental constraints \cite{Q,Okun} and form
composite dark matter species; (b) A Glashow's ``sinister'' heavy
tera-quark $U$ and tera-electron $E$, forming a tower of
tera-hadronic and tera-atomic bound states with ``tera-helium
atoms'' $(UUUEE)$ considered as dominant dark matter
\cite{Glashow,Fargion:2005xz}. (c) AC-leptons, predicted in the
extension \cite{5} of standard model, based on the approach of
almost-commutative geometry \cite{bookAC}, can form evanescent
AC-atoms, playing the role of dark matter
\cite{Khlopov:2006dk,5,FKS}.  (d) An elegant composite dark matter
solution \cite{KK} is possible in the framework of walking
technicolor models (WTC) \cite{Sannino:2004qp}. (e) Finally, stable
charged clusters $\bar u_5 \bar u_5 \bar u_5$ of (anti)quarks $\bar
u_5$ of 5th family can follow from the approach, unifying spins and
charges \cite{Norma}.

In all these models (see review in
\cite{Bled07,Khlopov:2006dk,Khlopov:2008rp}), the predicted stable
charged particles form neutral atom-like states, composing the dark
matter of the modern Universe. It offers new solutions for the
physical nature of the cosmological dark matter. The main problem
for these solutions is to suppress the abundance of positively
charged species bound with ordinary electrons, which behave as
anomalous isotopes of hydrogen or helium. This problem is
unresolvable, if the model predicts stable particles with charge -1,
as it is the case for tera-electrons \cite{Glashow,Fargion:2005xz}.
To avoid anomalous isotopes overproduction, stable particles with
charge $-1$ should be absent, so that stable negatively charged
particles should have charge $-2$ only.

In the asymmetric case, corresponding to excess of -2 charge
species, $X^{--}$, as it was assumed for $(\bar U \bar U \bar
U)^{--}$ in the model of stable $U$-quark of a 4th generation, as
well as can take place for $(\bar u_5 \bar u_5 \bar u_5)^{--}$ in
the approach \cite{Norma} their positively charged partners
effectively annihilate in the early Universe. Such an asymmetric
case was realized in \cite{KK} in the framework of WTC, where it was
possible to find a relationship between the excess of negatively
charged anti-techni-baryons $(\bar U \bar U )^{--}$ and/or
technileptons $\zeta^{--}$ and the baryon asymmetry of the Universe.

 After it is formed
in the Standard Big Bang Nucleosynthesis (SBBN), $^4He$ screens the
$X^{--}$ charged particles in composite $(^4He^{++}X^{--})$ {\it
O-helium} ``atoms''
 \cite{I}.
 For different models of $X^{--}$ these "atoms" are also
called ANO-helium \cite{lom,Khlopov:2006dk}, Ole-helium
\cite{Khlopov:2006dk,FKS} or techni-O-helium \cite{KK}. We'll call
them all O-helium ($OHe$) in our further discussion, which follows
the guidelines of \cite{I2}.

In all these forms of O-helium $X^{--}$ behave either as leptons or
as specific "heavy quark clusters" with strongly suppressed hadronic
interaction. Therefore O-helium interaction with matter is
determined by nuclear interaction of $He$. These neutral primordial
nuclear interacting objects contribute to the modern dark matter
density and play the role of a nontrivial form of strongly
interacting dark matter \cite{Starkman,McGuire:2001qj}. The active
influence of this type of dark matter on nuclear transformations
seems to be incompatible with the expected dark matter properties.
However, it turns out that the considered scenario is not easily
ruled out \cite{I,FKS,KK,Khlopov:2008rp} and challenges the
experimental search for various forms of O-helium and its charged
constituents. O-helium scenario might provide explanation for the
observed excess of positron annihilation line in the galactic bulge.
Here we briefly review the main features of O-helium dark matter and
concentrate on its effects in underground detectors. We refine the
earlier arguments \cite{I2,KK2} that the positive results of dark
matter searches in DAMA/NaI (see for review \cite{Bernabei:2003za})
and DAMA/LIBRA \cite{Bernabei:2008yi} experiments can be explained
by O-helium, resolving the controversy between these results and
negative results of other experimental groups.

The essential difference between O-helium and WIMP-like dark matter
is that cosmic O-helium is slowed down in elestic scattering with
matter nuclei and can not cause effects of recoil nuclei above the
threshold of underground detectors. However, strongly suppressed
inelastic interaction of O-helium, in which it is disrupted, $He$ is
emitted and $X^{--}$ is captured by a nucleus, changes the charge of
nucleus by 2 units. We argue that effects of immediate ionization
and rearrangement of electron shells is suppressed and that the
ionization signal in the range 2-6 keV can come in NaI detector with
sufficient delay after the OHe reaction, making this signal
distinguishable from much larger rapid energy release in this
reaction

\section{O-helium Universe}

Following \cite{I,lom,Khlopov:2006dk,KK,I2} consider charge
asymmetric case, when excess of $X^{--}$ provides effective
suppression of positively charged species.

In the period $100\s \le t \le 300\s$  at $100 \keV\ge T \ge T_o=
I_{o}/27 \approx 60 \keV$, $^4He$ has already been formed in the
SBBN and virtually all free $X^{--}$ are trapped by $^4He$ in
O-helium ``atoms" $(^4He^{++} X^{--})$. Here the O-helium ionization
potential is\footnote{The account for charge distribution in $He$
nucleus leads to smaller value $I_o \approx 1.3 \MeV$
\cite{Pospelov}.} \beq I_{o} = Z_{x}^2 Z_{He}^2 \alpha^2 m_{He}/2
\approx 1.6 \MeV,\label{IO}\eeq where $\alpha$ is the fine structure
constant,$Z_{He}= 2$ and $Z_{x}= 2$ stands for the absolute value of
electric charge of $X^{--}$.  The size of these ``atoms" is
\cite{I,FKS} \beq R_{o} \sim 1/(Z_{x} Z_{He}\alpha m_{He}) \approx 2
\cdot 10^{-13} \cm \label{REHe} \eeq Here and further, if not
specified otherwise, we use the system of units $\hbar=c=k=1$.

O-helium, being an $\alpha$-particle with screened electric charge,
can catalyze nuclear transformations, which can influence primordial
light element abundance and cause primordial heavy element
formation. These effects need a special detailed and complicated
study. The arguments of \cite{I,FKS,KK} indicate that this model
does not lead to immediate contradictions with the observational
data.

Due to nuclear interactions of its helium constituent with nuclei in
the cosmic plasma, the O-helium gas is in thermal equilibrium with
plasma and radiation on the Radiation Dominance (RD) stage, while
the energy and momentum transfer from plasma is effective. The
radiation pressure acting on the plasma is then transferred to
density fluctuations of the O-helium gas and transforms them in
acoustic waves at scales up to the size of the horizon.

At temperature $T < T_{od} \approx 200 S^{2/3}_3\eV$ the energy and
momentum transfer from baryons to O-helium is not effective
\cite{I,KK} because $$n_B \sv (m_p/m_o) t < 1,$$ where $m_o$ is the
mass of the $OHe$ atom and $S_3= m_o/(1 \TeV)$. Here \beq \sigma
\approx \sigma_{o} \sim \pi R_{o}^2 \approx
10^{-25}\cm^2\label{sigOHe}, \eeq and $v = \sqrt{2T/m_p}$ is the
baryon thermal velocity. Then O-helium gas decouples from plasma. It
starts to dominate in the Universe after $t \sim 10^{12}\s$  at $T
\le T_{RM} \approx 1 \eV$ and O-helium ``atoms" play the main
dynamical role in the development of gravitational instability,
triggering the large scale structure formation. The composite nature
of O-helium determines the specifics of the corresponding dark
matter scenario.

At $T > T_{RM}$ the total mass of the $OHe$ gas with density $\rho_d
= (T_{RM}/T) \rho_{tot} $ is equal to
$$M=\frac{4 \pi}{3} \rho_d t^3 = \frac{4 \pi}{3} \frac{T_{RM}}{T} m_{Pl}
(\frac{m_{Pl}}{T})^2$$ within the cosmological horizon $l_h=t$. In
the period of decoupling $T = T_{od}$, this mass  depends strongly
on the O-helium mass $S_3$ and is given by \cite{KK}\beq M_{od} =
\frac{T_{RM}}{T_{od}} m_{Pl} (\frac{m_{Pl}}{T_{od}})^2 \approx 2
\cdot 10^{44} S^{-2}_3 \g = 10^{11} S^{-2}_3 M_{\odot}, \label{MEPm}
\eeq where $M_{\odot}$ is the solar mass. O-helium is formed only at
$T_{o}$ and its total mass within the cosmological horizon in the
period of its creation is 
$$M_{o}=M_{od}(T_{od}/T_{o})^3 = 10^{37}\g.$$

On the RD stage before decoupling, the Jeans length $\lambda_J$ of
the $OHe$ gas was restricted from below by the propagation of sound
waves in plasma with a relativistic equation of state
$p=\epsilon/3$, being of the order of the cosmological horizon and
equal to $\lambda_J = l_h/\sqrt{3} = t/\sqrt{3}.$ After decoupling
at $T = T_{od}$, it falls down to $\lambda_J \sim v_o t,$ where $v_o
= \sqrt{2T_{od}/m_o}.$ Though after decoupling the Jeans mass in the
$OHe$ gas correspondingly falls down
$$M_J \sim v_o^3 M_{od}\sim 3 \cdot 10^{-14}M_{od},$$ one should
expect a strong suppression of fluctuations on scales $M<M_o$, as
well as adiabatic damping of sound waves in the RD plasma for scales
$M_o<M<M_{od}$. It can provide some suppression of small scale
structure in the considered model for all reasonable masses of
O-helium. The significance of this suppression and its effect on the
structure formation needs a special study in detailed numerical
simulations. In any case, it can not be as strong as the free
streaming suppression in ordinary Warm Dark Matter (WDM) scenarios,
but one can expect that qualitatively we deal with Warmer Than Cold
Dark Matter model.

Being decoupled from baryonic matter, the $OHe$ gas does not follow
the formation of baryonic astrophysical objects (stars, planets,
molecular clouds...) and forms dark matter halos of galaxies. It can
be easily seen that O-helium gas is collisionless for its number
density, saturating galactic dark matter. Taking the average density
of baryonic matter one can also find that the Galaxy as a whole is
transparent for O-helium in spite of its nuclear interaction. Only
individual baryonic objects like stars and planets are opaque for
it.

\section{Signatures of O-helium dark matter}
The composite nature of O-helium dark matter results in a number of
observable effects.
\subsection{Anomalous component of cosmic rays}
O-helium atoms can be destroyed in astrophysical processes, giving
rise to acceleration of free $X^{--}$ in the Galaxy.

O-helium can be ionized due to nuclear interaction with cosmic rays
\cite{I,I2}. Estimations \cite{I,Mayorov} show that for the number
density of cosmic rays 
$$ n_{CR}=10^{-9}\cm^{-3}$$ 
during the age of
Galaxy a fraction of about $10^{-6}$ of total amount of OHe is
disrupted irreversibly, since the inverse effect of recombination of
free $X^{--}$ is negligible. Near the Solar system it leads to
concentration of free $X^{--}$ $ n_{X}= 3 \cdot 10^{-10}S_3^{-1}
\cm^{-3}.$ After OHe destruction free $X^{--}$ have momentum of
order $p_{X} \cong \sqrt{2 \cdot M_{X} \cdot I_{o}} \cong 2 \GeV
S_3^{1/2}$ and velocity $v/c \cong 2 \cdot 10^{-3} S_3^{-1/2}$ and
due to effect of Solar modulation these particles initially can
hardly reach Earth \cite{KK2,Mayorov}. Their acceleration by Fermi
mechanism or by the collective acceleration forms power spectrum of
$X^{--}$ component at the level of $X/p \sim n_{X}/n_g = 3 \cdot
10^{-10}S_3^{-1},$ where $n_g \sim 1 \cm^{-3}$ is the density of
baryonic matter gas.

At the stage of red supergiant stars have the size $\sim 10^{15}
\cm$ and during the period of this stage$\sim 3 \cdot 10^{15} \s$,
up to $\sim 10^{-9}S_3^{-1}$ of O-helium atoms per nucleon can be
captured \cite{KK2,Mayorov}. In the Supernova explosion these OHe
atoms are disrupted in collisions with particles in the front of
shock wave and acceleration of free $X^{--}$ by regular mechanism
gives the corresponding fraction in cosmic rays.

If these mechanisms of $X^{--}$ acceleration are effective, the
anomalous low $Z/A$ component of $-2$ charged $X^{--}$ can be
present in cosmic rays at the level $X/p \sim n_{X}/n_g \sim
10^{-9}S_3^{-1},$ and be within the reach for PAMELA and AMS02
cosmic ray experiments.

\subsection{Positron annihilation and gamma lines in galactic
bulge} Inelastic interaction of O-helium with the matter in the
interstellar space and its de-excitation can give rise to radiation
in the range from few keV to few  MeV. In the galactic bulge with
radius $r_b \sim 1 \kpc$ the number density of O-helium can reach
the value $n_o\approx 3 \cdot 10^{-3}/S_3 \cm^{-3}$ and the
collision rate of O-helium in this central region was estimated in
\cite{I2}: $dN/dt=n_o^2 \sigma v_h 4 \pi r_b^3 /3 \approx 3 \cdot
10^{42}S_3^{-2} \s^{-1}$. At the velocity of $v_h \sim 3 \cdot 10^7
\cm/\s$ energy transfer in such collisions is $\Delta E \sim 1 \MeV
S_3$. These collisions can lead to excitation of O-helium. If 2S
level is excited, pair production dominates over two-photon channel
in the de-excitation by $E0$ transition and positron production with
the rate $3 \cdot 10^{42}S_3^{-2} \s^{-1}$ is not accompanied by
strong gamma signal. According to \cite{Finkbeiner:2007kk} this rate
of positron production for $S_3 \sim 1$ is sufficient to explain the
excess in positron annihilation line from bulge, measured by
INTEGRAL (see \cite{integral} for review and references). If $OHe$
levels with nonzero orbital momentum are excited, gamma lines should
be observed from transitions ($ n>m$) $E_{nm}= 1.598 \MeV (1/m^2
-1/n^2)$ (or from the similar transitions corresponding to the case
$I_o = 1.287 \MeV $) at the level $3 \cdot 10^{-4}S_3^{-2}(\cm^2 \s
\MeV ster)^{-1}$.
\subsection{OHe in the terrestrial matter}
The evident consequence of the O-helium dark matter is its
inevitable presence in the terrestrial matter, which appears opaque
to O-helium and stores all its in-falling flux.

After they fall down terrestrial surface the in-falling $OHe$
particles are effectively slowed down due to elastic collisions with
matter. Then they drift, sinking down towards the center of the
Earth with velocity \cite{KK} \beq V = \frac{g}{n \sigma v} \approx
80 S_3 A^{1/2} \cm/\s. \label{dif}\eeq Here $A \sim 30$ is the
average atomic weight in terrestrial surface matter, $n=2.4 \cdot
10^{24}/A$ is the number of terrestrial atomic nuclei, $\sigma $ is
the cross section Eq.(\ref{sigOHe}) of elastic collisions of OHe
with nuclei, $v=\sqrt{3T/A m_p}$ is thermal velocity of nuclei in
terrestrial matter and $g=980~ \cm/\s^2$. Due to strong suppression
of inelastic processes (see below) they can not affect significantly
the incoming flux of cosmic O-helium in terrestrial matter.

Near the Earth's surface, the O-helium abundance is determined by
the equilibrium between the in-falling and down-drifting fluxes.

The in-falling O-helium flux from dark matter halo is taken as in work \cite{KK}
with correction on the speed of Earth
$$
  F=\frac{n_{0}}{8\pi}\cdot |\overline{V_{h}}+\overline{V_{E}}|,
$$
where $V_{h}$-speed of Solar System (220 km/s), $V_{E}$-speed of
Earth (29.5 km/s) and $n_{0}=3 \cdot 10^{-4} S_3^{-1} \cm^{-3}$ is
the local density of O-helium dark matter. Due to thermalization of
O-helium in terrestrial matter the velocity distribution of cosmic
O-helium is not essential (though we plan to study this question in
the successive work).

At a depth $L$ below the Earth's surface, the drift timescale is
$t_{dr} \sim L/V$, where $V \sim 400 S_3 \cm/\s$ is given by
Eq.~(\ref{dif}). It means that the change of the incoming flux,
caused by the motion of the Earth along its orbit, should lead at
the depth $L \sim 10^5 \cm$ to the corresponding change in the
equilibrium underground concentration of $OHe$ on the timescale
$t_{dr} \approx 2.5 \cdot 10^2 S_3^{-1}\s$.

The equilibrium concentration, which is established in the matter of
underground detectors at this timescale, is given in the form similar to \cite{KK} by
\begin{equation}
    n_{oE}=\frac{2\pi \cdot F}{V} = \frac{n_{0}}{320 S_{3} A^{1/2}} \cdot
    |\overline{V_{h}}+\overline{V_{E}}|,
\end{equation}
where, with account for $V_{h} > V_{E}$, relative velocity can be
expressed as
$$
    |\overline{V_{o}}|=\sqrt{(\overline{V_{h}}+\overline{V_{E}})^{2}}=\sqrt{V_{h}^2+V_{E}^2+V_{h}V_{E}sin(\theta)} \simeq
$$
$$
\simeq V_{h}\sqrt{1+\frac{V_{E}}{V_{h}}sin(\theta)}\sim
V_{h}(1+\frac{1}{2}\frac{V_{E}}{V_{h}}sin(\theta)).
$$
Here $\theta=\omega (t-t_0)$ with $\omega = 2\pi/T$, $T=1yr$ and
$t_0$ is the phase. Then the concentration takes the form
\begin{equation}
    n_{oE}=n_{oE}^{(1)}+n_{oE}^{(2)}\cdot sin(\omega (t-t_0))
    \label{noE}
\end{equation}

So, there are two parts of the signal: constant and annual
modulation, as it is expected in the strategy of dark matter search
in DAMA experiment \cite{Bernabei:2008yi}.

Such neutral $(^4He^{++}X^{--})$ ``atoms" may provide a catalysis of
cold nuclear reactions in ordinary matter (much more effectively
than muon catalysis). This effect needs a special and thorough
investigation. On the other hand, $X^{--}$ capture by nuclei,
heavier than helium, can lead to production of anomalous isotopes,
but the arguments, presented in \cite{I,FKS,KK} indicate that their
abundance should be below the experimental upper limits.

It should be noted that the nuclear cross section of the O-helium
interaction with matter escapes the severe constraints
\cite{McGuire:2001qj} on strongly interacting dark matter particles
(SIMPs) \cite{Starkman,McGuire:2001qj} imposed by the XQC experiment
\cite{XQC}. Therefore, a special strategy of direct O-helium  search
is needed, as it was proposed in \cite{Belotsky:2006fa}.

\section{OHe in underground detectors}
\subsection {OHe reactions with nuclei}
In underground detectors, $OHe$ ``atoms'' are slowed down to thermal
energies and give rise to energy transfer $\sim 2.5 \cdot 10^{-4}
\eV A/S_3$, far below the threshold for direct dark matter
detection. It makes this form of dark matter insensitive to the
severe CDMS constraints \cite{Akerib:2005kh}. However, $OHe$ induced
nuclear transformations can result in observable effects.

One can expect two kinds of inelastic processes in the matter with
nuclei $(A,Z)$, having atomic number $A$ and charge $Z$ \beq
(A,Z)+(HeX) \rightarrow (A+4,Z+2) +X^{--}, \label{EHeAZ} \eeq and
\beq (A,Z)+(HeX) \rightarrow [(A,Z)X^{--}] + He. \label{HeEAZ} \eeq

The first reaction is possible, if the masses of the initial and
final nuclei satisfy the energy condition \beq M(A,Z) + M(4,2) -
I_{o}> M(A+4,Z+2), \label{MEHeAZ} \eeq where $I_{o} = 1.6 \MeV$ is
the binding energy of O-helium and $M(4,2)$ is the mass of the
$^4He$ nucleus. It is more effective for lighter nuclei, while for
heavier nuclei the condition (\ref{MEHeAZ}) is not valid and
reaction (\ref{HeEAZ}) should take place.

In the both types of processes energy release is of the order of
MeV, which seems to have nothing to do with the signals in the DAMA
experiment. However, in the reaction (\ref{HeEAZ}) such energy is
rapidly carried away by $He$ nucleus, while in the remaining
compound state of $[(A,Z)X^{--}]$ the charge of the initial $(A,Z)$
nucleus is reduced by 2 units and the corresponding transformation
of electronic orbits should take place, leading to ionization
signal. It was proposed in \cite{I2,KK2} that this signal comes with
sufficient delay $> 10^{-7} \s$ after emission of He and is in the
range 2-6 keV. We present below some refining arguments for this
idea.

\subsection {Mechanisms of ionization from OHe reactions with nuclei}

Owing to its atom-like nature, O-helium is polarized in nuclear
Coulomb field. Since the OHe components have opposite electric
charge, $X^{--}$ is attracted by the nucleus, while $He^{++}$ is
repelled. If the energy, release due to capture of $X^{--}$ by
nucleus, exceeds the binding energy, the OHe system is disrupted and
there are two options for the liberated He.

1. It can tunnel through the Coulomb potential barrier and bind in
the nucleus together with X.

2. It is emitted from the atom.

>From the calculations of Appendix 1 one can conclude that the first
possibility is suppressed and the most probable is the second
option, when He is pushed out from the atom.

\begin{figure}
    \centering
        \includegraphics[width=12cm]{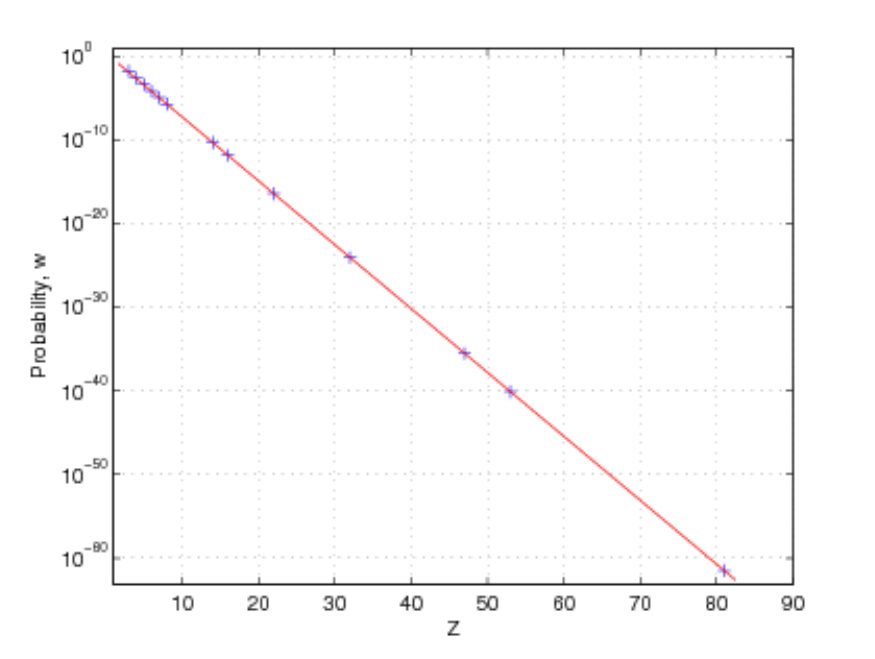}
\caption{\label{mk1fig1}%
The probability for He to
tunnel through the Coulomb potential barrier.}
\end{figure}

One should remark, that the probability of tunneling grows with the
decrease of the charge of nucleus $Z$ (see Fig.~\ref{mk1fig1}).

So for the lightest nuclei this probability is not negligible and
for Li it is on the level of few percents and for Be only one order
of magnitude smaller. It can provide a mechanism of anomalous
isotope production, which challenges the search for such isotopes
(e.g. for lithium $Li_{3}^{11+M_{X}}$).

Turning to the second case of emission of free He, let's estimate
the distance to the "break point", at which the potential energy
$U(r)$ of Coulomb interaction with nucleus (with the account for
screening by electron shells) becomes comparable with the binding
energy $I_{0}$ of OHe, which can be determined from the condition

$$
    U(r)=\frac{I_{o}}{2}.
$$

\begin{figure}
    \centering
        \includegraphics[width=12cm]{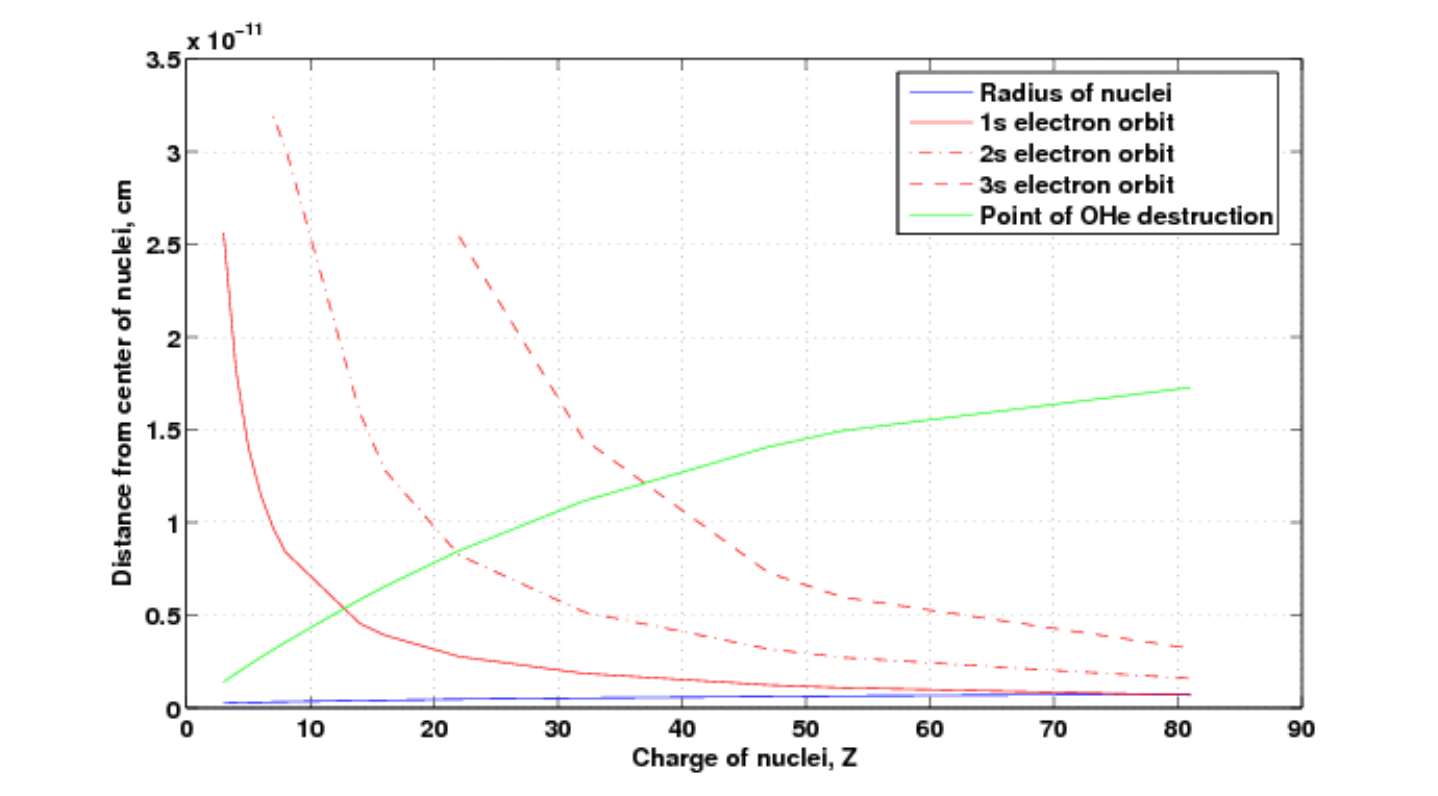}
\caption{\label{mk1fig2}%
Comparision of the distance between break point of OHe and nucleus with its size and with K-electron orbit of the atom.}
\end{figure}

This distance for various atoms is represented on Fig.~\ref{mk1fig2} in
comparison with the radius of K-electron orbit of the atom.

In particular for the iodine, this distance is of the order of
\begin{equation}
    r_{b}=\frac{2Z_{1}Z_{2}\alpha}{I_{o}}=1.5\cdot10^{-11}\cm,
\end{equation}

 In most of the atoms it is situated somewhere
between electron shells. However, as it has been shown on the
figure, for light nuclei the break point is between the nucleus and
K-electron orbit. It can lead to ionization of K-electron due to a
perturbation of radial nuclear Coulomb field and appearance of a
dipole component.

After breaking of the bound state, the both particles become free.
Since $r_b \ll r_{at}$, where $r_{at}$ is the size of atom, X
reaches the nucleus much quicker, than He leaves the atom. Therefore
the first ionization of atom can happen due to the dipole
perturbation of nuclear Coulomb field while He is present in the
atom. However, as it is shown in the Appendix 2, this effect is
negligible.

Immediate ionization is also possible due to the recoil momentum of
the nucleus. It causes drastic displacement of the center of Coulomb
field, what one can consider as a perturbation, which affects first
of all on the K-electron. The probability of ionization for
K-electron and  electron from the last outer shell is estimated in
Appendix 3. This estimation shows  that even the outer electron can
be emitted only in 1\% cases. The probability of ionization of
electrons from other levels is negligible. Therefore the
contribution from this mechanism is not essential. Similar
conclusion for the case of a WIMP-nucleus collision was obtained in
a detailed analysis of \cite{Bernabei:2007jz}

\begin{figure}
    \centering
        \includegraphics[width=12cm]{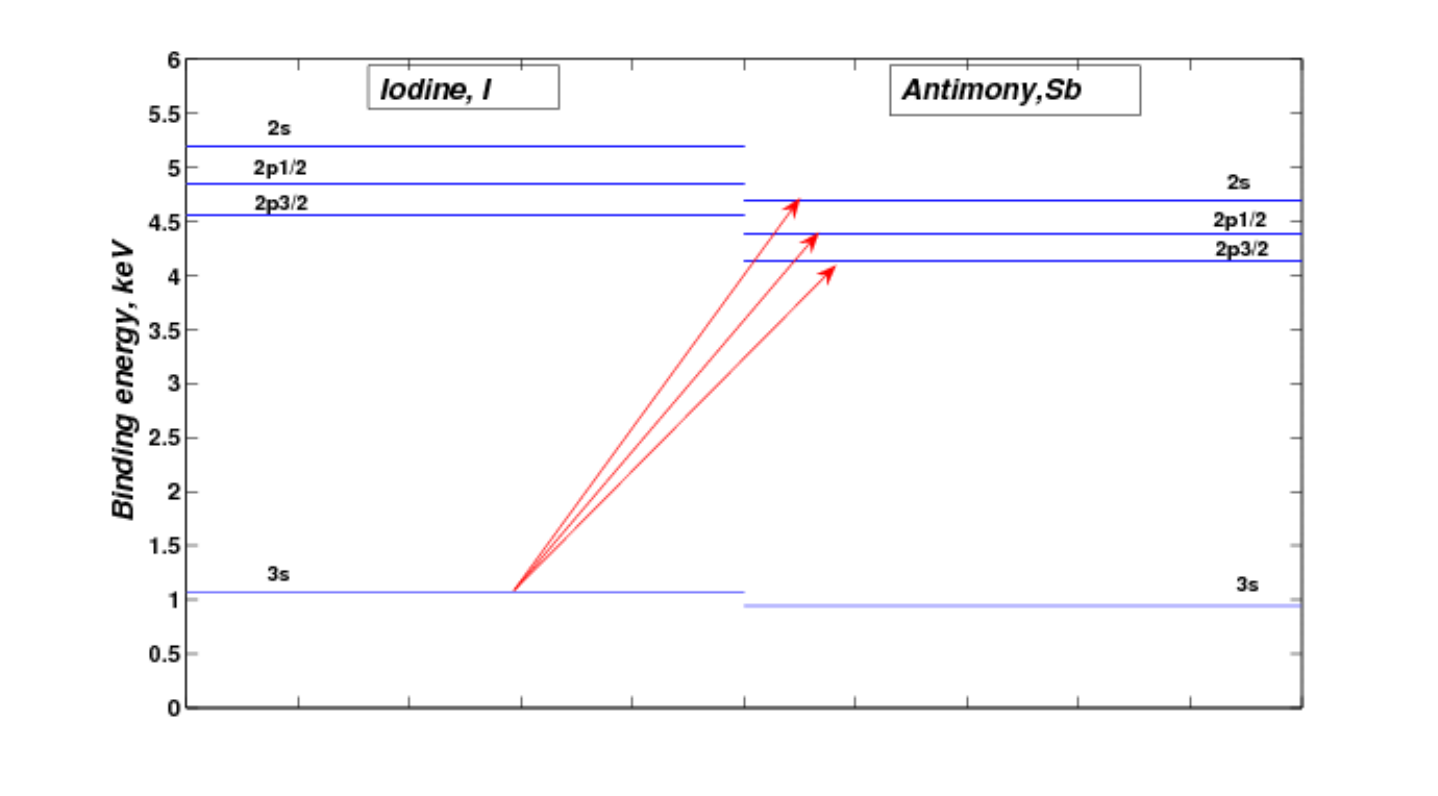}
\caption{\label{mk1fig3}%
Energy levels of electron shells in Antimony and Iodine.}
\end{figure}

Then the X particle enters the nucleus and the electrons begin to
feel the change of the Coulomb field. The nuclear charge decreases
by two units when the Iodine ($I_{53}^{127}$) nucleus converts into
anomalous Antimony ($Sb_{51}^{127+x}$), where $x$ is the mass of X
in atomic mass units. Due to change of the binding energy, electron
transitions with ionization and gamma ray emission take place. The
above arguments show that immediate ionization, which would be
inevitably masked by the large energy release from emitted He, is
suppressed. The structure of electronic shells can changes in a long
succession of atomic collisions, which can continue on scale up to
$10^{-3} \s$. The following transitions (look at Fig.~\ref{mk1fig3}) will give rise
to the signal in the range 2-6 keV. It refines the earlier arguments \cite{I2,KK2}.

\subsection {Ionization signal from OHe interaction in underground detectors}

To calculate the cross section of the inelastic reaction
(\ref{HeEAZ}) of $X^{--}$ capture by nucleus let's determine the
radius $R_d$, at which the field energy and the binding energy of
OHe are equal:
$$
    \frac{Z_X Z_{nuc}\cdot\alpha}{R_d}\cdot\frac{R_o}{R_d} = I_{o}.
$$
Here the radius of OHe $R_o$ is given by Eq.(\ref{REHe}), its
binding energy by Eq.(\ref{IO}), $Z_{nuc}$ - charge of initial
nuclei (for example, for Iodine $Z_{nuc}=51$) and $Z_X=Z_{He}=2$ is
the absolute value of charge of OHe components.

Then

\begin{equation}
    R_d^{2}=\frac{2 Z_X Z_{nuc}}{Z_{X}^{2}Z_{He}^2}\cdot\frac{1}{\alpha^{2}m_{He}^{2}}
\end{equation}

For rough estimation of the cross section we'll assume that the
inelastic process is determined by exchange of X in $t$-channel. It
gives a factor $(\frac{\Delta E}{m_{X}})^{2}$ in the probability of
the process (\ref{HeEAZ}), where $\Delta E = I_z-I_o$ and $I_z>I_o$
is the binding energy of X in the nucleus. Then the cross section
has the form

\begin{equation}
    \sigma_{total}=\sigma_{d}\cdot(\frac{\Delta E}{m_{X}})^{2},
\end{equation}

where $\sigma_{d}=\pi R_d^{2}$.

The relative probability for ionization signal in the range 2-6 keV,
taking place with delay $>2 \cdot 10^{-7} \s$ after emission of He
is taken into account by factor $f$. The actual value of this factor
is the subject of our further detailed analysis. Here we assume that
it's value is $0.01 < f < 0.1$.

Concentration of OHe in the matter of detector is given by
(\ref{noE}) and velocity of particles in thermal equilibrium inside
detector is $ V_{nuc}=\sqrt{\frac{3kT}{m_{nuc}}}.$ For Iodine it
equals $V/c=2.4\cdot 10^{-5}$ and for OHe $V/c=8.6\cdot10^{-6}\cdot
S_{3}^{-1/2}$.

So, the count rate in DAMA detector is

\begin{equation}
    N_{CR}=N_{CR}^{(1)}+N_{CR}^{(2)}\cdot sin(\omega(t-t_{0}))
    \label{rate0}
\end{equation}
$$
    N_{CR}^{(i)}=f n_{oE}^{(i)}\cdot |{\overrightarrow{V_{OHe}}-\overrightarrow{V_{nuc}}}| \cdot \sigma_{total} \cdot  \frac{N_{A}}{M},
$$
where $i=1,2$. Then total amount of events during the time $t \gg 2
\pi /\omega$ is determined by the constant part of the signal and is
given by

\begin{equation}
    N_{tot}= f n_{oE}^{(1)}\cdot |{\overrightarrow{V_{OHe}}-\overrightarrow{V_{nuc}}}| \cdot \sigma_{total} \cdot t \cdot
    \frac{N_{A}}{M},
    \label{rate}
\end{equation}

where $N_{A}$ - Avogadro's number,

$M$ - molar Iodine mass.

\begin{figure}
    \centering
        \includegraphics[width=12cm]{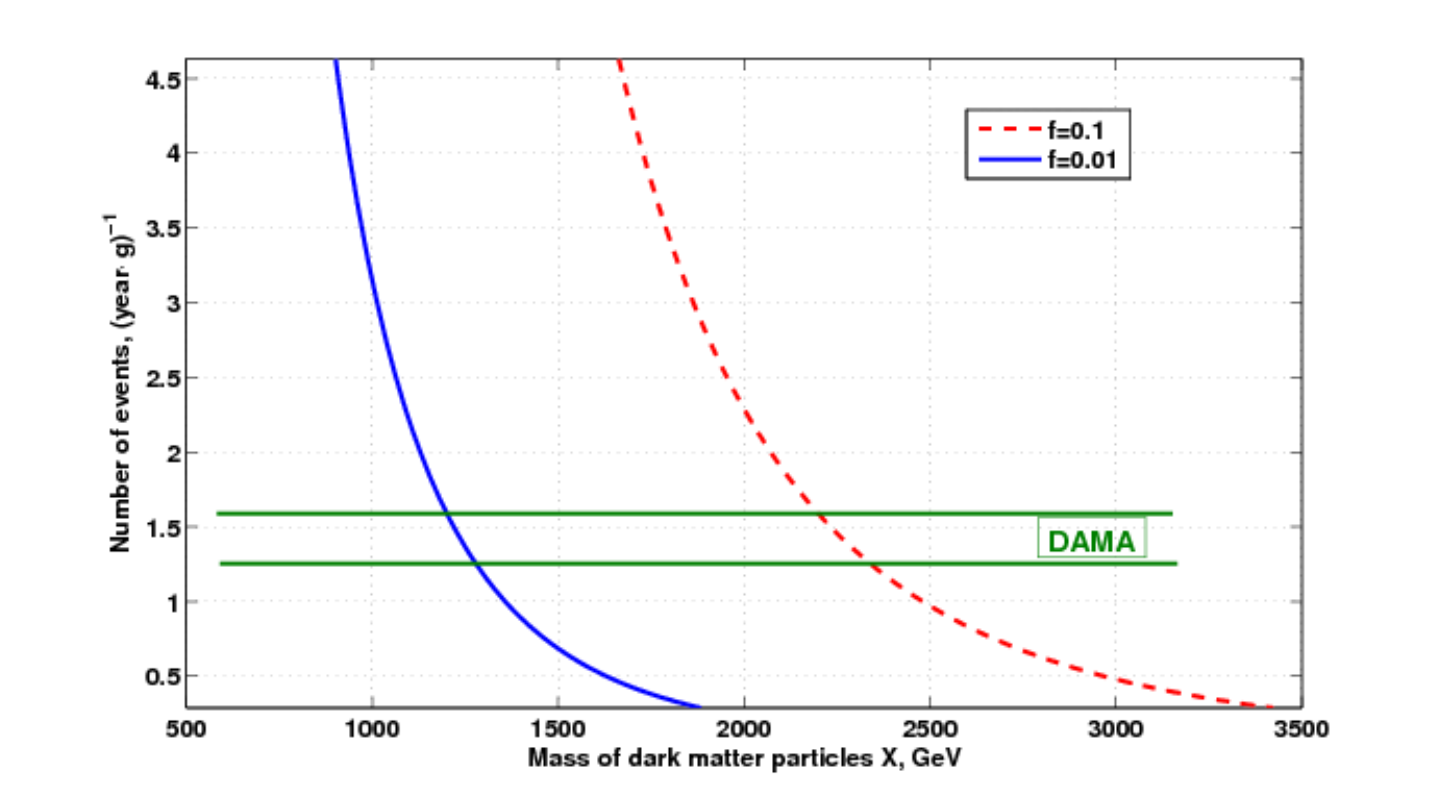}
\caption{\label{mk1fig4}%
Dependence for number of events per gramm of Iodine per year of work from the dark matter particle mass (for $f=0.1$ and $0.01$).}
\end{figure}

The part of dependence for the number of events per gramm per year
on the mass of X with the amount of the events that corresponds to
the observed amount of events 1.46 $\pm 3 \sigma$
\cite{Bernabei:2008yi} is given on Fig.~\ref{mk1fig4} (for factor $f=0.1$ and
$f=0.01$).

Thus, at $0.01<f<0.1$ X particles with masses of about 1-2 TeV can
explain the signal in DAMA experiment.

The Thallium concentration in NaI(Tl) detector is only 0.1
\% from Iodine concentration \cite{DAMA_app}, so the amount of
events induced by OHe reaction with this element will be by 1000
times less. Therefore one can neglect the effects of such reaction
with Thallium in the explanation of results from DAMA/NaI detector.

\subsection {No signal in other detectors}
The absence of the same result in other experiments (such as CDMS
\cite{Akerib:2005kh}) can follow from the difference in their
strategy.

For example in CDMS the working matter of detector is cooled down to
the extralow temperature that leads to the suppression of the number
of events (\ref{rate}) by two orders of magnitude. It was also shown
in \cite{KK2} that nuclear recoil from reactions (\ref{HeEAZ}) in
CDMS  is below the threshold of registration while the effects of
ionization, not accompanied by bolometric recoil are not considered
as events \cite{CDMS_app}.

Moreover this experiment is using semiconductor detectors. In these
conditions the nuclei of atom are surely fixed in the knot of
crystal lattice and electrons feel changing of the Coulomb potential
very slow. So the probability of ionization has additional factor of
suppression.

\section{Conclusions}

To conclude, the results of dark matter search in experiments
DAMA/NaI and DAMA/LIBRA can find explanation in the framework of
composite dark matter scenario without contradiction with negative
results of other groups. This scenario can be realized in different
frameworks, in particular in Minimal Walking Technicolor model or in
the approach unifying spin and charges and contains distinct
features, by which the present explanation can be distinguished from
other recent approaches to this problem \cite{Edward} (see also
review and more references in \cite{Gelmini}).

The mechanisms of ionization induced by OHe reactions with nuclei
were considered. It is argued that if in result of OHe interaction
with matter of DAMA detector, the energy release in ionization comes
in the range of 2-6 keV with sufficient delay, it can be
distinguished from from immediate large energy release due to X
capture by nucleus. Quantitative analysis of this explanation
implies detailed study of a possibility for delayed transitions in
atoms, perturbed by a rapid change of the charge of nucleus. An
analogy with rearrangement of atomic shells after $\alpha$ decay of
nucleus might be helpful in this analysis.

OHe concentration in matter of underground detectors follows the
change in the incoming cosmic flux with the relaxation time of few
minutes. It leads to annual modulations of the ionization signal
from OHe reactions.

The method to calculate the rate of OHe reactions was developed and
the calculated total amount of such events can to be consistent with
the results of DAMA/NaI and DAMA/LIBRA experiments for the mass
of OHe around 2 TeV. This method can be applied to the analysis of the
whole set of inelastic processes, induced by O-helium in matter.

An inevitable consequence of the proposed explanation is appearance
in the matter of DAMA/NaI or DAMA/LIBRA detector anomalous
superheavy isotopes of antinomy (Sb with nuclear charge $Z=53-2=51$)
and $10^3$ smaller amount of anomalous gold (Au with nuclear charge
$Z=81-2=79$), created in the inelastic process (\ref{HeEAZ}) and
having the mass roughly by $m_o$ larger, than ordinary isotopes of
these elements. If the atoms of these anomalous isotopes are not
completely ionized, their mobility is determined by atomic cross
sections and becomes about 9 orders of magnitude smaller, than for
O-helium. It provides conservation in the matter of detector of at
least 200 anomalous atoms per 1g, corresponding to the number of
events, observed in DAMA experiment. Therefore mass-spectroscopic
analysis of this matter can provide additional test for the O-helium
nature of DAMA signal. Similar mechanism can lead to presence of
anomalous magnesium and zinc in the matter of CDMS detector.

An interesting aspect of our results is the challenging possibility
of creation of anomalous isotopes of light elements like anomalous
lithium $Li_{3}^{11+M_{X}}$ (from usual Li bound with OHe and from B
bound with X), and of anomalous hydrogen $H_{1}^{7+M_{X}}$ (from
lithium bound with X).

\section*{Acknowledgments}

We would like to thank Organizers of Bled2008 for creative
atmosphere of the Workshop and to P. Belli, K.M. Belotsky, R. Bernabei,
A.M. Galper, J. Filippini, C. Kouva\-ris, F. Lebrun, 
N.S.Manko\v c Bor\v stnik and P. Picozza for stimulating discussions.

\section*{Appendix 1}
\begin{figure}
    \centering
        \includegraphics[width=10cm]{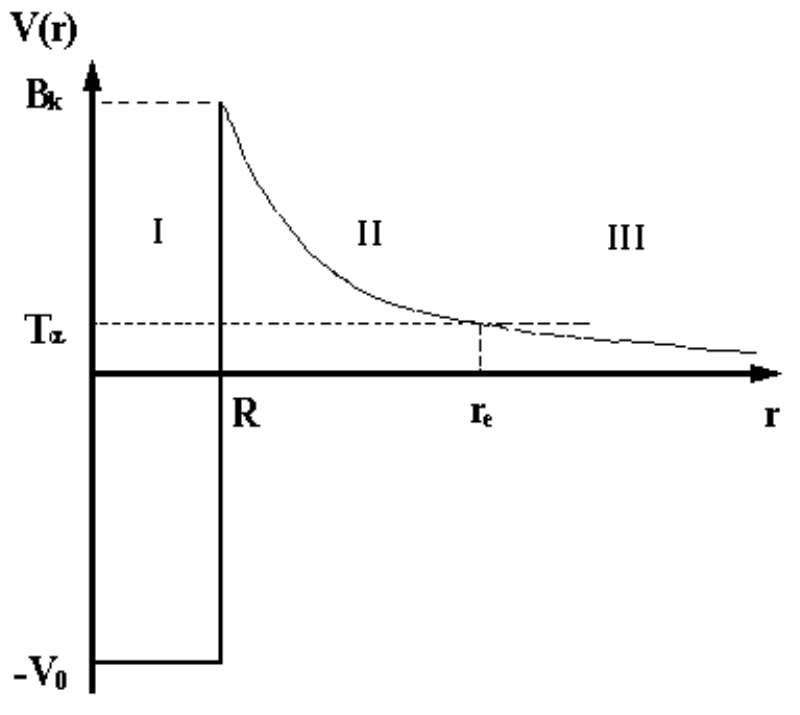}
\caption{\label{mk1fig5}%
Coulomb barrier of the nucleus.}
\end{figure}

 The potential barrier of nucleus is
presented on Fig.~\ref{mk1fig5},where $T_{\alpha}$ - kinetic energy of
$\alpha$-particle inside OHe, $r_{0}$ - characteristic distance of
feeling the Coulomb barrier, R - radius of nuclei.

The height of this barrier $B_{k}$ is given by
$$
    B_{k}=\frac{Z_{1}Z_{2}\alpha}{r_{0}A^{\frac{1}{3}}}
$$
and equal to 20.9 MeV for Iodine.

After the destruction of OHe, kinetic energy of helium equals to the
binding energy of OHe state, which we take $I_{o}=1.289 MeV$ with
the account for charge distribution in He \cite{Pospelov}. Effect of
nonzero velocity of OHe before destruction is only 0.1\% and can be
neglected.

 The probability of tunneling through the Coulomb barrier is given by
the formula

\begin{equation}
w \sim exp(\frac{-2}{\hbar}\cdot
\int_{R}^{r_{0}}\sqrt{2m(U(r)-T_{He})}dr)
\end{equation}
where $\alpha = B_{k} \cdot R$. In case of potential $U(r)\sim
\frac{\alpha}{r}$ the integral is equal to
$$
w \sim
exp(-\frac{2\alpha}{\hbar}\sqrt{\frac{2m}{T_{He}}}[arccos(\sqrt{\frac{T_{He}\cdot
r_{e}}{\alpha}}-\sqrt{\frac{T_{He}\cdot
r_{e}}{\alpha}(1-\frac{T_{He}\cdot r_{e}}{\alpha})})])
$$
and the probability is
$$w \sim exp(-40) \ll 1.$$

\section*{Appendix 2}

Effect of instantaneous ionization of atomic K-electron after
$X^{--}$ is captured by nucleus in reaction (\ref{HeEAZ}) is similar
to the same effect during $\alpha$-decay of a nucleus. Velocity of
the $\alpha$-particle is negligible as compared with the velocity of
K-electron, however the time, that it spend to leave the nucleus, is
negligible as compared with the period of electron orbit rotation.

Perturbation that produces ionization in this case is the deviation
of the combined nucleus and $\alpha$-particle field from ordinary
Coulomb field 1/r. In the result of this deviation appears dipole
momentum. On the other hand, the perturbation is efficient only
during the period, when $\alpha$-particle is at small distances from
the nucleus.

The initial state of the system is $(A+M+m,Z)=(A_{0},Z)$, then the
final state of the system is $[ (A+M,Z-q);(m,q) ] = [
(A_{1},Z-q);(m,q) ]$, where $q$ and $m$ are the charge and mass of
He, $Z$ and $A$ are the charge and mass of nucleus and $M$ is the
mass of X.

Dipole momentum equals to

\begin{equation}
    P=\frac{qA_{1}-(Z-q)m}{A_{0}}r_{0},
\end{equation}

 where $\overrightarrow{r_{0}}=\overrightarrow{r}_{nuclear}-\overrightarrow{r_{\alpha}}=\overrightarrow{V}t$
is the relative radius-vector of the nucleus and He and
$\overrightarrow{V}$ - relative velocity.

Dipole component of the field is given by

$$
    V=\frac{P\overrightarrow{z}}{r^{3}}.
$$
Here $\overrightarrow{z}$ is the direction along the velocity
 $\overrightarrow{V}$.

This component causes a perturbation for the electron on the orbit.
According to perturbation theory the probability of the transition
is determined by the matrix element of the transition from state 0
to the state with momentum k.

\beq
    V_{0k}=\int \psi_{k}^{*} V \psi_{0} dq = P \int \psi_{k}^{*} (\frac{z}{r^{3}}) \psi_{0} dq
\eeq

The equation of motion of the electron on the shell reads as

$$
    \ddot{z}=-\frac{Zz}{r^{3}}.
$$

Then

$$
    V_{0k}= P \int \psi_{k}^{*} (\ddot{z}) \psi_{0} dq =
    P\frac{(E-E_{0})^2}{Z} z_{0k}
$$

The probability for emission of electron from the shell to all the
finite states is given by $$\frac{dw}{dk}=2|\int
V_{0k}e^{i(E_{0}-E)t}dt|^2 =$$
\begin{equation}
    = 2|\int (\frac{qA_{1}-(Z-q)m}{A_{0}}) \overrightarrow{V}
    t \frac{(E-E_{0})^2}{Z}z_{0k} e^{i(E-E_{0})t} dt|^{2}
\end{equation}

With the use of a new multiplier $e^{-\lambda t}$, the integral is
two taken times by parts. Then the integral from the imaginary
exponent is taken and $\lambda \rightarrow 0$.

Thus

\begin{equation}
    \frac{dw}{dk}=\frac{1}{\pi}(\frac{qA_{1}-(Z-q)m}{A_{0}Z})^2|z_{0k}|^2
\end{equation}

Since the mass of the X $M \gg m$ the result is reduced to

\begin{equation}
    \frac{dw}{dk}=\frac{1}{\pi}(\frac{qA_{1}}{A_{0}Z})^2|z_{0k}|^2.
\end{equation}

Here $z_{0k}$ can be calculated if one takes into account that
$z=rcos(\theta)$. Then
$$
    |z_{0k}|^2=|r_{0k}|^2|cos(\theta)_{0k}|^2=\frac{1}{3}|r_{0k}|^2,
$$
where $|r_{0k}|^2$ could be calculated owing to radial function
$R_{k0}$
$$
    R_{k0}=\frac{2}{k^2}\sqrt{\frac{k!}{(k-1)!}}e^{-\frac{r}{k}}F(-k+1,2,r)=
    \frac{2}{k\sqrt{k}}e^{-\frac{r}{k}}F(-k+1,2,r)
$$
Function $F$ could be found from the integral
\begin{equation}
    J_{\alpha\gamma}^{\nu}=\int_{0}^{\infty}e^{-\lambda
    z}z^{\nu}F(\alpha,\gamma,kz)dz,
\end{equation}
where one can find J as
\begin{equation}
    J_{\alpha\gamma}^{\gamma + n
    -1}=(-1)^{n}\Gamma(\gamma)\frac{d^{n}}{d\lambda^{n}}[\lambda^{\alpha-\gamma}(\lambda-k)^{-\alpha}]
\end{equation}
In our case $\alpha=-k+1; \gamma=2; kz=r$.

In the result the final answer reads as

\begin{equation}
    dw=\frac{2^{11}(A_{1}-2Z+4)^2V^2}{3A_{0}^2Z^6(1+(\frac{k}{Z})^2)^5}\frac{1}{1-e^{\frac{2\pi
    Z}{k}}}e^{-4\frac{Z arctg(\frac{k}{Z})}{k}}
\end{equation}

In the limit of $M\rightarrow0$ this result is transformed to the
result, given in \cite{LL3}.

\begin{figure}
    \centering
        \includegraphics[width=12cm]{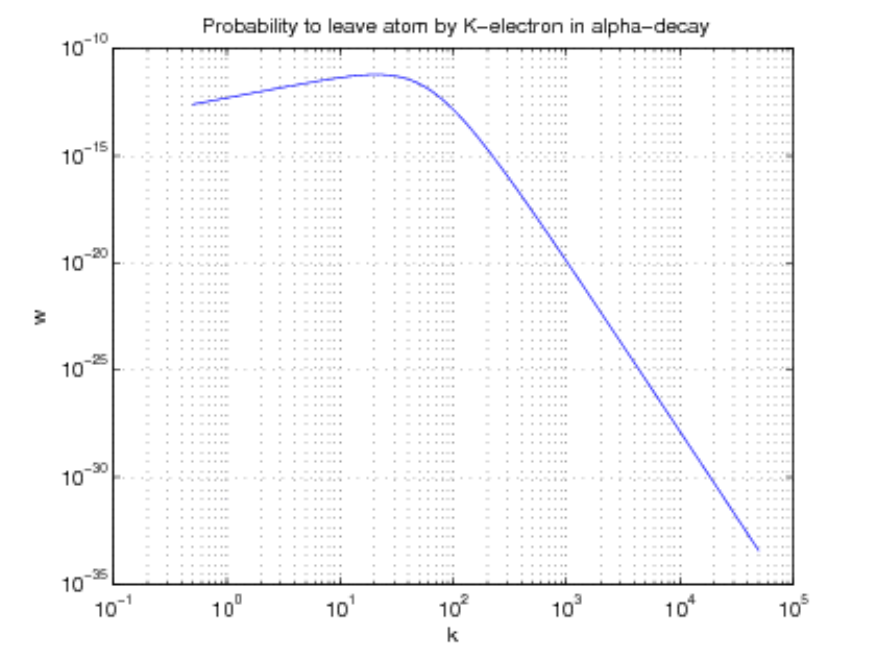}\\
        (a)\\
        \includegraphics[width=12cm]{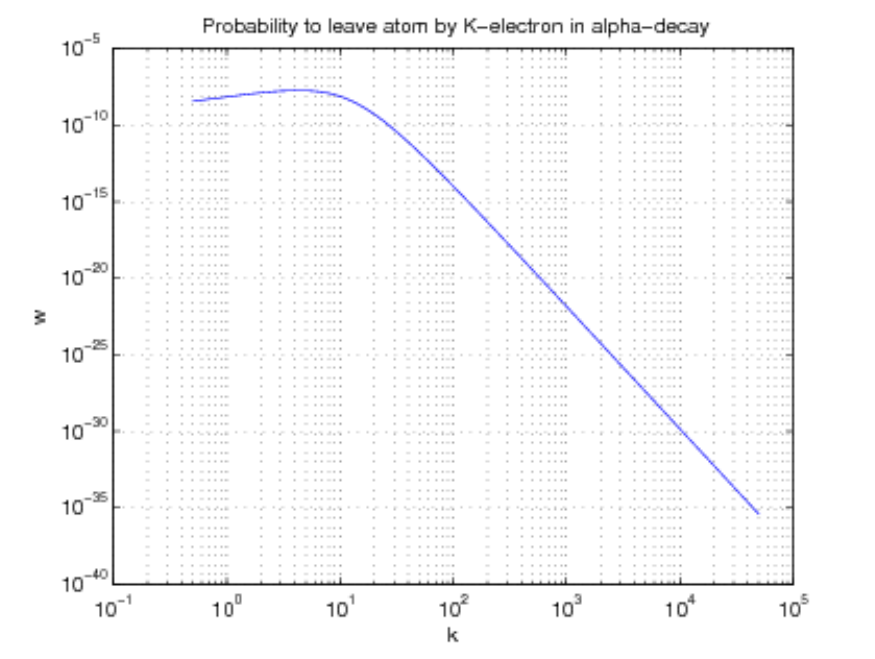}\\
        (b)
\caption{\label{mk1fig6}%
(a): Probability of the leaving Iodine atom by K-electron in the
studying process. (b): Probability of the leaving Natrium atom by K-electron in the studying process.}
\end{figure}

The probability distribution for K-electron ionization for Iodine
 and Natrium are presented correspondingly on Fig.~\ref{mk1fig6}(a)
 and (b) in the assumption that He
 leaves the nucleus with the velocity equal to 1/137).

The probabilities of K-electron emission from atoms of Iodine and
Natrium are equal to
 $2\cdot10^{-10}$ and $1.5\cdot10^{-7}$, respectively. Therefore the
 considered mechanism of ionization is not effective.

\section*{Appendix 3}

The probability of excitation and ionization of atom, when nucleus
acquires recoil momentum and obtains velocity \emph{V}.

The expansion of final wave functions through the initial ones is in
the form $\psi^{'}=\psi e^{-iq\sum r_{a}},$ where $q$ is momentum of
the recoil nucleus and $r_{a}$ is radius-vector of electrons.

So probability of the transition is:

\begin{equation}
    w_{k0}=|<k|e^{-iq\sum r_{a}}|0>|^{2}=1-w_{00}=1-|\int \psi^{2} e^{-iq\sum r_{a}} dV|^{2}
\end{equation}

The integral can be taken analytically in the approximation $qr_{n}
<< 1$ for the wave functions from hydrogen atom.

Then

\begin{equation}
    w_{k0}=1-\frac{1}{(1+\frac{1}{4}q^2r_{n}^2)^4},
\end{equation}
where $a$ is a radius of the considered electron shell,

    $$
    \left\{
        \begin{aligned}
            \frac{m (V_{n}^{e})^2}{2}=I_{n} \\
            mV_{n}^{e}r_{n}=n\hbar \\
        \end{aligned}
    \right.
    $$
$\varepsilon$ is the binding energy of the electron on shell n and

$V_{n}^{e}$ is its velocity.

The radius of the electron shell is equal to

\begin{equation}
    r_{n}=\frac{\hbar}{\sqrt{2m I_{n}}}.
\end{equation}

Then
\begin{equation}
    qr_{n}=\frac{mV_{nuc}}{\hbar} \cdot \frac{\hbar}{\sqrt{2m \varepsilon_{n}}}=\sqrt{\frac{m}{2\varepsilon_{n}}} V_{nuc}
\end{equation}

The probability of ionization for K-electron and  electron from the
last outer shell are given in table.

\begin{center}
\begin{tabular}{|l|l|l|}
\hline
Antimony $Sb_{51}^{127}$ & $qa$ & w \\
\hline
1s & $1.8 \cdot 10^{-3}$ & $3.5 \cdot 10^{-6}$ \\
\hline
5p & 0.109 & 0.01 \\
\hline
\end{tabular}
\end{center}

%
\def\snmbA{{\cal A}}
\def\snmbB{{\cal B}}
\author{D. Lukman${}^1$, N.S. Manko\v c Bor\v stnik${}^1$ and H. B. Nielsen${}^2$}
\title{Spin Connection Makes Massless Spinor Chirally Coupled to 
Kaluza-Klein Gauge Field After Compactification 
of $M^{1+5}$ to $M^{1+3} \times$ Infinite Disc Curved on $S^2$}
\institute{%
${}^1$ Department of Physics, FMF, University of Ljubljana\\
Jadranska 19, 1000 Ljubljana, Slovenia\\
${}^2$ Department of Physics, Niels Bohr Institute\\
Blegdamsvej 17,Copenhagen, DK-2100}
 
\titlerunning{Spin Connection Makes Massless Spinor Chirally Coupled}
\authorrunning{D. Lukman, N.S. Manko\v c Bor\v stnik and H. B. Nielsen}
\maketitle

\begin{abstract} 
One step towards realistic Kaluza-Klein-like theories is presented for a toy model 
of a spinor in $d=(1+5)$ compactified on an infinite disc with the zweibein which makes a disc 
curved on $S^2$ and with the spin connection 
field  which allows on such a sphere only one massless spinor state of a particular charge, 
which couples the spinor chirally to the corresponding Kaluza-Klein gauge field. 
In refs.~\cite{snmb1:hnkk06,snmb1:hn07} we achieved masslessness of spinors with  the appropriate 
choice of a boundary on a finite disc, in this paper the masslessness is achieved with 
the choice of a spin connection 
field on a curved infinite disc.  
\end{abstract}

\section{Introduction}
\label{snmb1:introduction}

Kaluza-Klein-like theories, which are extremely elegant in the unification of 
all the interactions, have difficulties with masslessness of spinors after the compactification 
of some of the dimensions. The Approach unifying spins and charges (proposed by one of 
us---S.N.M.B.) is a Kaluza-Klein-like 
theory~\cite{snmb1:norma92,snmb1:norma93,snmb1:norma94,snmb1:norma95,snmb1:Portoroz03,snmb1:pikanorma06,snmb1:gmdn06,snmb1:gmdn07}, 
which is besides unifying spins and charges into only the spin  offering also the 
mechanism for generating families. This approach     
can be accepted as the theory showing a new way beyond the Standard model of the 
electroweak and colour interactions only after solving many open problems, among which is also  
the problem of the masslessness of spinors after the break of the starting symmetry 
in $d=(1+(d-1))$, $d \ge 14$ up 
to the symmetry assumed by the Standard model before the break of the 
electroweak symmetry in $d=(1+3)$. 

Some of the open problems, common to all the Kaluza-Klein-like theories, are discussed 
in the paper of Witten~\cite{snmb1:witten} 
within the eleven-dimensional supergravity in a transparent way, leaving a strong  
impression that there is no hope, that the Kaluza-Klein-like 
theories can ever lead to the 
''realistic'' (observable) theory, since there is almost no hope for masslessness of quarks and 
leptons at the low energy level and no hope for the appearance of families. 
Many an open question of the Kaluza-Klein-like theories 
stays open also in other theories, like in theories of membranes, which  assume that 
the dimension of space is more than  1+3. The question of masslessness of spinors at low energies 
as well as the appearance of families is also in these theories a  hard not yet solved problem.

One of us is trying for long to develop the  Approach unifying spins and charges so that 
spinors which carry in $d\ge 4$ nothing but two kinds of the spin (no charges), would manifest 
in $d=(1+3)$ all the properties assumed by the Standard model. The Approach  
proposes in $d=(1 + (d-1))$  a simple starting action for spinors  with the two kinds of the 
spin generators: the Dirac one, which takes care of the spin and the charges, and the second one, 
anticommuting with the Dirac one, which generates families~\footnote{To understand the 
appearance of the two kinds of the spin generators we invite the reader to look at the 
refs.~\cite{snmb1:pikanorma06,snmb1:holgernorma2002,snmb1:holgernorma2003}.}. A spinor couples in $d=1+13$ 
to only the vielbeins and (through two kinds of the spin generators to) the 
spin connection fields.  
Appropriate breaks of the starting symmetry lead to the 
left handed quarks and leptons in $d=(1+3)$, which carry the weak charge while the 
right handed ones are weak chargeless. The Approach might have the right answer to  
the questions about the origin of families of quarks and leptons, about the explicit 
values of their masses and mixing matrices  
as well as about the masses of the scalar  and the weak gauge fields, about the dark 
matter candidates, and about the break of the discrete symmetries\footnote{  
There are many possibilities in the Approach unifying spins and charges for breaking 
the starting symmetries to those of the Standard model. These problems were studied 
in some crude approximations in refs.~\cite{snmb1:gmdn06,snmb1:gmdn07}. It was also studied~\cite{snmb1:hnm06}  
how does the Majorana  mass of spinors depend on the dimension of space-time 
if spinors carry only the spin and no charges. We have proven 
that only in even dimensional spaces of  $d=2$ 
modulo $4$ dimensions (i.e.\ in $d=2(2n+1),$ $n=0,1,2,\cdots$) spinors  (they are allowed 
to be in families) of one handedness and with no conserved 
charges gain no  Majorana mass.}. 

In the refs.~\cite{snmb1:hnkk06,snmb1:hn07} we demonstrated that the appropriate boundary condition   
ensures that a Weyl spinor in $d=(1+5)$ stays massless after the break 
of the starting symmetry to the symmetry of the flat disc with the boundary and the $(1+3)$ 
 space, carrying one charge only~\footnote{Let us remind the reader that 
after the second quantization procedure the oppositely charged anti-particle appears 
anyhow.} and coupling chirally to the corresponding gauge fields.   

In this paper we study a similar toy model as in the refs.~\cite{snmb1:hnkk06,snmb1:hn07}: 
a Weyl spinor in $d=(1+5)$, which breaks into $M^{1+3}$ and this time to  an infinite disc, 
which  the zweibein  curves  into $S^2$, while the chosen spin 
connection field allows on $S^2$ only one massless state of only one charge, since for 
this  spin the spin connection field and the zweibein cancel each other. 
Then the charge of 
the massless spinor couples it  to the corresponding gauge field. 
In $d=2$ the spin connection field and the zweibein (with no spinor sources) 
namely decouple from each other 
(there are not enough indices to make them coupled~\cite{snmb1:dhnproc04,snmb1:deser95}). 

\begin{figure}
\centering
\includegraphics{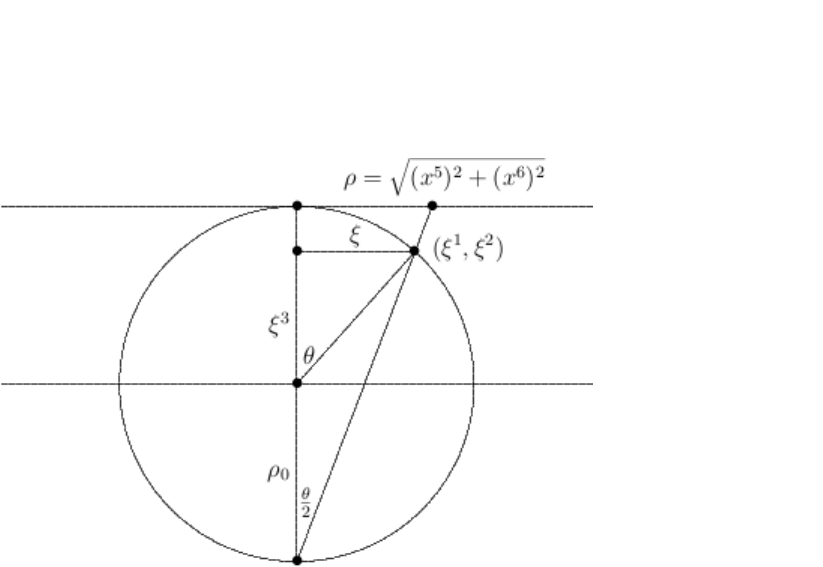}
\caption{The disc is curved on the sphere $S^2$. 
\label{snmb1:discgrav}}
\end{figure}

We take (as we did in ref.~\cite{snmb1:hnkk06,snmb1:hn07}) for the covariant momentum of a spinor  
\begin{eqnarray}
p_{0 a} &=& f^{\alpha}{\!}_{a}p_{0 \alpha}, \quad p_{0 \alpha} \psi = 
p_{ \alpha} - \frac{1}{2} S^{cd} 
\omega_{cd \alpha}. 
\label{snmb1:covp}
\end{eqnarray}
A spinor carries in $d\ge 4$ nothing but 
a spin and interacts accordingly with only the gauge fields of the corresponding 
generators of the infinitesimal transformations (of translations and  the 
Lorentz transformations in the space of spinors), that is with 
vielbeins $f^{\alpha}{\!}_{a}$ \footnote{$f^{\alpha}{}_{a}$ are inverted vielbeins to 
$e^{a}{}_{\alpha}$ with the properties $e^a{}_{\alpha} f^{\alpha}{\!}_b = \delta^a{\!}_b,\; 
e^a{\!}_{\alpha} f^{\beta}{\!}_a = \delta^{\beta}_{\alpha} $. 
Latin indices  
$a,b,..,m,n,..,s,t,..$ denote a tangent space (a flat index),
while Greek indices $\alpha, \beta,..,\mu, \nu,.. \sigma,\tau ..$ denote an Einstein 
index (a curved index). Letters  from the beginning of both the alphabets
indicate a general index ($a,b,c,..$   and $\alpha, \beta, \gamma,.. $ ), 
from the middle of both the alphabets   
the observed dimensions $0,1,2,3$ ($m,n,..$ and $\mu,\nu,..$), indices from 
the bottom of the alphabets
indicate the compactified dimensions ($s,t,..$ and $\sigma,\tau,..$). 
We assume the signature $\eta^{ab} =
diag\{1,-1,-1,\cdots,-1\}$.
} and  spin connections $\omega_{ab\alpha}$ (the gauge fields of 
$S^{ab}= \frac{i}{4}(\gamma^a \gamma^b - \gamma^b \gamma^a)$).
The corresponding Lagrange density 
for   a Weyl spinor has the form
${\cal L}_{W} = \frac{1}{2} [(\psi^{\dagger} E \gamma^0 \gamma^a p_{0a} \psi) + 
(\psi^{\dagger} E \gamma^0\gamma^a p_{0 a}
\psi)^{\dagger}]$, leading to
\begin{eqnarray}
{\cal L}_W &=& \psi^{\dagger}  \gamma^0 \gamma^a E \{f^{\alpha}_a p_{\alpha} +
\frac{1}{2E} \{p_{\alpha},f^{\alpha}_a E\}_-   -\frac{1}{2} S^{cd}  \omega_{cda}) 
 \}\psi,
\label{snmb1:weylL}
\end{eqnarray}
with $ E = \det(e^a{\!}_{\alpha}), $ where  
\begin{eqnarray}
  \omega_{cda} &=& \Re e \;\omega_{cda}, \;\; {\rm if \;\; c,d,a\;\; all\;\; different} \nonumber\\
               &=& i\,\Im m\; \omega_{cda},\;\; \rm{otherwise}.
\label{snmb1:p0s}
\end{eqnarray}
Let us have no gravity in $d=(1+3)$ ($f^{\mu}_m = \delta^{\mu}_m$ and  
$\omega_{mn\mu}=0$ for $ m,n=0,1,2,3, \mu =0,1,2,3 $) and let us make  a choice of  a 
zweibein on our disc (a two dimensional infinite plane with the rotational symmetry around the 
axes perpendicular to the plane)
\begin{eqnarray}
e^{s}{}_{\sigma} = f^{-1}
\begin{pmatrix}1  & 0 \\
 0 & 1 \end{pmatrix},
f^{\sigma}{}_{s} = f
\begin{pmatrix}1 & 0 \\
0 & 1 \end{pmatrix},
\label{snmb1:fzwei}
\end{eqnarray}
with 
\begin{eqnarray}
\label{snmb1:f}
f &=& 1+ (\frac{\rho}{2 \rho_0})^2= \frac{2}{1+\cos \vartheta},\nonumber\\ 
x^5 &=& \rho \,\cos \phi,\quad  x^6 = \rho \,\sin \phi, \quad E= f^{-2}.
\end{eqnarray}
The last relation follows  from $ds^2= 
e_{s \sigma}e^{s}{\!}_{\tau} dx^{\sigma} dx^{\tau}= f^{-2}(d\rho^{2} + \rho^2 d\phi^{2})$.
The zweibein curves the infinite disc on  the $S^2$ sphere with the radius $\rho_{0}$. We make a 
choice  of the spin connection field
\begin{eqnarray}
  \omega_{st \sigma} &=& i \varepsilon_{st}\; \frac{4F x_{\sigma}}{\rho}
  \frac{f-1}{\rho f}= -i \varepsilon_{st}\, \frac{F\,\sin \vartheta}{\rho_0}\, (\cos \phi,\sin \phi), \: s=5,6,\; \sigma=(5),(6), \nonumber\\ 
\label{snmb1:omegas}
\end{eqnarray}
which for  the choice $0 <2F \le 1$ allows only one massless spinor of a particular charge on
$S^2$, as we shall see in sect.~\ref{snmb1:equations}. In the  particular case that $2F=1$ the 
spin connection term $- S^{56} \omega_{56 \sigma} $ 
compensates the term  $\frac{1}{2Ef} \{ p_{\sigma},E f \}_-$ for the left handed spinor  
with respect to $d=1+3$, while for the spinor 
of the opposite handedness  the 
spin connection term doubles the term $\frac{1}{2Ef} \{p_{\sigma},E f \}_-$.  
$\phi$ determines the angle of rotations around  the axis through the two poles of a sphere, 
while $\rho = 2 \rho_0 \, \sqrt{\frac{1- \cos \vartheta}{1+ \cos \vartheta}}$, where 
$\tan \frac{\vartheta}{2} = \frac{\rho}{2\rho_0}$, as can be read on Fig.~\ref{snmb1:discgrav}. 
We shall see in sect.~\ref{snmb1:equations} that in the presence of the spin connection 
field from Eq.(\ref{snmb1:omegas})  the covariant derivative 
$\frac{\partial}{\partial \phi}$, 
becomes $\frac{\partial}{\partial \phi}- 2iF(1- \cos \vartheta)$.

Such a choice of vielbeins and spin connection fields manifests the isometry, which 
leaves the form of the fields on $S^2$ unchanged. The infinitesimal coordinate 
transformations manifesting this symmetry are: $x^{'\mu}= x^{\mu}, $ $x^{'\sigma}= x^{\sigma} + 
\phi_{A} \,K^{A \sigma}$, with $\phi_{A}$ the parameter of rotations 
 around the axis which goes through both poles and with the 
infinitesimal generators of rotations around this axis $M^{(5)(6)}(= x^{(5)} p^{(6)}- 
x^{(6)} p^{(5)} + S^{(5)(6)})$
\begin{eqnarray}
 K^{A \sigma}= K^{(56) \sigma} = -i M^{(5)(6)} x^{\sigma} =
 \varepsilon^{\sigma}{}_{\tau} x^{\tau}, 
\label{snmb1:killings}
\end{eqnarray}
with $\varepsilon^{\sigma}{}_{\tau}= -1 = - \varepsilon_{\tau}{}^{\sigma}, \, 
\varepsilon^{(5) (6)}=1. $ The operators $K^{A}_{ \sigma}=f^{-2} 
\varepsilon_{\sigma \tau} x^{\tau}$ fulfil the Killing relation 
$$K^{A}_{ \sigma, \tau} + 
\Gamma^{\sigma'}{}_{\sigma \tau} K^{A}_{ \sigma'} + K^{A}_{ \tau, \sigma} + 
\Gamma^{\sigma'}{}_{\tau \sigma} K^{A}_{ \sigma'}=0$$ (with $\Gamma^{\sigma'}{}_{\sigma \tau}= - 
\frac{1}{2} \, g^{\rho \sigma'} (g_{\tau \rho,\sigma} + g_{\sigma \rho,\tau} - 
g_{\sigma \tau,\rho})$).

The equations of motion for spinors (the Weyl equations) which follow from the Lagrange density 
(Eq.\ref{snmb1:weylL}) are then
\begin{eqnarray}
&&\{E\gamma^0 \gamma^m p_m + E f \gamma^0 \gamma^s \delta^{\sigma}_s  ( p_{0\sigma} 
+  \frac{1}{2 E f}
\{p_{\sigma}, E f\}_- )\}\psi=0,\quad {\rm with} \nonumber\\
&& p_{0\sigma} = p_{\sigma}- \frac{1}{2} S^{st}\omega_{st \sigma},
\label{snmb1:weylE}
\end{eqnarray}
with $f$ from  Eq.(\ref{snmb1:f})
and with $ \omega_{st \sigma}$ from Eq.(\ref{snmb1:omegas}).
Taking into account that 
$$\gamma^a p_{0a}\gamma^b p_{0b}= p_{0a} p_{0}{}^a - 
i S^{ab} S^{cd}\, {\cal R}_{abcd} + S^{ab}\,{\cal T}^{\beta}{}_{ab} \,p_{0 \beta}$$ 
we find for the Riemann tensor and the torsion 
\begin{eqnarray}
&&{\cal R}_{abcd} = f^{\alpha}_{[a} f^{\beta}_{b]}(\omega_{cd \beta, \alpha} 
- \omega_{ce \alpha} \omega^{e}{}_{d \beta} ), \nonumber\\
&& {\cal T}^{\beta}{}_{ab}= f^{\alpha}_{[a} (f^{\beta}_{b]})_{, \alpha} + \omega_{[a}{}^{c}{}_{b]} 
f^{\beta}_{c}. 
\label{snmb1:RT}
\end{eqnarray}
 $[a\,\,b]$ means the antisymmetrization over the two indices $a$ and $b$. 
 From Eq.(\ref{snmb1:RT}) we read that to the torsion on $S^2$ both, the zweibein 
 $f^{\sigma}_{\tau}$  and the spin connection $\omega_{st \sigma}$,  
 contribute. While we have on $S^2$ for ${\cal R}_{\sigma \tau} = 
 f^{-2} \eta_{\sigma \tau} \frac{1}{\rho^2} $ and correspondingly for the curvature 
 ${\cal R}= \frac{-2}{(\rho_0)^2}$, we find for the torsion 
 ${\cal T}^{s}{}_{t s'} = {\cal T}^{s}{}_{t \sigma} f^{\sigma}_{s'}$ with 
 %
 ${\cal T}^{5}{}_{ss} =  
 0 = {\cal T}^{6}{}_{ss},\quad s=5,6, $ 
 ${\cal T}^{5}{}_{65} = - {\cal T}^{5}{}_{56} =   -(f_{,6} + \frac{4iF (f-1)}{\rho^2 } x_5), $
 ${\cal T}^{6}{}_{56} = -  
  {\cal T}^{6}{}_{65} = - f_{,5} + \frac{4iF (f-1)}{\rho^2 } x_6.$
  %
  The torsion ${\cal T}^2 = {\cal T}^{s}{}_{t s'} {\cal T}_{s}{}^{t s'} $ 
  is for our particular choice of the zweibein and spin connection fields from Eq.\ref{snmb1:weylL} 
  correspondingly   equal to  
  $-\frac{2 \rho^2}{ (\rho_0)^4} (1-  (2F)^2)$ (which is for the choice $2F=1$ equal to zero.).
  
  We assume that the action for the gravitational field 
  (which could hopefully give the desired solutions of equations of motion for 
  the zweibein and the spin connection field in the presence of many spinor sources and what we have tried to 
  find but with no success up to now) is 
   linear in the Riemann scalar ${\cal R} = {\cal R}_{abcd}\eta^{ac}\eta^{bd}$ 
   and is in the lowest order with respect to the torsion 
($\beta {\cal T}^2 = \beta_a {\cal T}^{a}{}_{bc} {\cal T}_{a}{}^{bc} 
   + \beta_b {\cal T}^{a}{}_{bc} {\cal T}^{b}{}_{a}{}^{c}   
   + \beta_c {\cal T}^{a}{}_{ac} {\cal T}^{b}{}_{b}{}^{c}$) of Eq.\ref{snmb1:RT}  
  \begin{eqnarray}
  {\cal S} = \int \; d^d x \, (E \alpha {\cal R} + E \beta {\cal T}^2 + {\cal L}_W ).
  \label{snmb1:action}
  \end{eqnarray}
  The fermion part ${\cal L}_W$ is presented in Eq.(\ref{snmb1:weylL}) (and must include when searching  
for the desired spin connection and zweibein  many spinors). 

\section{Equations of motion  for spinors and the solutions}
\label{snmb1:equations}

The equations of motion (\ref{snmb1:weylE}) for a spinor   
in $(1+5)$-dimensional space, which breaks into  
$M^{(1+3)}$ $\times S^2$, let the spinor  "feel" the zweibein 
$f^{\sigma}{\!}_{s}= 
\delta^{\sigma}{\!}_{s} f(\rho), \,f(\rho)= 1+ (\frac{\rho}{2 \rho_0})^2 =
\frac{2}{1+ \cos \vartheta}$   and 
the spin connection 
$$\omega_{st \sigma}= 4iF  \varepsilon_{st}\frac{x_{\sigma}}{\rho}\, 
 \frac{f-1}{\rho f}= iF \frac{\sin \vartheta }{\rho_0}\, (\cos \phi, \sin \phi).$$ 
The  solution for a spinor in $d=(1+5)$ should be written as a superposition
of all  four ($2^{6/2 -1}$) states of a single Weyl representation. (We kindly ask the 
reader to see the technical details  about how to write 
a Weyl representation 
in terms of the Clifford algebra objects after making a choice of the Cartan subalgebra,  
for which we take: $S^{03}, S^{12}, S^{56}$ in the refs.~\cite{snmb1:holgernorma2002,snmb1:hn07}. )
In our technique \cite{snmb1:holgernorma2002} one spinor representation---the four 
states, which all are the eigenstates of the chosen Cartan subalgebra---are  
the following four products of projections $\stackrel{ab}{[k]}$ and nilpotents 
$\stackrel{ab}{(k)}$: 
\begin{eqnarray}
\varphi^{1}_{1} &=& \stackrel{56}{(+)} \stackrel{03}{(+i)} \stackrel{12}{(+)}\psi_0,\nonumber\\
\varphi^{1}_{2} &=&\stackrel{56}{(+)}  \stackrel{03}{[-i]} \stackrel{12}{[-]}\psi_0,\nonumber\\
\varphi^{2}_{1} &=&\stackrel{56}{[-]}  \stackrel{03}{[-i]} \stackrel{12}{(+)}\psi_0,\nonumber\\
\varphi^{2}_{2} &=&\stackrel{56}{[-]} \stackrel{03}{(+i)} \stackrel{12}{[-]}\psi_0,
\label{snmb1:weylrep}
\end{eqnarray}
where  $\psi_0$ is a vacuum state.
If we write the operators of handedness in $d=(1+5)$ as $\Gamma^{(1+5)} = \gamma^0 \gamma^1 
\gamma^2 \gamma^3 \gamma^5 \gamma^6$ ($= 2^3 i S^{03} S^{12} S^{56}$), in $d=(1+3)$ 
as $\Gamma^{(1+3)}= -i\gamma^0\gamma^1\gamma^2\gamma^3 $ ($= 2^2 i S^{03} S^{12}$) 
and in the two dimensional space as $\Gamma^{(2)} = i\gamma^5 \gamma^6$ 
($= 2 S^{56}$), we find that all four states are left handed with respect to 
$\Gamma^{(1+5)}$, with the eigenvalue $-1$, the first two states are right handed and the second two 
 states are left handed with respect to 
$\Gamma^{(2)}$, with  the eigenvalues $1$ and $-1$, respectively, while the first two are 
left handed 
and the second two right handed with respect to $\Gamma^{(1+3)}$ with the eigenvalues $-1$ and $1$, 
respectively. 
Taking into account Eq.(\ref{snmb1:weylrep}) we may write~\cite{snmb1:hn07} the most general wave function  
$\psi^{(6)}$ obeying Eq.(\ref{snmb1:weylE}) in $d=(1+5)$ as
\begin{eqnarray}
\psi^{(6)} = \snmbA \,{\stackrel{56}{(+)}}\,\psi^{(4)}_{(+)} + 
\snmbB \,{\stackrel{56}{[-]}}\, \psi^{(4)}_{(-)}, 
\label{snmb1:psi6}
\end{eqnarray}
where $\snmbA$ and $\snmbB$ depend on $x^{\sigma}$, while $\psi^{(4)}_{(+)}$ 
and $\psi^{(4)}_{(-)}$  determine the spin 
and the coordinate dependent parts of the wave function $\psi^{(6)}$ in $d=(1+3)$ 
\begin{eqnarray}
\psi^{(4)}_{(+)} &=& \alpha_+ \; {\stackrel{03}{(+i)}}\, {\stackrel{12}{(+)}} + 
\beta_+ \; {\stackrel{03}{[-i]}}\, {\stackrel{12}{[-]}}, \nonumber\\ 
\psi^{(4)}_{(-)}&=& \alpha_- \; {\stackrel{03}{[-i]}}\, {\stackrel{12}{(+)}} + 
\beta_- \; {\stackrel{03}{(+i)}}\, {\stackrel{12}{[-]}}. 
\label{snmb1:psi4}
\end{eqnarray}
Using $\psi^{(6)}$ in Eq.(\ref{snmb1:weylE}) we recognize the following expressions as the mass terms:  
$\frac{\alpha_+}{\alpha_-} (p^0-p^3) - \frac{\beta_+}{\alpha_-} (p^1-ip^2)= m,$ 
$\frac{\beta_+}{\beta_-} (p^0+p^3) - \frac{\alpha_+}{\beta_-} (p^1+ip^2)= m,$ 
$\frac{\alpha_-}{\alpha_+} (p^0+p^3) + \frac{\beta_-}{\alpha_+} (p^1-ip^2)= m,$
$\frac{\beta_-}{\beta_+} (p^0-p^3) + \frac{\alpha_-}{\beta_+} (p^1-ip^2)= m.$ 
(One can notice that for massless solutions  ($m=0$) the $\psi^{(4)}_{(+)}$ 
and $\psi^{(4)}_{(-)}$ 
decouple.) 
Taking into account that $S^{56} \stackrel{56}{(+)}= \frac{1}{2} \stackrel{56}{(+)}$, while 
$S^{56} \stackrel{56}{[-]}= -\frac{1}{2} \stackrel{56}{[-]}$, 
we end up with the equations of motion
 for $\snmbA$ and $\snmbB$ as follow 
\begin{eqnarray}
-2i\,f\, (\frac{\partial}{\partial z} + \frac{\partial \ln \sqrt{Ef}}{\partial z} -    
\frac{e^{-i\phi}}{\rho} G) \, \snmbB  
+ m \;\snmbA =0,\nonumber\\
-2i\,f\, (\frac{\partial}{\partial \bar{z}} + \frac{\partial\ln \sqrt{Ef}}{\partial \bar{z}}
+ \frac{e^{i\phi}}{\rho} G)\, \snmbA  
+ m \;\snmbB =0,  
\label{snmb1:equationm56gen}
\end{eqnarray}
where $z: = x^5 + i x^6 = \rho \,e^{i\phi}$, $\bar{z}: = x^5 - i x^6 = \rho \, e^{-i\phi}$ 
and $\frac{\partial}{\partial z}   = \frac{1}{2}\,(\frac{\partial}{\partial x^5} - i 
\frac{\partial}{\partial x^6}) = \frac{e^{-i\phi}}{2} \;
(\frac{\partial}{\partial \rho} - \frac{i}{\rho}\,\frac{\partial}{\partial \phi} ) $, 
$\frac{\partial}{\partial \bar{z}} = \frac{1}{2}\, (\frac{\partial}{\partial x^5} + i 
\frac{\partial}{\partial x^6}) = \frac{e^{ i\phi}}{2} \;
(\frac{\partial}{\partial \rho} + \frac{i}{\rho}\,\frac{\partial}{\partial \phi} ) $. 
Eq.(\ref{snmb1:equationm56gen}) can be rewritten as follows
\begin{eqnarray}
&&-if \, e^{-i \phi}\, (\frac{\partial}{\partial \rho} - \frac{i}{\rho}\,(\frac{\partial}{\partial \phi} -  
 i 2G ) + \frac{\partial}{\partial \rho} \ln {\sqrt{Ef}}) \snmbB + m \snmbA = 0,  
\nonumber\\
&&-if \, e^{i \phi}\, (\frac{\partial}{\partial \rho} + \frac{i}{\rho}\,(\frac{\partial}{\partial \phi} - 
 i 2G ) + \frac{\partial}{\partial \rho} \ln {\sqrt{Ef}}) \snmbA + m \snmbB= 0,
\label{snmb1:equationm56gen1}
\end{eqnarray}
with $G=4F \frac{f-1}{f}(=2F(1-\cos \vartheta))$. 
Having the rotational symmetry around the axis perpendicular to the plane of the fifth and the sixth 
dimension we require that $\psi^{(6)}$ is the eigenfunction of the total angular momentum
operator $M^{56}$
\begin{eqnarray}
M^{56}\psi^{(6)}= (n+\frac{1}{2})\,\psi^{(6)}, \quad M^{56} = x^5 p^6-x^6 p^5  + 
S^{56}.
\label{snmb1:mabx}
\end{eqnarray}
Let $\snmbA=\snmbA_n(\rho) \,\rho^n \, e^{i n \phi}  $ and $\snmbB= \snmbB_n(\rho)\,\rho^{-n}\,
e^{i n \phi}$. 

Let us treat first the massless case ($m=0$). Taking into account that $\frac{G}{\rho} = 
\frac{\partial}{\partial \rho} \ln f^{2F}$ and that $E=f^{-2}$ it  follows  
\begin{eqnarray}
\frac{\partial \, \ln (\snmbB \,f^{-F -1/2})}{\partial \rho}&=&0,\nonumber\\
\frac{\partial \, \ln (\snmbA \,f^{F -1/2})}{\partial \rho}&=&0.
\label{snmb1:masslesseq}
\end{eqnarray}
We get correspondingly the solutions
\begin{eqnarray}
\snmbB_n &=& \snmbB_0 \, e^{in \phi}\, \rho^{-n} f^{F+1/2}, \nonumber\\
\snmbA_n &=& \snmbA_0 \, e^{in \phi}\, \rho^{n} f^{-F+1/2}. 
\label{snmb1:masslesseqsol}
\end{eqnarray}
Requiring that only normalizable (square integrable) solutions are acceptable 
\begin{eqnarray}
2\pi \, \int^{\infty}_{0} \,E\, \rho d\rho \snmbA^{\star}_{n} \snmbA_{n} && < \infty, \nonumber\\
2\pi \, \int^{\infty}_{0} \,E\, \rho d\rho \snmbB^{\star}_{n} \snmbB_{n} && < \infty, 
\label{snmb1:masslesseqsolf}
\end{eqnarray}
it follows 
\begin{eqnarray}
&&{\rm for}\; \snmbA_{n}: -1 < n < 2F, \nonumber\\
&&{\rm for}\; \snmbB_{n}: 2F < n < 1, \quad n \;\; {\rm is \;\; an \;\;integer}.
\label{snmb1:masslesseqsolf1}
\end{eqnarray}
Eq.(\ref{snmb1:masslesseqsolf1}) tells us that the strength $F$ of the spin connection field 
$\omega_{56 \sigma}$ can make a choice between the two massless solutions $\snmbA_n$ and $\snmbB_n$: 
For $0< 2F \le 1$ the only massless solution is the left handed spinor with respect to 
$(1+3)$
\begin{eqnarray}
\psi^{(6)}_{0} ={\cal N}_0  f^{-F+1/2} 
\stackrel{56}{(+)}\psi^{(4)}_{(+)}.
\label{snmb1:Massless}
\end{eqnarray} 
It is the eigenfunction  of $M^{56}$ with the eigenvalue $1/2$. 
No right handed 
solution is allowed for $0< 2F \le 1$.
For the  particular choice  $2F=1$ the spin connection field $-S^{56} \omega_{56\sigma}$ 
compensates the term $\frac{1}{2Ef} \{p_{\sigma}, Ef \}_- $ and the  left handed spinor
with respect to $d=1+3$ becomes a constant with respect to $\rho $ and $\phi$.

For $2F=1$ it is easy to find also all the massive solutions of Eq.(\ref{snmb1:equationm56gen1}).
To see this let us rewrite Eq.(\ref{snmb1:equationm56gen1}) in terms of the parameter $\vartheta$. 
Taking into account that 
$f= \frac{2}{1+\cos \vartheta}$, $\omega_{56 \sigma}= 
-iF \frac{\sin \vartheta}{\rho_0} \;(\cos \phi, \sin \phi)$
and assuming that $\snmbA=\snmbA_n(\rho) \, e^{i n \phi}  $ and 
$\snmbB= \snmbB_{n+1}(\rho)\, e^{i (n+1) \phi}$, which guarantees that the states will be  
the eigenstates of $M^{56}$, it follows
\begin{eqnarray}
&&(\frac{\partial}{\partial \vartheta} +  \frac{n+1 -(F+1/2)(1-\cos \vartheta)}{\sin \vartheta} ) 
\snmbB_{n+1} + i \tilde{m} \snmbA_n = 0,  
\nonumber\\
&&(\frac{\partial}{\partial \vartheta} +  \frac{-n +(F-1/2)(1-\cos \vartheta)}{\sin \vartheta} ) 
\snmbA_{n} + i \tilde{m} \snmbB_{n+1} = 0,
\label{snmb1:equationm56theta}
\end{eqnarray}
with $\tilde{m}=\rho_0 m$. For the particular choice of $2F= 1$ the equations simplify to
\begin{eqnarray}
&&(\frac{\partial}{\partial \vartheta} +  \frac{n +\cos \vartheta}{\sin \vartheta} ) 
\snmbB_{n+1} + i \tilde{m} \snmbA_n = 0,  
\nonumber\\
&&(\frac{\partial}{\partial \vartheta} - \frac{n }{\sin \vartheta} ) 
\snmbA_{n} + i \tilde{m} \snmbB_{n+1} = 0,
\label{snmb1:eqthetaF}
\end{eqnarray}
from where we obtain 
\begin{eqnarray}
&&\{\frac{1}{\sin{\vartheta}} \frac{\partial}{\partial \vartheta}(\sin{\vartheta} 
\frac{\partial}{\partial \vartheta} ) + [\tilde{m}^2 + \frac{(-n^2-1-2n 
\cos \vartheta)}{\sin^2\vartheta}]\} \snmbB_{n+1} =0, 
\nonumber\\
&&\{\frac{1}{\sin \vartheta} \frac{\partial}{\partial \vartheta}(\sin \vartheta 
\frac{\partial}{\partial \vartheta} ) + [\tilde{m}^2 - \frac{n^2}{\sin^2\vartheta}]\} \snmbA_{n} =0.
\label{snmb1:sphtheta}
\end{eqnarray}
From above  equations 
we see that for $\tilde{m}=0$, that is for the massless 
case, the only solution 
with $n=0$ exists, which is $Y^{0}_{0}$, the spherical harmonics, which is a constant (in  
agreement with our discussions  above). 
All the massive solutions have $\tilde{m}^2= l(l+1), \, l=1,2,3,..$ and $-l\le n \le l$.  
Legendre polynomials are the solutions for $\snmbA_{n}= P^{l}_n$, 
as it can be read from the second of the equations Eq.(\ref{snmb1:sphtheta}), while 
we read from the second equation of Eq.(\ref{snmb1:eqthetaF}) that $$\snmbB_{n+1}= 
\frac{i}{\sqrt {l(l+1}}\, (\frac{\partial}{\partial \vartheta} - \frac{n }{\sin \vartheta} ) 
P^{l}_{n}.$$

Accordingly  the massive solution  
with the mass equal to $m  = l (l+1)/\rho_0$ (we use the units in which $c=1=\hbar$) 
and the eigenvalues of $M^{56}$ 
((Eq.\ref{snmb1:mabx}))---which is 
the charge as we shall see later---equal to $(\frac{1}{2}+n)$, with $-l \le n \le l$, $l=1,2,..$, 
are
\begin{eqnarray}
\psi^{(6)\tilde{m}^2=l(l+1)}_{n+1/2} &=& {\cal N}^{l}_{n+1/2} \{ 
\stackrel{56}{(+)} \psi^{(4)}_{(+)} + 
\frac{i}{\sqrt{l(l+1)}}\, \stackrel{56}{[-]} \psi^{(4)}_{(-)}\, e^{i\phi}
(\frac{\partial}{\partial \vartheta} \, -\frac{n}{\sin \vartheta})\}
Y^{l}_{n})\nonumber\\ 
\label{snmb1:knsol}
\end{eqnarray}
with $Y^{l}_{n}$, which are the spherical harmonics. 
Rewriting the mass operator $\hat{m}= \gamma^0 \gamma^s f^{\sigma}_{s} (p_{\sigma} - 
S^{56} \omega_{56 \sigma} + \frac{1}{2Ef} \{p_{\sigma}, Ef\}_-)$ as a function of 
$\vartheta $ and $\phi$  
\begin{eqnarray}
\rho_0 \hat{m} &=& i \gamma^0\, \{\stackrel{56}{(+)} e^{-i\phi} 
(\frac{\partial}{\partial \vartheta} \, -\frac{i}{\sin \vartheta}
\frac{\partial}{\partial \phi } \, - \frac{1-\cos \vartheta}{\sin \vartheta}) +
\stackrel{56}{(-)} e^{i\phi} 
(\frac{\partial}{\partial \vartheta} \, + \frac{i}{\sin \vartheta}
\frac{\partial}{\partial \phi } ) \}, \nonumber\\
\label{snmb1:m}
\end{eqnarray}
one can easily show that when  applying $\rho_0 \hat{m}$ and $M^{56}$ 
on $\psi^{(6)\tilde{m}^2=k(k+1)}_{n+1/2}$, for $-k \le n \le k$,  one obtains
from Eq.(\ref{snmb1:knsol})
\begin{eqnarray}
\label{snmb1:aigenmass}
\rho_0 \hat{m}\, \psi^{(6)\tilde{m}^2=k(k+1)}_{n+1/2} &=&
k(k+1) \psi^{(6)\tilde{m}^2=k(k+1)}_{n+1/2}, \nonumber\\ 
M^{56}\, \psi^{(6)\tilde{m}^2= (n+1/2)k(k+1)}_{n+1/2} &=&
(n+1/2) \psi^{(6)\tilde{m}^2=k(k+1)}_{n+1/2}.
\end{eqnarray}
A  wave packet,  which is the eigen function of $M^{56}$ with the eigenvalue 
$1/2$,  for example,  can be written as
\begin{eqnarray}
\psi^{(6)}_{1/2} &=&\sum_{k=0,  \infty} C_{1/2}^k \;\,
{\cal N}_{1/2} \{ 
\stackrel{56}{(+)} \psi^{(4)}_{(+)} + (1 - \delta^{k}_0)
\frac{i}{\sqrt{k(k+1)}}\, \stackrel{56}{[-]} \psi^{(4)}_{(-)}\,e^{i\phi}
\frac{\partial}{\partial \vartheta} \}
Y^{k}_{0}. \nonumber\\
\label{snmb1:gaussian}
\end{eqnarray}
The expectation value of the mass operator $ \hat{m}$ on such a wave packet is 
$$\sum_{k=0,  \infty} C_{1/2}^{k*} C_{1/2}^{k} \sqrt{k(k+1)}/\rho_0.$$ 

It remains to comment the meaning of the exclusion of the south pole on $S^2$, since the 
disc with the zweibein equal to $f= \frac{2}{1+ \cos \vartheta}$ looks like $S^2$ up to the 
southern pole. 

To start from the southern pole one must rewrite  Eq.(\ref{snmb1:eqthetaF}) 
and the second equation of Eq.(\ref{snmb1:sphtheta}) so that $\vartheta$ is replaced by 
$(\pi - \vartheta)$  

\begin{eqnarray}
&&(\frac{\partial}{\partial (\pi-\vartheta)} +  
\frac{- n +\cos (\pi - \vartheta)}{\sin(\pi - \vartheta)} ) 
(-)\snmbB_{-n+1} + i \tilde{m} \snmbA_{-n} = 0,  
\nonumber\\
&&(\frac{\partial}{\partial(\pi- \vartheta)} - \frac{-n }{\sin \vartheta} ) 
\snmbA_{-n} + i \tilde{m} (-)\snmbB_{-n+1} = 0,
\label{snmb1:equationm56thetaFissouth}
\end{eqnarray}
and
\begin{eqnarray}
&&\{\frac{1}{\sin(\pi- \vartheta)} \frac{\partial}{\partial (\pi- \vartheta)}
(\sin (\pi- \vartheta )
\frac{\partial}{\partial (\pi - \vartheta)} ) + [\tilde{m}^2 - 
\frac{(-n)^2}{\sin^2 (\pi- \vartheta)}]\} \snmbA_{-n} =0. \nonumber\\
\label{snmb1:sphsouth}
\end{eqnarray}
Since $\snmbA_{-n} (\pi- \vartheta)= P^{l}_{-n} (\pi-\vartheta) = (-1)^{l+n} 
P^{l}_{n} (\vartheta)$  
are the solutions of Eq.(\ref{snmb1:sphsouth}) and since $P^{l}_{-n} (\pi-\vartheta)= (-)^{l+2n} 
P^{l}_{n}(\theta)$, the solutions of Eq.(\ref{snmb1:sphsouth}) coincide with the solutions 
of Eq.(\ref{snmb1:sphtheta}). Correspondingly also the solutions 
for $(-)\snmbB_{-n+1} (\pi - \vartheta)= \frac{i}{\tilde{m}} 
(\frac{\partial}{\partial(\pi- \vartheta)} - \frac{-n }{\sin \vartheta} ) 
\snmbA_{-n} (\pi - \vartheta)$ coincide with the solutions of $\snmbB_{n+1} (\vartheta)$, 
which proves 
that the one missing point on $S^2$ makes no harm.

\section{Spinors and the gauge fields in $d=(1+3)$}
\label{snmb1:properties1+3}
To study how do spinors couple to the Kaluza-Klein gauge fields in the case of 
$M^{(1+5)}$, ``broken'' to 
$M^{(1+3)} \times S^2$ with the radius of $S^2$ equal to  $\rho_0$ and with 
the spin connection field 
$\omega_{st \sigma} = i4F \varepsilon_{st} \frac{x_{\sigma}}{\rho}\frac{f-1}{\rho f}$
we first look for (background) gauge gravitational fields, which preserve the rotational symmetry 
around the axis through the northern and southern pole.
Requiring that the symmetry determined by the Killing vectors of Eq.(\ref{snmb1:killings}) 
(following ref.~\cite{snmb1:hnkk06}) with $f^{\sigma}{}_{s} = f \delta^{\sigma}_{s}, f^{\mu}{}_s=0, 
e^{s}{}_{\sigma}= f^{-1} \delta^{s}_{\sigma}, e^{m}{}_{\sigma}=0,$ is preserved, we find 
for the background vielbein field  
\begin{eqnarray}
e^a{}_{\alpha} = 
\begin{pmatrix}\delta^{m}{}_{\mu}  & e^{m}{}_{\sigma}=0 \\
 e^{s}{}_{\mu} & e^s{}_{\sigma} \end{pmatrix},
f^{\alpha}{}_{a} =
\begin{pmatrix}\delta^{\mu}{}_{m}  & f^{\sigma}{}_{m} \\
0= f^{\mu}{}_{s} & f^{\sigma}{}_{s} \end{pmatrix},
\label{snmb1:f6}
\end{eqnarray}
with 
\begin{eqnarray}
\label{snmb1:background}
f^{\sigma}{}_{m} &=& K^{(56)\sigma} B^{(5)(6)}_{\mu} f^{\mu}{}_{m} = 
\varepsilon^{\sigma}{}_{\tau} x^{\tau} A_{\mu} \delta^{\mu}_{m}, \nonumber\\
e^{s}{}_{\mu} &=& - \varepsilon^{\sigma}{}_{\tau} x^{\tau} A_{\mu} e^{s}{}_{\sigma}, 
\end{eqnarray}
 $ 
s=5,6; \sigma = (5),(6)$.  
Requiring that correspondingly the only nonzero torsion fields are those from 
Eq.(\ref{snmb1:RT}) 
we find for the spin connection fields 
\begin{eqnarray}
\omega_{st \mu} =  \varepsilon_{st}  A_{\mu},
\quad \omega_{sm \mu} = 
\frac{1}{2}f^{-1}\varepsilon_{s \sigma } x^{\sigma} \delta^{\nu}{}_{m} F_{\mu \nu},
\label{snmb1:omega6}
\end{eqnarray}
$F_{\mu \nu}= A_{[\nu,\mu]}$. 
 The $U(1)$ gauge field $A_{\mu}$ depends only on $x^{\mu}$.
All the other components of the spin connection fields, except (by the 
Killing symmetry preserved)  $\omega_{st\sigma}$ from Eq.(\ref{snmb1:weylE}), are zero, 
since for simplicity we allow no gravity in
$(1+3)$ dimensional space. 
The corresponding nonzero torsion fields ${\cal T}^{a}{}_{bc}$ are presented in 
Eq.(\ref{snmb1:RT}), all the other components are zero. 

To determine the current, which couples the spinor to the Kaluza-Klein gauge fields 
$A_{\mu}$, we
analyze (as in the refs.~\cite{snmb1:hnkk06,snmb1:hn07}) the spinor action (Eq.(\ref{snmb1:weylL}))
\begin{eqnarray}
{\cal S} &=& \int \; d^dx  \bar{\psi}^{(6)} E \gamma^a p_{0a} \psi^{(6)} =\nonumber\\
&& \int \, d^dx  \bar{\psi}^{(6)} \gamma^s  p_{s} \psi^{(6)}+ \nonumber \\  
&& \int \, d^dx  \bar{\psi}^{(6)} \gamma^m \delta^{\mu}{}_{m} p_{\mu} \psi^{(6)} + \nonumber\\
&& \int \, d^dx   \bar{\psi}^{(6)} \gamma^m  \delta^{\mu}{}_{m} A_{\mu} 
(\varepsilon^{\sigma}{}_{\tau} x^{\tau}
 p_{\sigma} + S^{56}) \psi^{(6)} + \nonumber\\
&& {\rm \; terms } \propto  x^{\sigma} \,{\rm or } \propto   x^{5}  x^{6}.
\label{snmb1:spinoractioncurrent}
\end{eqnarray}
 Here $\psi^{(6)}$ is a spinor state  in $d=(1+5)$ after the break of $M^{1+5}$ 
 into $M^{1+3} \times $ $S^2$.
 $E$ is for $f^{\alpha}{}_{a}$ from Eq.(\ref{snmb1:f6}) equal to $f^{-2}$. 
The  term in the second row in Eq.(\ref{snmb1:spinoractioncurrent}) is the mass term  
(equal to zero for the massless spinor), the term in the third row is the kinetic term, 
together with the term in the fourth row  defines  
the  covariant derivative $p_{0 \mu}$ in $d=(1+3)$.  
The terms in the last row  contribute nothing when the integration over 
the disk (curved into a sphere $S^2$) is performed, since they all 
are proportional to $x^{\sigma}$ or to $ x^{5} x^{6}\;$ 
($-\gamma^{m} \,\frac{1}{2}S^{sm} \omega_{s m n} = -\gamma^{m}\,\frac{1}{2}\,f^{-1}
F_{m n}  \varepsilon_{s \sigma} x^{\sigma}$ and $-\gamma^m \,f^{\sigma}{}_{m}\frac{1}{2}
\,S^{st} \omega_{st \sigma}= 
\gamma^m A_m   x^{5}x^{6} S^{st} \varepsilon_{s t} \frac{4iF(f-1)}{f \rho^2}$).

We end up with the current in $(1+3)$
\begin{eqnarray}
j^{\mu} = \int \;E  d^2x \bar{\psi}^{(6)} \gamma^m \delta^{\mu}{}_{m} M^{56}  \psi^{(6)}.
\label{snmb1:currentdisk}
\end{eqnarray}
 The charge in $d=(1+3)$ is  proportional to the total 
angular momentum  $M^{56} =L^{56} + S^{56}$ around the axis from the southern to the 
northern  pole of $S^2$, but since for the choice of 
$  2 F =1$ (and for any $0 < 2F \le 1 $) in Eq.(\ref{snmb1:masslesseqsolf1}) only a left 
handed massless spinor exists,  
with the angular momentum zero, the charge of a massless 
spinor in $d=(1+3)$ is equal to  $1/2$.

The Riemann scalar is for 
the vielbein of Eq.(\ref{snmb1:f6}) equal to 
$${\cal R}= -\frac{1}{2} \rho^2 f^{-2} F^{mn}F_{mn}.$$ 
If we integrate the Riemann scalar 
over the fifth and the sixth dimension, we get $-\frac{8\pi}{3} (\rho_0)^4 F^{mn}F_{mn}$.

\section{Conclusions}
\label{snmb1:conclusion}

We presented in this letter a toy model of a left handed spinor  
carrying in $d=1+5$ nothing but a spin, with the symmetry of 
$M^{(1+5)}$, which breaks to $M^{(1+3)} \times$ the infinite disc with the zweibein, which curves 
the disc on $S^2$ ($f= 1+ (\frac{\rho}{2 \rho_0})^2$, with $\rho_0$ the radius of $S^2$), 
and with the spin connection field on the disc equal to $\omega_{st \sigma} = \varepsilon_{st}\,
i 4F \frac{f-1}{\rho f} \frac{x_{\sigma}}{\rho}, \sigma=(5),(6); s,t=5,6,$ which allows for 
$0 < 2F \le 1 $ one massless spinor of the charge $1/2$ and of the left handedness with respect to 
$d=(1+3)$. This spinor state couples chirally to the 
corresponding Kaluza-Klein gauge field. 
There are infinite many massive states, which are at the same time 
the eigenstates of $M^{56}= x^{5} p^{6}- x^{6} p^{5} + S^{56}$, with the eigen values 
$n+1/2$, carrying the Kaluza-Klein charge $n+1/2$. 
For the choice of $2F=1$ the massive states have the mass equal to $k(k+1)/\rho_0, k=1,2,3,..$,
 with $-k \le n \le k$. 
We found the expression for the massless eigenstate  and for 
the particular choice of $2F=1$ also for all the massive states.  

We therefore found an example, in which the internal gauge fields---spin connections 
and zweibeins---allow only one massless state, that is the spinor of 
one handedness and of one charge with respect to 
$d=1+3$ space. Since for the zweibein  curving the infinite disc on $S^2$,    
the spin connection  field  $\omega_{st \sigma} = 
i 4F \frac{f-1}{\rho f} \frac{x_{\sigma}}{\rho}, $ with any $2F$ 
fulfilling the condition 
$\;0< 2F \le 1$  ensures that a massless spinor state of only one handedness and one 
charge in $d=(1+3)$ exists (only one massless state is normalizable),  it is not 
a fine tuning what we propose. To find simple solutions for the massive states, we made 
a choice  of  $2F= 1$. The massless state is in this case a constant with respect to 
the two angles on $S^2$, while  the angular dependence of the massive states, 
with the masses equal to $l(l+1)/\rho_0$, 
are expressible with the spherical harmonics $Y^{l}_{n},\;\; -l \le n \le l$, and 
with the 
$e^{i\phi} \, \frac{i}{\sqrt{l(l+1)}}\; (\frac{\partial}{\partial \vartheta} - 
\frac{i}{\sin \vartheta} ) Y^{l}_{n}$ (Eq.(\ref{snmb1:knsol})).

We do not explain either how does the break of the $M^{(1+5)}$ to $M^{(1+3)} \times $ the 
infinite disc, with the zweibein which curves the disc on $S^2$,  occur or what does make the 
spin connection field in the 
radial direction and of the strength, which allows spinors of only one handedness  
on $S^2$.

If the break of the starting symmetry $M^{1+5}$ occurs spontaneously 
because of the many body effects, 
like it is a many spinors state, then the spin connection field must be proportional to the 
number of spinors (it might be of different angular momentum each) and accordingly 
quantized. WE have tried to prove that such a spin connection field can occur due to 
the many body spinor functions, describing spinors in $d=1+5$. We could not found such 
a many body state. We even proved that such a field can not be generated by only 
 spinors.

Let us conclude the paper by pointing out that while in the two papers~\cite{snmb1:hnkk06,snmb1:hn07} 
we  achieved the masslessness 
of a spinor, its mass protection and the chiral coupling to the 
corresponding Kaluza-Klein gauge field  after a break of a symmetry from 
$d=1+5$ to $d=(1+3)$, with the choice of the boundary condition on a flat (finite)
disk (without explaining where does such a boundary condition come from), 
in this letter the massless spinor and its chiral coupling to the 
corresponding Kaluza-Klein gauge field is  achieved by the choice of the 
appropriate spin connection and zweibein fields (whose origin we were not able to derive).

We do not discuss the problem of the families in this paper. We kindly ask the 
reader to take a look on the 
refs.~\cite{snmb1:norma92,snmb1:norma93,snmb1:norma94,snmb1:norma95,snmb1:Portoroz03,snmb1:pikanorma06,snmb1:gmdn06,snmb1:gmdn07}
where the proposal for solving the problem of families is presented.

\section{Acknowledgement } One of the authors (N.S.M.B.) 
would like to warmly thank Jo\v ze Vrabec 
for  fruitful discussions.

\author{R. Mirman\thanks{sssbbg@gmail.com}}
\title{Some Obvious Matters of Physics That Are Not Obvious}
\institute{%
14U\\
155 E 34 Street\\
New York, NY  10016
}

\titlerunning{Some Obvious Matters of Physics That Are Not Obvious}
\authorrunning{R. Mirman}
\maketitle

\begin{abstract}
There are some well-known properties of our universe whose reasons are clear but strangely not well-known. Physicists seem to believe that they hold because God wants them. Actually it is usually because geometry wants them. We summarize these here; detailed discussion and proofs were given long ago~(\cite{rm1nmb}; \cite{rm1nm2}; \cite{rm1nm3}; \cite{rm1nm4}; \cite{rm1ia}; \cite{rm1gf}; \cite{rm1ml}; \cite{rm1pt}; \cite{rm1qm}; \cite{rm1cnfr}; \cite{rm1bna}; \cite{rm1bnb}; \cite{rm1bnc}; \cite{rm1op}; \cite{rm1cg}; \cite{rm1imp}; \cite{rm1rn}). 
\end{abstract}

\section{Why we cannot expect gravitation to have weird properties}\label{rm1sgv}

General relativity seems to have unphysical solutions like closed time-like curves, wormholes, ... . This does not follow and is quite unlikely: the Einstein equation is a necessary condition for a gravitational field but not sufficient. There are additional requirements~\cite{rm1ml} and before we can conclude that there are fields with strange properties we must show that all conditions are satisfied. This is implausible for weird fields.  Abnormal solutions imply that not all conditions are satisfied. And there are other problems.

These do not imply anything wrong with general relativity --- it is almost certainly correct. It just means that it is applied incorrectly. 

What are other conditions~(\cite{rm1ml}; \cite{rm1bna}; \cite{rm1imp})?

The field must be produced, else it does not exist. What produces a gravitational field? A sphere, a star, dust? But there are no spheres, stars, dust. These are merely collections of protons, neutrons, electrons and such --- which are what creates and is acted upon. Such a collection must give a strange field. However these objects are governed by quantum mechanics. The uncertainty principle applies. Can a collection of such objects produce strangeness? Before it is claimed that there are closed timelike curves, wormholes, ..., it must be shown that there is a collection of quantum mechanical objects capable of producing them. 

Would we expect a single proton, a single electron, to give closed timelike curves? If not why would we expect a collection to? This implies that the formalism is being used incorrectly. This can be tricky because we often think in ways different than the ones nature thinks in, like using classical physics as a formalism while nature uses quantum mechanics, or using large objects while nature produces gravitational fields from collections of quantum mechanical ones. Using the proper formalism is essential. 

If such single objects, atoms and subsets, cannot form wormholes, or any other such strange things, then for them to be believed to exist it must be shown that sets of such elementary objects can give them. How many objects does it take? What determines this number? This does seem implausible does it not? The difficulty is that we take the source of the gravitational field as a macroscopic body, but there are none, only collections of electrons, protons and such. Hence we must either show that such objects are possible or conclude that they are purely the result of misuse of the formalism, like the use of macroscopic sources. 

There is another condition which is especially interesting since it requires that general relativity be the theory of gravity (thus the quantum theory of gravity, as it so clearly is~(\cite{rm1ml})). All properties of gravitation come from it. This has been discussed in depth, with all the mathematics shown and proven~(\cite{rm1ml}; \cite{rm1bna}). Here we summarize. 

A physical object, like a gravitational field, must be a representation basis state of the transformation group of geometry, the Poincar\'e group. (The Poincar\'e group is the transformation group, not the symmetry group, although it is interesting that it is the symmetry group also~(\cite{rm1imp}, sec.~VI.2.a.ii, p.~113)). 

To clarify consider the rotation group and an object with spin up. Its statefunction (a better term than wavefunction since nothing waves) gives the spin as up. A different observer sees the spin at some angle, thus a different statefunction. The statefunction of the first must be transformed to give that of the second. Thus for each set of coordinates there is a statefunction and these are transformed into each other when the coordinates are. For each rotation there is a transformation of the statefunction. Moreover the product of two transformations must correspond to the product of the two rotations that they go with. Also a rotation, being a group element, can be written as a product of two, or ten, or 1000, or in any of an infinite number of ways. Each such product has a product of transformations on the statefunction going with it, with each term in the product of transformations corresponding to a term in the product of rotations. Thus the transformations on the statefunction form a representation of the rotation group, and each statefunction generated from any one by such a transformation is a basis state of the rotation group representation.  

This does not require that space or physics be invariant under the group. Rotations are a property of geometry whether space is invariant under them or not. Thus a state can be written as a sum of rotation basis states (spherical harmonics) and is taken into another such sum by a rotation. Each term in the latter is a sum of terms of the former (with coefficients functions of the angles). Each term is a sum only of terms from the same representation (states of angular momentum 1 go only into states of angular momentum 1, and so on). This is true whether space is invariant under rotations or not (say there is a direction, simulated by the vertical, that is different). An up state may with time go into a down one, but that is irrelevant since these (mathematical) transformations are considered at a single time. Also no matter how badly symmetry is broken there cannot be an object with spin-${1\over 3}$ or $\pi$. These would not be true if we expanded in unitary group states. The rotation group is a property of our (real) geometry. 

It is only a subgroup. The transformation group of space thus of the fields is the Poincar\'e group. Statefunctions (including those of gravity, the connections) must be basis states of it. The Poincar\'e group is an inhomogeneous group so very different from the simple rotation group. Gravitation is massless. The entire analysis depends on this.

Massless and massive representations are much different. The little group of massive representations is semisimple (the rotation group), while that of massless ones is solvable. Thus massless objects have difficulty in coupling to massive ones. There are only three that can. Scalars apparently can. Helicity 1 gives electromagnetism (with its properties completely determined). For helicity 2 the indices do not match. Fortunately the formalism gives a nonlinear condition, the Bianchi identities, that allow gravitation to interact with massive objects. Gravitation must be nonlinear else it could not couple, so could not exist. Einstein's equation then follows from the formalism, but is not all of it. 

A supposed gravitational field must be shown to form a representation basis state of a massless helicity-2 representation of the Poincar\'e group or it is not a gravitational field. Unless ones with strange properties are shown to be that then they are results of the wrong or incomplete formalism, so nonexistent. 

Since the Poincar\'e group is inhomogeneous the momentum operators (the Hamiltonian is one) must commute. There would be many problems if not~(\cite{rm1ml}, sec.~6.3.8, p.~110). It must be checked for a proposed field that the momenta commute on it. 

The proper way to find fields is thus to find functions satisfying these properties --- extremely difficult. To see if a field can be produced we must find if the momentum operators of the entire system commute. These consist of three sets of terms, for the field, for massive matter and for the interactions. Thus we have to find a (quantum mechanical) distribution of matter which, with the fields it produces, gives these operators, and such that they commute. 

It is likely to be very rare that we can do this. Great caution is required; we cannot jump to conclusions about the existence of strange solutions. 

To illustrate the importance of proper formalism, properly applied, we consider other related topics~(\cite{rm1ml}). 

Are there "graviton"'s~(\cite{rm1ml},sec.~11.2.2, p.~187)? We are used to taking electromagnetic fields as sets of photons so try to apply it to gravity. But electromagnetism is linear, gravitation nonlinear. What is a photon? It is not a little ball, a ridiculous idea. If we Fourier expand an electromagnetic potential (a solution of the equations) each term is a solution. Each term is then a photon. A solution is a sum of solutions. If we do the same for a field that is a solution of the gravitational equations the terms are not solutions. A gravitational field is a collection of "graviton"s each producing a collection of "graviton"s, each ... . Obviously the concept is useless. Consider a gravitational wave extending over a large part of the universe. That single wave is a "graviton". The concept is not likely useful. 

Are there magnetic monopoles~(\cite{rm1ml},sec.~7.3, p.~131)? Maxwell's equations have an asymmetry. But these are classical, so irrelevant. Quantum electrodynamics does not have such an asymmetry. There is no hole to be filled and, using the correct formalism, there is no way a magnetic monopole can act on a charge. There are no magnetic monopoles. 

What is the value of the cosmological constant? In Einstein's equation one side is a function of space, the other a constant (obvious nonsense), that is one side is a function of a massless representation, the other a momentum-zero representation. This is like equating a scalar and a vector. The cosmological constant is trivially 0, unfortunately else gravitation would have a fascinating property: a wave would be detected not only an infinitely long time before arrival but before emission~(\cite{rm1ml}, sec.~8.1.4, p.~139).

For small fields the term multiplying the cosmological constant can be taken as constant, so putting it in the equation sets a variable equal to a constant.

Are there Higg's bosons? Gauge transformations are the form Poincar\'e transformations take for massless objects, and these only~(\cite{rm1ml}, sec.~3.4, p.~43). This is explained in one paragraph~(\cite{rm1imp}, sec.~E.2.1, p.~445). They cannot be applied to massive objects because of the mathematics, not because of some new field. People are entranced by gauge invariance and decided to apply it to objects where it cannot hold. This is like deciding that orbital angular momentum is integral so spins must be. They are not so there must be some new field that makes them half-integral. But the mathematics gives both types of spin, does not allow spin-${1\over 3}$, gives gauge invariance for massless objects, and does not allow it for massive ones. This is a result of the mathematics, not of some new field. There are no Higgs bosons.

\section{Uncertainty principles for gravitation}\label{rm1su1}

Gravity is described quantum mechanically by its statefunction, $\Gamma$. This means that there are uncertainty principles for the gravitational field. What are these? Why do uncertainty principles arise? Essentially objects are wavepackets. These can be Fourier analyzed into terms of the form (schematically) $\sum A(p)exp(ipx + i\nu t)$. The more terms, the more values of the momentum $p$, the narrower is the wavepacket, the narrower the range of $p$'s which contribute significantly, the wider the packet. This, in the well-known manner, gives the uncertainty principle. 

For gravity it is similar, except that gravitation is nonlinear. This means that solutions of Einstein's equation cannot be added, as solutions of Schr\"{o}dinger's equation can. So for an electron we can construct wavepackets, including ones with minimum uncertainty. For gravitation we cannot. A gravitational field is a wavepacket. It is spread over space so there is an uncertainty in position, and being a wavepacket also one in momentum. By the same argument the smaller the extent of the field the more momentum values must give significant contributions. There is thus an uncertainty principle, and it is stronger than $\delta$p$\delta$q = 1. This holds for the minimum wavepacket ~(\cite{rm1mz}, p.~156), which a gravitational field is not, and no gravitational field can be constructed to be one. Such a field would not obey Einstein's equation. The uncertainty for gravitation is stronger than in other cases --- because gravitation is nonlinear. 

Also the Gaussian form does not hold for the gravitational field, so there is not one hump, but many. 

We leave open the question whether a minimum uncertainty gravitational field can be constructed. It would have to obey Einstein's equation, the commutativity conditions, and also be shown to give the minimum. This would be a difficult mathematical problem especially because there seems no general formula for gravitational fields. 

Can we measure in a way to violate uncertainty? We leave open the general analysis giving just a few remarks. The way to measure the gravitational field is to put a small mass in and see how it behaves. But the mass then produces its own field and the narrower we make the mass wavepacket the more momentum states we must include. The more momentum states the greater the uncertainty of the extra field. And the total field cannot be separated into the original field and the induced one because gravitation is nonlinear. Thus the measurement must give an uncertainty. We leave open the question whether these procedures give the same uncertainty, which is difficult because there is no known way of calculating a minimum uncertainty for gravitation.

For a more formal analysis we start with the expectation value of the commutator~(\cite{rm1mz}, p.~154)
$<|[p,q]|>$. 
This gives that the product of the uncertainties (schematically)
\begin{equation}\delta p \delta q \ge {1 \over 4} <|[p,q]|>. \end{equation}
For x and p the right-hand side is a constant, 1 (in the proper units). However in general, including for gravitation, it is not, and is greater than 1. The uncertainty for gravitation is greater than the minimum and depends on the statefunction. For gravitation there are uncertainty principles but strong ones and ones that cannot be given in general since they depend on the statefunction. 

\section{Dirac's equation}\label{rm1d1}

Why does Dirac's equation hold? Despite an all too prevalent belief it is not some strange property of nature. It is a trivial property of geometry. 

Considering only space transformations, ignoring interactions and internal symmetry, objects (thus free) belong to states of the Poincar\'{e} group. This has two invariants (like the rotation group has one, the total angular momentum). For a massive object these are the mass and spin in the rest frame. Knowing these the object is completely determined. Thus two equations, not one, are needed to determine an object. For spin-${1 \over 2}$, only, these two can be replaced by one, Dirac's equation. Why is this? The momentum, $p_{\mu}$ is a four-vector. There is another four-vector, $\gamma_{\mu}$. Thus $\gamma_{\mu}p_{\mu}$ is an invariant. It is a property of the object, and we give that property the name mass. Thus
\begin{equation}\gamma_{\mu}p_{\mu} = m, \end{equation}
which is Dirac's equation. It gives the mass of the object, and the spin, ${1 \over 2}$. This is only possible because of the $\gamma_{\mu}$'s. These form a Clifford algebra and there is (up to inversions) only one for each dimension. This is then the reason for Dirac's equation, and only for a single spin. 

\section{Nobody noticed? Highly unlikely! --- the irrationale for string theory}\label{rm1st1}

String theory is designed to solve the problems caused by point particles. However there is nothing in any formalism that even hints at particles, let alone point particles. Where did this idea of particles come from? Could it really be that thousands of physicists are wasting their careers to solve the problems caused by particles with not a single one even noticing that there are none? What objects are is discussed elsewhere\cite{rm1imp}. This also has a rigorous proof, verified by others, that physics is possible only in dimension 3+1 so string theory must be wrong. Don't the dots on the screen in, say, the double slit experiment show that objects are points? Of course not, they are consequences of conservation of energy. See\cite{rm1rn} and also\cite{rm1cnfr}. There are infinities in intermediate steps of a particular approximation scheme, but they are all gone by the end. If a different scheme was used the idea of infinities would never have arisen. The laws of physics are not determined by physicists' favorite approximation method. Thus string theory is a mathematically impossible theory, in violent disagreement with experiment, carefully designed to solve the terrible nonexistent problems caused by nonexistent particles. Perhaps that is why physicists are so enthusiastic about it. 

\section{There are no Higgs; The reason for gauge transformations}\label{rm1hs1}

Why is there gauge invariance? Despite the opinion of many physicists it is not because God likes it. Rather it is the form Poincar\'e transformations take for massless objects and are possible for these only. This has been discussed in depth previously~(\cite{rm1ml}) although it can be explained in one obvious paragraph~(\cite{rm1imp}). Consider an electron and photon with momenta parallel and spins along the momenta (so parallel). There are transformations that leave the momenta unchanged, changing the spin direction of the electron, but cannot change that of the photon. Electromagnetic waves are transverse. (This is required by the Poincar\'e group, not God). Thus there are transformations acting on the electron but not on the photon, which is impossible. What are these transformations? Obviously gauge transformations. And that is exactly what the Poincar\'e group gives; all their properties follow. They are not possible for massive objects but are a required property of massless ones. 

The belief in Higgs bosons comes from the belief that all objects are invariant under gauge transformations, which strongly disagrees with experiment. Instead of giving that belief up it is kept, because physicists are emotionally attached to it, and a new field, that of Higgs bosons, is introduced to give objects mass. However as gauge transformations are the form Poincar\'e transformations take for massless objects and are possible only for these they cannot be applied to massive objects and it makes no sense to so apply them. That would be like saying that since orbital angular momentum has integer values all angular momenta has. Since this is not true a new field is introduced to produce half-integer values. That would make no sense and neither do Higgs bosons. There are no Higgs bosons. 

\section{Inertia}\label{rm1is1}
Some people are confused about inertia regarding it as a force or as something caused by distant matter in the universe. Why is there inertia? A consequence of it is that the velocity of an object does not change unless there is a force acting on it. Suppose that this were not true. Then objects would just move randomly, starting and stopping for no reason, moving erratically, unpredictably. There would be no law. But if there were no laws how can we say that inertia is due to distant matter? That would be meaningless since it would be impossible to predict or explain anything. Explanations like a fictitious force or distant matter would be meaningless. Nothing could be said. There has to be inertia otherwise there could be no physics. Physicists like to take the obvious and develop convoluted and impossible theories to explain what is beyond explanation. This explains nothing about physics but much about physicists.

\section{Theories necessary and nonsensical, and theories that are just nonsensical}\label{rm1ns1}

Classical physics, and Bohr's theory of the atom, are both nonsense, mathematically inconsistent, simply absurd\cite{rm1gf}. But they are essential. They are needed as steps to correct theories, and classical physics is essential for our civilization. String theory however is just nonsense, mathematically inconsistent, completely absurd. Why, what is the difference? Why do they work, of course only to some degree~(\cite{rm1imp})? 

The variables in classical physics, like position and momentum, are wrong, a correct theory cannot be built upon them. Yet correctly interpreted they can be used for a correct theory. They are actually operators, or expectation values of these. And friction is a phenomenological function summarizing the correct variables, the electromagnetic or gravitational fields.

Thus Newton's second law is a relationship between the second derivative of an expectation value and a function determined by the fields acting on the object. In that sense it is correct, but it cannot be pushed too far. It is purely phenomenological. 

What about Bohr's theory? Of course there are no orbits. What the theory gives are the regions of maximum probabilities. Here luck is very much involved. For the hydrogen atom there are simple rules for these. As a result it is possible to find these regions of maximum probabilities using Bohr's rules even though the way of guessing them is nonsense. 

Bohr's theory is nonsense, but essential. It worked, and physics advanced, only through great luck. Otherwise the advance of physics would have been greatly slowed. Look at nuclear physics, where there is no such luck.

But string theory, for example, no matter how interpreted, has no relationship to physics. It is just nonsense.

Does the vacuum have energy? Can particles pop out of the vacuum to change the solutions of equations? Where did such absurd ideas come from? One approximation scheme for solving the equations of quantum electrodynamics is perturbation theory. In it are terms which have been called vacuum expectation values. Does it have anything to do with the vacuum? Of course not. Just because someone gave it a name that includes the word vacuum physicists, who get very confused because of names (like quantum mechanics), decided that since the word vacuum is part of the name it must have something to do with a property of the vacuum~(\cite{rm1imp}). Because of that silly mistake they have invented a large set of (religious) beliefs about the properties of the vacuum. Of course all these beliefs are ridiculous. That is why physicists believe them so strongly. If a different approximation method was used, or a different name was given, these absurd beliefs would never have occurred. And if someone suggested them they would have been laughed at. It is an interesting psychological question why physicists, and journalists, accept such nonsense instead of laughing at it. Perhaps they enjoy being crackpots (and being laughed at). Undoubtedly they often are. 

\section*{Acknowledgements}

This discussion could not have existed without Norma Manko\v c Bor\v stnik.

\cleardoublepage
\chapter*{\Huge DISCUSSION SECTION}
\addcontentsline{toc}{chapter}{Discussion Section}
\newpage

\cleardoublepage

\title{Discussions on the Puzzles of the Dark Matter
Search~\thanks{These discussions took place through video conference organized by the
Virtual institute of astroparticle physics, whose activities are presented in this proceedings.}}
\author{G. Bregar${}^1$, J. Filippini${}^{2}$, M. Khlopov${}^{3}$,
  N.S. Manko\v c Bor\v stnik${}^1$, A.Mayorov${}^{4}$ and E.Soldatov${}^{4}$}
\institute{%
${}^1$Department of Physics, FMF, University of Ljubljana\\
 Jadranska 19, 1000 Ljubljana, Slovenia\\ 
${}^{2}$Department of Physics, University of California, Berkeley, CA 94720, USA\\
${}^{3}$Moscow Engineering Physics Institute (National Nuclear Research University), 115409 Moscow, Russia;\\   
Centre for Cosmoparticle Physics "Cosmion" 125047 Moscow, Russia;\\
 APC laboratory 10, rue Alice Domon et L\'eonie Duquet
\\75205 Paris Cedex 13, France\\
${}^{4}$Moscow Engineering Physics Institute (National Nuclear Research University)\\ 115409 Moscow, Russia}

\authorrunning{Video Conference Participants}
\titlerunning{Discussions on the Puzzles of the Dark Matter Search}
\maketitle

 \begin{abstract}
 The possibility that the two experiments: DAMA/NaI-LIBRA~\cite{mk2disc:rita0708}
 and CDMS~\cite{mk2disc:cdms} measure the fifth family baryons  predicted
 by the approach unifying spin and charges, proposed by N.S. Manko\v c Bor\v stnik, is discussed
 from the point of view of the measurements of the DAMA/NaI-LIBRA~\cite{mk2disc:rita0708} and CDMS~\cite{mk2disc:cdms}
 experiments.
 While the DAMA/NaI-LIBRA experiment very clearly sees not only the signal but also the annual
 modulation of the signal,  CDMS sees no signal. N. Manko\v c Bor\v stnik and G. Bregar, by 
 estimating what is happening when the fifth family neutron hits the first family nucleon, 
 predict that CDMS will in the near future see the fifth family baryons, if the fifth family baryons
 are what  DAMA/NaI-LIBRA measures.  M.Khlopov, A.Mayorov and E.Soldatov are proposing
 alternative scenario, which is applicable for the fifth family clusters and formulates 
 the conditions, under which the results
 of both experiments can be explained.
 In these discussions J. Filippini from CDMS collaboration helped a lot to 
 clarify what is happening in the measuring procedure of 
 both experiments, if they do measure
 a heavy family cluster with small enough scattering cross section on ordinary nuclei.
 \end{abstract}

\section{Introduction}

In this proceedings the talk of Norma Susana Manko\v c Bor\v stnik
(with G. Bregar) is presented, which analyzes the
possibility that the new stable family, predicted by the approach
unifying spins and charges, proposed by N.S. Manko\v c Bor\v
stnik~\cite{mk2disc:rpn}, (the approach is offering the mechanism for generating family, 
the only one in the literature, which is not assuming the number of 
families, as the standard model of the electroweak and colour interactions 
also does) is offering the solution for the open question of the cosmology and the
elementary particle physics concerning the origin of the dark matter. 
 Namely, the approach predicts more than three known
families. It predicts two times four in the Yukawa couplings (from
the point of view of the age of our universe) decoupled families.
The first very rough estimations show~\cite{mk2disc:rpn,mk2disc:Norma} that the
fourth family might be seen at LHC, while the fifth family is, as a
stable family,  a candidate to form the dark matter clusters. The
analyze of S.N.M.B. and G.B. (presented in this proceedings) of what
is happening in the measuring apparatus of both experiments, if they measure 
the fifth family neutrons, still
leaves the possibility that CDMS and DAMA/NaI-LIBRA experiments do
measure the clusters of the fifth family quarks open, predicting
that in this case the CDMS (or any other similar experiment) will in
the near future see the events, triggered by  the fifth family
clusters. The discussions below
  with J. Filippini helped a lot  to clarify  
  the uncertainties in the approximate evaluations of what the two experiments measure~\cite{mk2disc:Norma}.

On the other hand M. Khlopov, A. Mayorov and E. Soldatov (M.K. talk
is in this proceedings), found a possible solution for the DAMA/CDMS
controversy based on the scenario of composite dark
matter~\cite{mk2disc:rpm} and noted that for the excess of anti-u quarks of
the fifth family predicted by the approach unifying spins and
charges proposed by N.S.Manko\v c Bor\v stnik~\cite{mk2disc:rpn})
their scenario can be  realized, with no contradiction with the DAMA/CDMS 
experiments even if CDMS sees no events and DAMA does. 

\section{Discussions}

These discussions took place through videoconference organized by the
Virtual institute of astroparticle physics, whose activities are presented in this
proceedings~\cite{mk2disc:MaximVIA}.

To clarify the CDMS data analysis Jeff Filippini was asked questions
during VIA discussions. Bellow the questions and the answers are
presented. There were also two talks taking place during the
discussions: The one of N.S. Mank\v c Bor\v stnik and the one of M.
Khlopov and the contributions of G. Bregar, A. Mayorov and E.
Soldatov. Contributions are included in the two talks of this
Proceedings while the talk of N.S. Manko\v c Bor\v stnik and M.
Khlopov can be found on the website http://www.cosmovia.org
\cite{mk2disc:Norma,mk2disc:Maxim}

The two starting questions for J. Filippini:

\begin{itemize}

\item What is the real exposure time in your experiment?  It is not
clear for us what means effective exposure 121.3 kg-d, averaged over
recoil energies and weighted for WIMPs of a definite mass.

\item Are we right to consider as  a real exposure  the  number of
germanium and silicon nuclei, active in the period of 
measurements?

\end{itemize}

The talk of N.S. Manko\v c Bor\v stnik entitled "Does the dark
matter consist of baryons of a new  heavy  stable family  predicted
by the approach unifying spins and charges?", which can be found on
website \cite{mk2disc:Norma}, took then place.

Then the talk of M. Khlopov, which can be found on website
\cite{mk2disc:Maxim} was presented.

Then the three contributions
of G. Bregar, A. Mayorov and E. Soldatov, included now in the proceedings as a part of two talks,
were presented.

The J. Fillippini answers:

\begin{itemize}

\item
The raw Ge exposure (mass times good running time) for this run is 421
kg-days.  Our various data quality and event selection cuts reduce our
fiducial exposure for WIMP search substantially, and the efficiency
(signal acceptance) of these cuts varies with recoil energy.   The raw
exposure and efficiency function (Figure 2) describe our sensitivity,
but we can't give a single number which expresses our exposure for all
WIMP masses.

The goal of the "WIMP-spectrum-averaged exposure" is to express the
effects of the cuts (including their energy dependencies) to give an
equivalent exposure in a single number.  Our value of 121 kg-days at
60 GeV/c2 means that our sensitivity (assuming no background) is
equivalent to that of a "perfect" Ge experiment with 100
acceptance and the same energy thresholds (10-100 keV in recoil
energy) attempting to detect a 60 GeV/c2 WIMP.  To compute this we
convolve the energy-dependence of our signal acceptance with the
expected recoil spectrum of 60 GeV WIMPs.  Our experiment and this
"perfect" experiment would set the same limit on a 60 GeV WIMP if no
events were observed.

Since the expected WIMP recoil energy spectrum varies with WIMP mass
and our signal acceptance varies with recoil energy, this equivalence
is only true for one particular WIMP mass.  The equivalent exposure of
our experiment will be slightly different for different WIMP masses,
but since our efficiency is nearly flat with recoil energy the
equivalent exposure will not vary much.

\end{itemize}

Questions for J. Fillipini and his answers:

\begin{itemize}

\item Does your exclusion curve means that particles with mass 10 TeV
and cross section $10^{-34} {\rm cm}^{2}$ are excluded?

The answer: That's correct, up to the usual uncertainties from the WIMP halo
model.  The WIMP recoil spectrum goes roughly as $~e^(-E/E0)$, so heavy
WIMPs still should produce plenty of low-energy recoils that we could
detect.  The WIMP-nucleus cross section becomes essentially
independent of WIMP mass when the WIMP is much heavier than the target
nucleus.  The incident flux of WIMPs goes inversely as the WIMP mass
(since the total density is fixed), so our limit curve asymptotically
goes as ~M.  At 10 TeV our upper limit should be a few times $10^{-42}
{\rm cm}^2$.

The WIMP cross section on a nucleus is proportional to $A^2 $ (from  the
coherence across the various protons and neutrons in the nucleus) and
to $mu^2$ (the square of the reduced mass: $1/mu = 1/M_W + 1/M_N$).  There
are also form factor corrections, but we'll ignore those for now.

We traditionally plot the WIMP's cross section on a single proton,
$\sigma_P$, since this allows for fairer comparisons between different
experiments.  Because of the above scaling factors, $\sigma_{Ge} = \sigma_{P}
\, (A^2) \, (mu_{Ge}/mu_{P})^2$.  From our plot, the WIMP-proton cross
section is $\sigma_{P} = 6.6 \cdot e-44$ ${\rm cm}^2$.  The coherence factor
($A^2$) is $~72^2$
~5200 and the reduced mass factor is ($mu_{Ge}/mu_{P})^2 ~ 1040$.  This
gives $\sigma_{Ge} = 3.5 \cdot e-37 {\rm cm}^2$ for the cross section on a Ge atom.

Following through your computation (putting in an expected incident
velocity of $(\sqrt{(3/2)}\, 220 = 270) {\rm km/s}$), I get a net rate of R ~ 1 / (43
kg-d).  For zero observed events, a $90 \%$ upper limit corresponds to an
expectation of 2.3 events.  From the above computation, this should
happen for an exposure of ~100 kg-days.  The form factor and full halo
model probably account for the remaining $20\%$, but the order of
magnitude is correct.

The factor of $~10^7$ comes from the ratio of total cross sections
for Ge atoms versus single protons.

\item
(Norma): Can it be that your ”cleaning procedure” of the noise disregards
3?, 14?, 25?, 28? events? Which is the higher value you could agree with?\\
Does your ”cleaning procedure” of the noise depend on the assumed
recoiled energy?
To which extend is your way of "cutting away the
noise"  similar to DAMA?

The answer:  WIMPs are
often assumed to be stationary on average within the galactic frame,
but individually they must be moving at velocities comparable to the
sun's (more similarly to case 2b).  Most toy models of the dark matter
halo give an exponentially-declining distribution of recoil energies;
this exponential spectrum makes the choice of energy threshold
especially important.  Some of these additional effects become less
important to the overall event rate as the WIMP becomes sufficiently
heavy, but they are very important to the changes of spectrum which
determine the an annual modulation signal.

One useful resource for checking the results of the usual framework is
a set of web tools posted by the ILIAS consortium in Europe:

http://pisrv0.pit.physik.uni-tuebingen.de/darkmatter/

 From these pages you can quickly check event rates and recoil energy
spectra for various dark matter particle masses and cross sections
incident upon different targets; you can also vary the halo model
parameters and annual modulation phase (theta) somewhat.  I have not
checked the accuracy of this site's calculations personally, but I
expect them to be good.

For now, I'll skip to your questions ...

 It seems very unlikely to me that CDMS's analysis procedure could
have missed such a large number of events, assuming coherent
scattering on a nucleus.  The performance of our analysis is measured
directly using an in situ calibration with a neutron source, which
gives us confidence in our result.  There is a $~5\%$ chance that a model
predicting 3 events could have produced nothing in our experiment due
simply to statistical fluctuations, but much larger numbers are very
hard to believe.

The main implicit assumption we're making (and that most other
experiments in this field make) is that recoils induced by dark matter
particles behave similarly to nuclear recoils of the same energies
induced by neutrons.  This assumption could be wrong: a particle which
deposited significant electromagnetic energy, for example, could fall
outside of our signal region and be missed.  We only claim validity
for our limit curve within the usual spin-independent WIMP framework:
a massive, neutral particle scattering coherently from an atomic
nucleus.  Within this framework, I know of no effect which could
change our limit by more than, say, $~10\%$.

Other frameworks require different analyses, of course.  We have
looked for signatures of axion-like particles and low-mass WIMPs in
our data, and these results will be made available in upcoming months.

 The performance of our analysis cuts does vary with recoil
energy.  As you can see from Figure 2 of our preprint, however, the
signal acceptance does not change much over the 10-100 keV range.

 CDMS and DAMA handle our backgrounds and noise quite differently.
Both experiments have cuts in place to exclude events due to readout
noise, and these are not so very different in principle.  DAMA does
very little beyond this, trusting the annual modulation signal to be
visible above the remaining background.  CDMS imposes a series of
further cuts upon our data to reject electron recoil events, and I do
not believe these have direct analogues in the DAMA analysis.

In principle, any dark matter signal should be visible as both a
modulation and an excess event rate.  For now, we can only say that
CDMS and DAMA have results which are inconsistent within some
frameworks for dark matter (albeit the traditionally popular ones).
The DAMA/LIBRA modulation could be due to an unusual systematic, or it
could be a sort of dark matter we weren't looking for.  CDMS and
similar experiments will continue to push the bounds of sensitivity
for WIMP dark matter in the usual mass range ($>$ a few GeV), while
further analyses and new experiments explore more of the possible
parameter space (very low masses, non-WIMPs, etc.).

\item
(Maxim): I propose to put the present discussion and to continue
discussion of Puzzles of dark matter searches on Forum of Virtual
Institute of Astroparticle physics \cite{mk2disc:Forum}.
\end{itemize}

\author{A. Kleppe}
\title{Scattering With Very Heavy Fermions}
\institute{%
SCAT, Oslo, Norway}

\authorrunning{A. Kleppe}
\titlerunning{Scattering With Very Heavy Fermions} 
\maketitle 

\begin{abstract} 
Discussion concerns the evaluation of the colour, the weak and the "nuclear" interactions
among the fifth family quarks, among the fifth family baryons and  among the 
fifth family and ordinary baryons.
\end{abstract}
 
We have been discussing how the cross section for scattering of the two fifth family quarks (or a quark and an antiquark)
depends on the mass  of quarks, if the average kinetic energy of the colliding members of the fifth family quarks
is of the same order of magnitude as their mass (the temperature, when they collide, is namely  $k_b T= m_{q_5} c^2$).
There are four families of fermions of lower masses (the three observed and the fourth with the masses of quarks at
around 200 GeV).

\section{Generic scattering processes}

We want to know how the cross sections depend on the quark masses when the average kinetic energy of the colliding members is of the same order of magnitude as their (very heavy) masses.

\begin{figure}
  \centering
  \includegraphics[width=10cm]{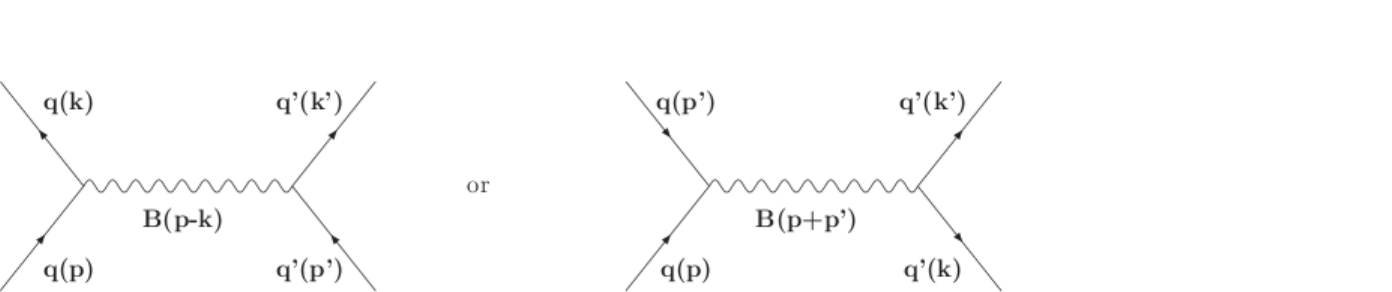}
\caption{\label{ak-disc-fig1}%
Generic scattering process.}
\end{figure}

We should in principle take all the interactions into account, but the strong interaction is the dominating one.
I am no QCD expert, so I start by looking at some generic scattering processes, like shown in figure~\ref{ak-disc-fig1}

where ${\bf{q}}$, ${\bf{q'}}$ and ${\bf{\bar{q}}}$, ${\bf{\bar{q'}}}$ are quarks and antiquarks, respectively, and $B_{\mu}$ is some vector boson.

With a Lagrangian of the form
${\cal{L}}={\bar{q}}\gamma^{\mu}q B_{\mu}$, where $B_{\mu}$ is the (generic) vector boson, and the amplitudes corresponding to the diagrams in Fig.~\ref{ak-disc-fig1} are 
\begin{eqnarray}
& & M_1 \sim \bar{u_s}(k)\gamma^{\mu} u_r(p){\cal{P}}_{\mu \rho}(p-k)\bar{u_s'}(k')\gamma^{\rho} u_r'({p'}),\nonumber\\
& & M_2 \sim \bar{v_r'}({p'})\gamma^{\mu} u_r(p){\cal{P}}_{\mu \rho}(p+p')\bar{u_s'}(k')\gamma^{\rho} v_s(k)\nonumber
\end{eqnarray}
where the $u$:s and $v$:s are spinors, $s,r,s',r'$ are spins, and ${\cal{P}}_{\mu \rho}$ are the boson propagators.
The corresponding cross sections are  
\begin{equation}
d\sigma=\frac{1}{v_{rel}(2\pi)^3}{\frac{|M_{j}|^2}{2p_02p'_0}}\partial^4(p+p'-k-k')\frac{d^3kd^3k'}{2k_02k'_0}
\end{equation}
where $M_{j}$ are the transition amplitudes, $j=1,2$.

For particles in a Lorentz frame, moving collinearly (e.g. the lab system or the Center of Mass system). In such a frame
$v_{rel}$ is given by
\begin{equation}
v_{rel} = \frac{{\sqrt{(pp')^2-m^2m^{'2}}}}{p_0 p'_0}
\end{equation}
where $m$ = $m_q$ and $m'$ = $m_{q'}$.

The diagrams in Fig.~\ref{ak-disc-fig1} encompass electromagnetic, weak and strong interactions. We are mainly interested in the strong interaction diagrams, but the question is the form of the gluon propagators, since unlike
the electromagnetic and weak interactions that can be handled pertubatively, in strong interaction all orders may be of comparable magnitude (i.e. the contributions from the lowest order diagrams are no longer overwhelmingly dominating).

\section{The cross sections}
In order to calculate the cross sections for the prosesses in Fig.~\ref{ak-disc-fig1}, we initially use the standard Lagrangian
\begin{eqnarray}
{\cal{L}}=\sum_{j=1}^{n_f}\bar{q_j}(i{\displaystyle{\not}D-m_j)q_j-\frac{1}{4}G_{\mu\nu}G^{\mu\nu}}
\end{eqnarray}
where where $n_f$ is the number of families, $q_j$ is the quark (Dirac) field, $${\displaystyle{\not}D=(\partial_{\mu}-igA_{\mu})\gamma^{\mu}},$$ 
$A_{\mu}$ is the gluon field and $G_{\mu\nu}G^{\mu\nu}$ the gluon field strength; and
the quark mass parameter $m_j$ depends on the renormalization scheme and the scale parameter.

We follow the usual procedures with averaging over initial spins and summing the final spins,  
\begin{eqnarray}
& &\frac{1}{4}\sum_{s,r,s',r'} |M_1|^2 \sim 
 \frac{1}{4}\sum_{s,r,s',r'}
|\bar{u}_s(k)\gamma^{\mu} u_r(p){\cal{P}}_{\mu \rho}(p-k)\bar{u}_s'(k')\gamma^{\rho} u_r'({p'})|^2 =\nonumber\\ 
 & &= \frac{1}{4} Tr[(\frac{\displaystyle{\not} k +m}{2m})\gamma^{\mu}(\frac{\displaystyle{\not} p +m}{2m})\gamma^{\eta}]
\nonumber\\
& & Tr[(\frac{\displaystyle{\not} k' +m'}{2m'})\gamma^{\rho}(\frac{\displaystyle{\not} {p'} +m'}{2m'})\gamma^{\phi}]{\cal{P}}_{\mu \rho}(p-k){\cal{P}}_{\eta\phi}^{\dagger}(p-k)=\nonumber\\
& &= \frac{1}{4}\frac{{\cal{P}}_{\mu \rho}(p-k){\cal{P}}_{\eta\phi}^{\dagger}(p-k)}{m^2m'^2}[k^{\mu}p^{\eta}+k^{\eta}p^{\mu}+g^{\mu \eta}(m^2-kp)]\nonumber\\
 & & [k'^{\rho}p'^{\phi}+k'^{\phi}{p'}^{\rho}+g^{\rho \phi}(m'^2-k'{p'})]
\end{eqnarray}
and
\begin{eqnarray}
& &\frac{1}{4}\sum_{s,r,s',r'} |M_2|^2 \sim \\
           & &\sim \frac{1}{4}\frac{{\cal{P}}_{\mu \rho}(p+p'){\cal{P}}_{\eta\phi}^{\dagger}(p+p')}{m^2m'^{'2}}
 [{p'}^{\mu}p^{\phi}+{p'}^{\phi}p^{\mu}-g^{\mu \phi}(pp'+m^2)]\nonumber\\
 & & [k'^{\rho}k^{\eta}+k'^{\eta}k^{\rho}+g^{\rho \eta}(kk'+{m'}_q^2)],\nonumber
\end{eqnarray}
with the corresponding cross sections 
\begin{eqnarray}
& &d\sigma_1=\frac{1}{4v_{rel}(2\pi)^3}{\frac{|M_1|^2}{2p_02p'_0}}\partial^4(p+p'-k-k')\frac{d^3kd^3k'}{2k_02k'_0}=\nonumber\\
& &=\frac{1}{4v_{rel}(2\pi)^3}{\cal{P}}_{\mu \rho}(p-k){\cal{P}}_{\eta\phi}^{\dagger}(p-k)\frac{[k^{\mu}p^{\eta}+k^{\eta}p^{\mu}+g^{\mu \eta}(m^2-kp)]}{m^2m'^2}\otimes\nonumber\\
& &\otimes{\frac{[k'^{\rho}p'^{\phi}+k'^{\phi}{p'}^{\rho}+g^{\rho \phi}(m'^2-k'{p'})]}{ 2p_02p'_0}}\partial^4(p+p'-k-k')\frac{d^3kd^3k'}{2k_02k'_0}
\end{eqnarray}
and
\begin{eqnarray}
& &d\sigma_2=\frac{1}{4v_{rel}(2\pi)^3}{\frac{|M_2|^2}{2p_02p'_0}}\partial^4(p+p'-k-k')\frac{d^3kd^3k'}{2k_02k'_0}=\nonumber\\
& &=\frac{1}{4v_{rel}(2\pi)^3}{\cal{P}}_{\mu \rho}(p+{p'}){\cal{P}}_{\eta\phi}^{\dagger}(p+{p'})\frac{[{p'}^{\mu}p^{\phi}+{p'}^{\phi}p^{\mu}-g^{\mu\phi}(p{p'}+m^2)]}{m^2m'^2}\otimes\nonumber\\
& &\otimes{\frac{[k'^{\rho}k^{\eta}+k'^{\eta}k^{\rho}+g^{\rho \eta}(kk'+{m'}_q^2)]}{ 2p_02p'_0}}\partial^4(p+p'-k-k')\frac{d^3kd^3k'}{2k_02k'_0}
\end{eqnarray}

In calculating the first process, we go to the lab system, where ${\bf{p'}}$ = 0, and 
\begin{eqnarray} 
pp' = (p_0,{\bf{p}})(p'_0,{\bf{0}}) = p_0p'_0,\nonumber
\end{eqnarray} 
so
\begin{eqnarray}
v_{rel}=\frac{|{\bf{p}}|}{p_0}\nonumber,
\end{eqnarray} 
and the cross section reads
\begin{eqnarray}
& &d\sigma_1=\frac{p_0}{4(2\pi)^3|{\bf{p}}|}{\frac{|M_1|^2}{2p_02p'_0}}\partial^4(p+p'-k-k')\frac{d^3kd^3k'}{2k_02k'_0}=\nonumber\\
& &=\frac{p_0}{4(2\pi)^3|{\bf{p}}|}{\cal{P}}_{\mu \rho}(p-k){\cal{P}}_{\eta\phi}^{\dagger}(p-k)\frac{[k^{\mu}p^{\eta}+k^{\eta}p^{\mu}+g^{\mu \eta}(m^2-kp)]}{m^2m'^2}\otimes\nonumber\\
& &\otimes{\frac{[k'^{\rho}p'^{\phi}+k'^{\phi}{p'}^{\rho}+g^{\rho \phi}(m'^2-k'{p'})]}{ 2p_02p'_0}}\partial^4(p+p'-k-k')\frac{d^3kd^3k'}{2k_02k'_0}
\end{eqnarray}
For the second process, we go to the center of mass system, i.e. 
${\bf{p'}}$ = -${\bf{p}}$, 
so
\begin{eqnarray}
v_{rel}=|{\bf{p}}|\frac{(p_0+p'_0)}{p_0p'_0}\nonumber,
\end{eqnarray} 
and 
\begin{eqnarray}
& &d\sigma_2=\frac{p_0p'_0}{{4|{\bf{p}}|}(p_0+p'_0)(2\pi)^3}{\frac{|M_2|^2}{2p_02p'_0}}\partial^4(p+p'-k-k')\frac{d^3kd^3k'}{2k_02k'_0}=\nonumber\\
& &=\frac{p_0p'_0}{4|{\bf{p}}|(p_0+p'_0)(2\pi)^3}{\cal{P}}_{\mu \rho}(p+{p'}){\cal{P}}_{\eta\phi}^{\dagger}(p+{p'})\nonumber\\
 & & \frac{[{p'}^{\mu}p^{\phi}+{p'}^{\phi}p^{\mu}-g^{\mu\phi}(p{p'}+m^2)]}{m^2m'^2}\otimes\nonumber\\
& &\otimes{\frac{[k'^{\rho}k^{\eta}+k'^{\eta}k^{\rho}+g^{\rho \eta}(kk'+{m'}_q^2)]}{ 2p_02p'_0}}\partial^4(p+p'-k-k')\frac{d^3kd^3k'}{2k_02k'_0}
\end{eqnarray}

\section{Gluon propagators}

There are several methods to obtain the gluon propagator by using non-pertur\-bative
methods, like in lattice field theory or the Dyson-Schwinger equations (an infinite system of non-linear coupled integral equations 
relating the different Green functions of a quantum field theory). 
There are many different solutions for both methods, and in many cases it is useful to introduce a dynamical gluon mass \cite{corn}.

In a general covariant gauge, the gluon propagator can be expressed \cite{burden}
\begin{eqnarray}
g²D_{\mu\nu}(Q)=(\delta_{\mu\nu}-Q_{\mu}Q_{\nu}/Q^2)\Delta(Q^2)+g^2 \xi Q_{\mu}Q_{\nu}/Q^2
\end{eqnarray}
where $\xi$ is a gauge fixing parameter.

We merely want to know how the cross sections
depend on the quark masses. For $d\sigma_2$ we go to the CM system and represent the gluon propagator by a a generic expression, neglecting the subtleties of transversal and longitudinal terms, $P_{\alpha\beta}\sim (\delta_{\alpha\beta}-Q_{\alpha}Q_{\beta}/Q²)$. This gives, for $Q=p+p'=k+k'$, 

\begin{eqnarray}
& &d\sigma_2\sim\frac{p_0p'_0}{4|{\bf{p}}|(p_0+p'_0)(2\pi)^3}(\delta_{\mu \rho}-\frac{Q_{\mu}Q_{\rho}}{Q^2})(\delta_{\eta\phi}-\frac{Q_{\eta}Q_{\phi}}{Q^2})
\nonumber\\
 & & \frac{[{p'}^{\mu}p^{\phi}+{p'}^{\phi}p^{\mu}-g^{\mu\phi}(p{p'}+m^2)]}{m^2m^{'2}}\otimes\nonumber\\
& &\otimes{\frac{[k'^{\rho}k^{\eta}+k'^{\eta}k^{\rho}+g^{\rho \eta}(kk'+{m'}^2)]}{ 2p_02p'_0}}\partial^4(p+p'-k-k')\frac{d^3kd^3k'}{2k_02k'_0}=\nonumber\\
& &=\frac{2[(pk)(p'k')+(pk')(p'k)]+Q^2(m'^2-m^2)-Q^4}{64|{\bf{p}}|(p_0+p'_0)(2\pi)^3m^2m'^2}\nonumber\\
 & & \partial^4(p+p'-k-k')\frac{d^3kd^3k'}{k_0k'_0}
\end{eqnarray}
Likewise for the first process in Fig.~\ref{ak-disc-fig1} (in lab system).  
In the limit of very heavy quark masses, it is perhaps more realistic to use 
the static HQET Lagrangian below.

This is just an attempt to get a grip on the form of these (very basic) cross sections, well aware of the fact that  
QCD is very different at different energy scales, and that the form of the gluon propagators therefore need a careful consideration.

In fact, QCD is really a multi-scale theory:
\begin{eqnarray}
& &m \ll \Lambda \ll m\nonumber\\
& &m = m_u, m_d , m_s \nonumber\\
& &m = m_c, m_b, m_t,...\nonumber\\ 
& &\Lambda \sim 700 MeV\nonumber
\end{eqnarray}

There are two more scales: for the ultraviolet cutoff
$(\pi/a)$ and the infrared cutoff (${L^{-1}}$).
In principle we have
${{L^{-1}}}\ll m \ll \Lambda \ll  m_b \ll \pi/a$,        
but in practice 
${L^{-1}}\ll m \ll \Lambda \ll  m_b \sim \pi/a$.        

A multi-scale problem can be dealt with by introducing some scale-separating scheme (like an effective field theory), this is precisely what is done in lattice QCD.
An effective field theory separates short-distance effects from long-distance effects by introducing a separation scale to place a boundary between “long” and “short”, and also introducing
new fields that to describe the long-distance part, and then equating the effective theory and the underlying theory, i.e.                                     
                                              $L_{underlying} = L_{effective}$.

The electroweak effective Hamiltonian is an example of this. The underlying theory is the
Standard Model, and the effective Hamiltonian is a theory of
photons, gluons, and the five lighter quarks.

Another example is the non-relativistic effective theory for heavy quarks, like the heavy-quark effective theory (HQET) and non-relativistic QCD. Heavy-quark fields in underlying theory has four components; in effective theory two components.
These can be derived to all orders in QED and QCD perturbation theory.

\section{Heavy Quark Effective Theory}
The Heavy Quark Effective Theory, HQET, is an effective theory that is obtained from QCD by performing a $1/m$ expansion, where $m$ is the heavy quark mass. In the infinite mass limit the heavy quark acts like a static colour source. The momentum $p_q$ of the heavy quark scales with the mass, therefore one uses the quark velocity $v$ as the kinematic parameter. 
Taking the heavy quark momentum to be $p_q=mv+p'=m(v+p'/m)$, where $p'$ is the part of the momentum that does not scale with the mass, we can express the heavy quark field (in full QCD) as 
\begin{eqnarray}
& &Q(x)=e^{-imvx}[1+(\frac{1}{2m +iv D})i\displaystyle{\not}{D}_{\perp}]q_{v}\nonumber\\
& & =e^{-imvx}[1+\frac{1}{2m}\displaystyle{\not}{D}_{\perp} +(\frac{1}{2m})^2(-iDv)\displaystyle{\not}{D}_{\perp}+...]q_{v}
\end{eqnarray}
and the Lagrangian
\begin{eqnarray}
& &{\cal{L}}=\bar{q}_{v}(ivD)q_{v}+\bar{q}_{v}i\displaystyle{\not}{D}_{\perp}(\frac{1}{2m +iv D}) i\displaystyle{\not}{D}_{\perp}q_{v}=\\
& &=\bar{q}_{v}(ivD)q_{v}+\frac{1}{2m}\bar{q}_{v}(i\displaystyle{\not}{D}_{\perp})^2iq_{v}+(\frac{1}{2m})^2\bar{q}_{v}(i\displaystyle{\not}{D}_{\perp})(-ivD)(i\displaystyle{\not}{D}_{\perp})q_{v}+..\nonumber
\end{eqnarray}
where $D$ is the covariant derivative of QCD, and $1/m$ is the mass of the heavy quark field $Q$ in full QCD, and $q_v$ is the static heavy quark moving with velocity $v$, corresponding to the upper components of the full field, since

$P_{(+)}q_v=q_v$, $P_{(-)}q_v=0$, $P_{(\pm)}=(\displaystyle{\not}{v}\pm 1)/2$. The leading terms of these expansions define the static limit with the static Lagrangian   
\begin{equation}
{\cal{L}}_{stat}=\bar{q}_{v}(ivD)q_{v}
\end{equation}
that describes the heavy degrees of freedom. 

In the case when bottom and charm are perceived as be heavy, the static Lagrangian for both quarks is written

\begin{equation}
{\cal{L}}_{stat}=\bar{b}_{v_b}(ivD)b_{v_b}+\bar{c}_{v_c}(ivD)c_{v_c}
\end{equation}
where $b_{v_b}$ and $c_{v_c}$ are the bottom and charm quarks moving with velocities $v_c$ and $v_b$, respectively. Notice that the quark masses do not appear in the Lagrangian, which means that the Lagrangian has a heavy flavour symmetry which allows rotations of the $b$- and $c$-fields into each other.

Both spin directions of the heavy quark couple in the same way to the gluons as well, so the Lagrangian has a symmetry under rotations of the heavy quark spin, so all the heavy hadron states that move with velocity $v$ fall into spin-symmetry doublets in the infinite mass limit.

\section{How to proceed}
We should discuss which processes should be considered, i.e. what scattering processes (involving heavy fermions) would be characteristic within this scenario.
We should moreover discuss what gluon propagators we should use, and also consider the electroweak contribution to the cross sections.

\cleardoublepage
\chapter*{\Huge PRESENTATION OF \\
VIRTUAL INSTITUTE OF ASTROPARTICLE PHYSICS \\
AND\\
 BLED 2008 WORKSHOP VIDEO CONFERENCE}
\addcontentsline{toc}{chapter}{Presentation of Virtual Institute of
  Astroparticle Physics and Bled 2008 Workshop Video Conference}
\newpage
\cleardoublepage


\author{M.Yu. Khlopov$^{1,2,3}$}
\title{Scientific-Educational Complex\,---\,Virtual Institute of Astroparticle Physics}
\institute{%
$^{1}$Moscow Engineering Physics Institute (National Nuclear Research University), 115409 Moscow, Russia \\
$^{2}$ Centre for Cosmoparticle Physics "Cosmion" 125047 Moscow, Russia \\
$^{3}$ APC laboratory 10, rue Alice Domon et L\'eonie Duquet \\75205
Paris Cedex 13, France}

\authorrunning{M.Yu. Khlopov}
\titlerunning{Scientific-Educational Complex\,---\,VIA}
\maketitle

\begin{abstract}
Virtual Instutute of Astroparticle Physics (VIA) has evolved in a
unique
 multi-functional complex, aimed to combine various forms of collaborative
 scientific work with programs of education on distance.
The activity on VIA website includes regular video conferences with
systematic basic courses and lectures on various issues of
astroparticle physics, library of their records and presentations, a
multilingual forum. VIA virtual rooms are open for meetings of
scientific groups and for individual work of supervisors with their
students. The format of a VIA video conference was used in the program
of Bled Workshop to discuss the puzzles of dark matter searches.
\end{abstract}

\section{Introduction}
Studies in astroparticle physics link astrophysics, cosmology and
particle physics and involve hundreds of scientific groups linked by
regional networks (like ASPERA/ApPEC \cite{mk2viaaspera}) and national
centers. The exciting progress in precision cosmology, in
gravitational wave astronomy, in underground, cosmic-ray and
accelerator experiments promise large amount of a new information
and discoveries in coming years. Theoretical analysis of these
results will have impact on the fundamental knowledge on the
structure of microworld and Universe and on the basic, still
unknown, physical laws of Nature (see e.g. \cite{mk2viabook} for review).

It is clear that the effectiveness of the work depends strongly on
the number of groups involved in this activity, on the information
exchange rate and on the overall coordination. An international
forum, be it virtual, which can join all the groups and coordinate
their efforts would give a boost to this cooperation. Particularly
this is important for isolated scientific groups and scientists from
small countries which can contribute a lot to this work being a part
of the large international collaboration. A possibility of education
on distance, involving young people from all over the world, is another
important aspect of this activity.

A good example of such
kind of structure  is an International Virtual Observatory
\cite{mk2viaIVOA}, created in 2002. It has demonstrated the work
effectiveness and fruitful cooperation of many organizations all
over the world. Problems of Virtual Laboratories were discussed in
\cite{mk2viaHut:2007ke}.

In the proposal \cite{mk2viaKhlopov:2008vd} it was suggested to organize a
Virtual Institute of Astroparticle Physics (VIA), which can play the
role of such unifying and coordinating structure for astroparticle
physics. Starting from the January of 2008 the activity of the
Institute takes place on its website \cite{mk2viaVIA} in a form of regular
weekly video conferences with VIA lectures, covering all the
theoretical and experimental activities in astroparticle physics and
related topics. The library of records of these lectures and their
presentations is now accomplished by multi-lingual forum. Here the
general structure of VIA complex and the format of its
video conferences are stipulated to clarify the way in which VIA
discussion of puzzles of dark matter searches took place in the
framework of Bled Workshop.
\section{The structure of VIA complex}
\begin{figure}
    \begin{center}
        \includegraphics[scale=0.38]{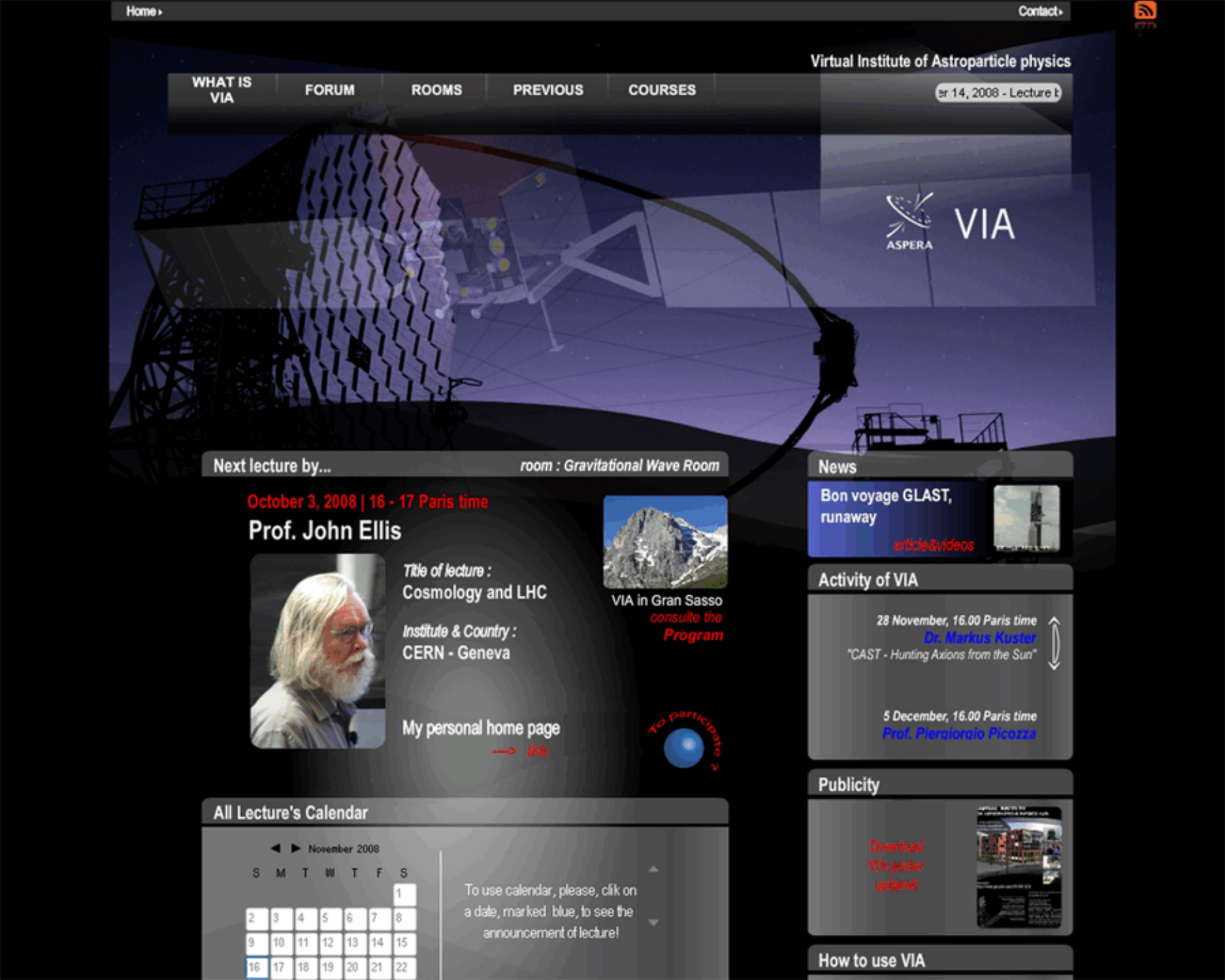}
        \caption{The home page of VIA site}
        \label{mk2viaa}
    \end{center}
\end{figure}

The structure of VIA complex is illustrated on Fig. \ref{mk2viaa}.
The home page, presented on this figure, contains the information on
VIA activity and menu, linking to directories (along the upper line
from left to right): with general information on VIA (What is VIA),
to Forum, to VIA virtual lecture hall and meeting rooms (Rooms), to
the library of records and presentations of VIA lectures and courses
(Previous) and to contact information (Contacts). The announcement
of the next Virtual meeting, the calender with the program of future
lectures and courses together with the links to VIA news and posters
as well as the instructions How to use VIA are also present on the
home page. The VIA forum, now being ready to operate, is intended to cover the
topics: beyond the standard model, astroparticle physics, cosmology,
gravitational wave experiments, astrophysics, neutrinos. Presently
activated in English, French and Russian with trivial extension
to other languages, the Forum represents a first
step on the way to multi-lingual character of VIA complex and its activity.

\section{VIA lectures and virtual meetings}
\begin{figure}
    \begin{center}
        \includegraphics[scale=0.38]{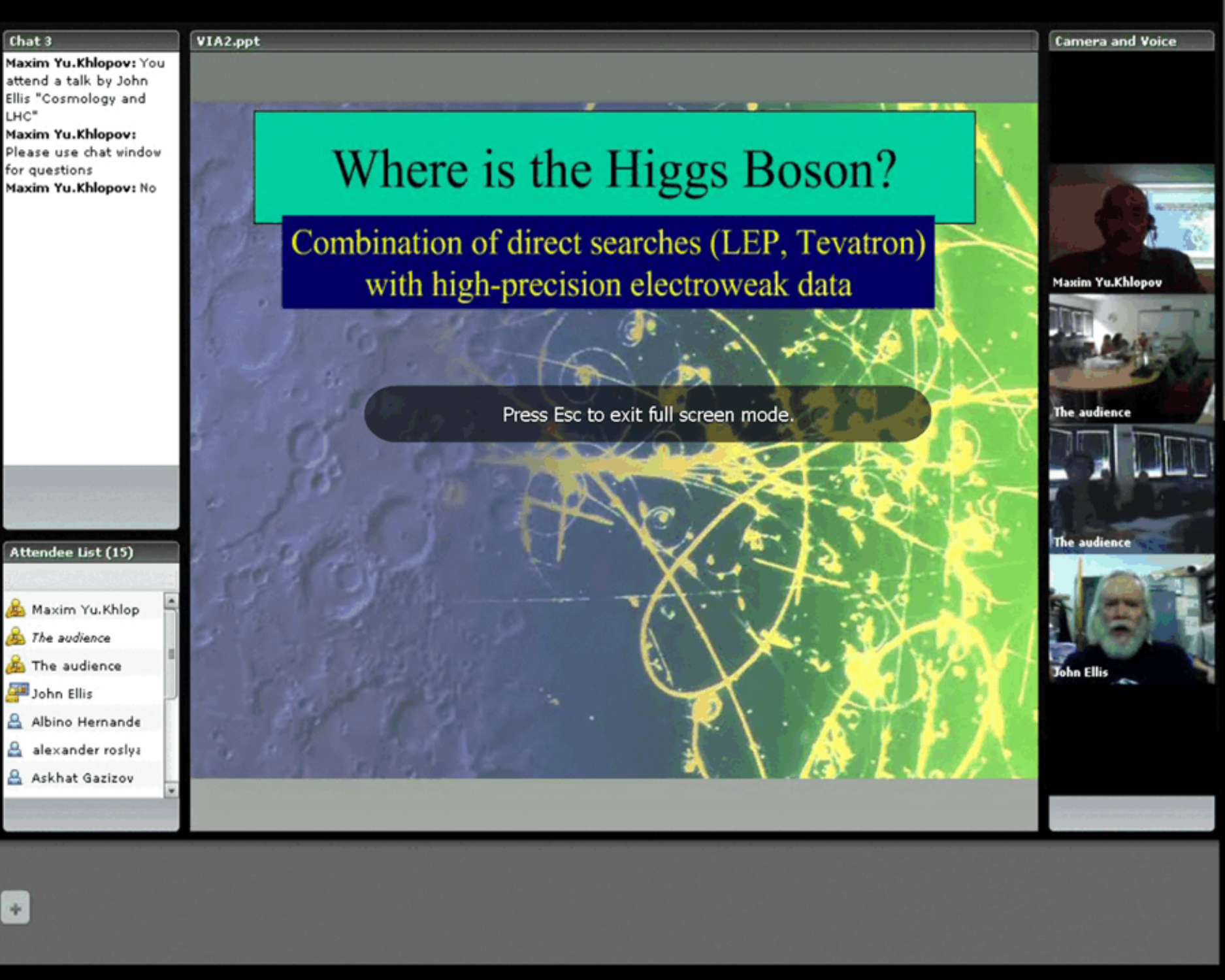}
        \caption{Video conference with lecture by John Ellis, which he gave from his office in CERN,
        Switzerland, became a part of the program of
           XIII Summer Institute on Astroparticle physics in Gran Sasso, Italy}
        \label{mk2viab}
    \end{center}
\end{figure}

First tests of VIA system, described in \cite{mk2viaKhlopov:2008vd}, involved various
systems of video conferencing. They included skype, VRVS, EVO, WEBEX, marratech
and adobe Connect. In the result of these tests the adobe Connect system
was chosen and properly acquired. Its advantages are: relatively easy use for participants,
a possibility to make presentation in a video contact between presenter and audience,
a possibility to make high quality records and edit them, removing from records occasional
and rather rare disturbances of sound or connection, to use a whiteboard facility for discussions,
the option to open desktop and to work online with texts in any format.
The regular form of VIA meetings assumes that their time and Virtual room are announced in advance.
Since the access to the Virtual room is strictly controlled by administration,
the invited participants should enter the Room as Guests, typing their names,
and their entrance and successive ability to use video and audio system is authorized by the Host
of the meeting.
The format of VIA lectures and discussions is shown on Fig.~\ref{mk2viab}.

The ppt file of presentation is uploaded in the system in advance and then demonstrated in the central window.
Video images of presenter and participants appear in the right window, while in the lower left window the list of all the attendees
is given. To protect the quality of sound and record, the participants are required to switch out their audio system during presentation
and to use upper left Chat window for immediate comments and urgent questions. The Chat window can be also used by participants, having no microphone,
 for questions and comments during Discussion. In the end of presentation the central window can be used for a whiteboard utility
 as well as the whole structure of windows can be changed, e.g. by making full screen the window with the images of participants of discussion.

\section{\label{mk2viaBled} VIA Discussion Session at Bled Workshop}

\begin{figure}
    \begin{center}
        \includegraphics[scale=0.38]{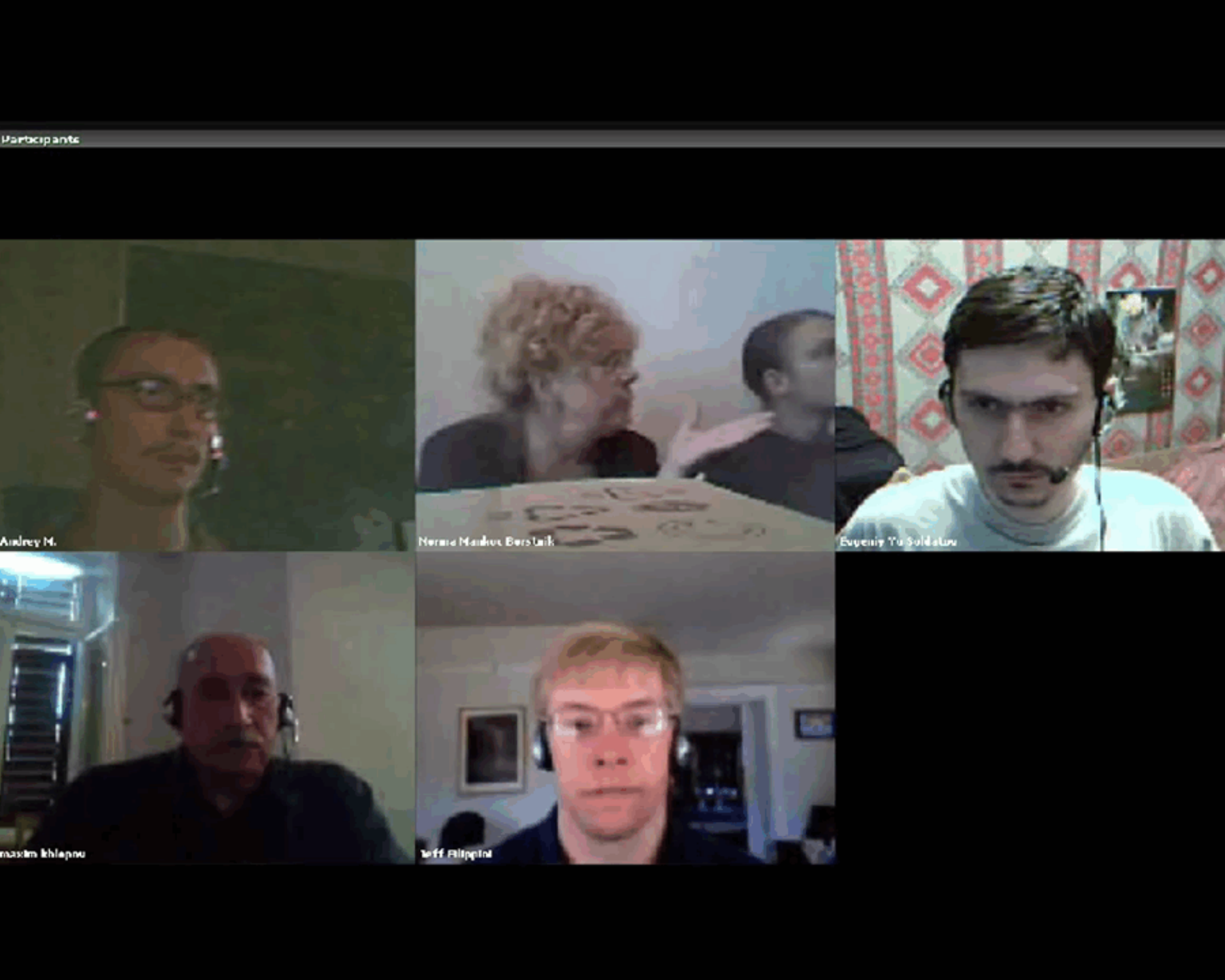}
        \caption{Bled Conference Discussion Bled-Moscow-Berkeley}
        \label{mk2viac}
    \end{center}
\end{figure}

VIA discussion session took place in the framework of Bled Workshop on 24 July. It contained presentations: by N.S. Manko\v c Bor\v stnik about dark matter candidates
following from her approach, unifying spins and charges, by G. Bregar about the possibility to explain the results of dark matter searches by some of these candidates,
M. Yu. Khlopov about the composite dark matter scenario and by A.Mayorov and E. Soldatov about the application of this scenario to solution of the controversy between
the results of DAMA and CDMS experiments. The content of these presentations can be found in the contributions \cite{mk2viaBled}. To clarify the possibilities
to explain the positive results of DAMA/NaI and DAMA/Libra experiments without contradiction with strong constraints, which follow from the results of
CDMS experiment, J.Filippini (Berkeley,USA) from CDMS collaboration took part in Discussion (Fig. \ref{mk2viac}).
His arguments can be found in the Discussion section of this volume.

In spite of technical problem for some participants the main root of virtual meeting, which was organized in Bled,Slovenia
and involved J.Filippini in Berkeley, USA and A.Mayorov, E.Soldatov and other participants from Moscow, Russia was stable
during all the 3 hours of the video conference.
\section{Conclusions}
The exciting experiment of VIA Discussion at Bled Workshop, the three days of permanent online transmissions
and distant participation in the Gran Sasso Summer Institute on Astroparticle physics, the VIA interactive form of Seminars in Moscow and
in Pisa with participation and presentation on distance, the stable
regular weekly video conferences with VIA lectures and the solid library of their records and presentations,
creation of multi-lingual VIA Internet forum, regular basic courses and individual work on distance with students of MEPhI
prove that the Scientific-Educational complex of Virtual Institute of Astroparticle physics can provide regular
communications between different groups and scientists, working in
different scientific fields and parts of the world, get the first-hand information on the newest scientific results,
as well as to support various educational programs on distance. This activity
would easily allow finding mutual interest and organizing task
forces for different scientific topics of astroparticle physics. It
can help in the elaboration of strategy of experimental particle,
nuclear, astrophysical and cosmological studies as well as in proper
analysis of experimental data. It can provide young talented people from all over the world to get the highest level education,
come in direct interactive contact with the
world known scientists and to find their place in the fundamental research. To conclude the VIA complex is in operation and
ready for a wide use and applications.
\section*{Acknowledgements}
 The initial step of creation of VIA was
 supported by ASPERA. I am grateful to S.Katsanevas for permanent stimulating support, to P.Binetruy, J.Ellis, F.Fidecaro,
 M.Pohl for help in development of the project, to K.Belotsky and K.Shibaev for assistance in educational
 VIA program and to A.Mayorov and E.Soldatov for cooperation in its applications, to Z.Berezhiani and all Organizers of XIII Summer School on Astroparticle Physics in Gran Sasso
 for cooperation in the use of VIA for online distant participation and to D.Rouable for help in technical realization and support of VIA complex.
 I express my gratitude to N.S. Manko\v c Bor\v stnik, G.Bregar, D. Lukman and all
 Organizers of Bled Workshop for cooperation in the exciting experiment of
 VIA Discussion Session at Bled.


\backmatter

\thispagestyle{empty}
\parindent=0pt
\begin{flushleft}
\mbox{}
\vfill
\vrule height 1pt width \textwidth depth 0pt
{\parskip 6pt

{\sc Blejske Delavnice Iz Fizike, \ \ Letnik~9, \v{s}t. 2,} 
\ \ \ \ ISSN 1580-4992

{\sc Bled Workshops in Physics, \ \  Vol.~9, No.~2}

\bigskip

Zbornik 11. delavnice `What Comes Beyond the Standard Models', 
Bled, 15.~-- 25.~julij 2008

Proceedings to the 11th workshop 'What Comes Beyond the Standard Models', 
Bled, July 15.--25.,  2008

\bigskip

Uredili Norma Manko\v c Bor\v stnik, Holger Bech Nielsen in Dragan Lukman 

Publikacijo sofinancira Javna agencija za raziskovalno dejavnost Republike Slovenije 

Brezpla\v cni izvod za udele\v zence 

Tehni\v{c}ni urednik Vladimir Bensa

\bigskip

Zalo\v{z}ilo: DMFA -- zalo\v{z}ni\v{s}tvo, Jadranska 19,
1000 Ljubljana, Slovenija

Natisnila BIROGRAFIKA BORI v nakladi 150 izvodov

\bigskip

Publikacija DMFA \v{s}tevilka 1724

\vrule height 1pt width \textwidth depth 0pt}
\end{flushleft}


\end{document}